%% file: main_arxiv.tex
\documentclass[a4paper,12pt]{article}

\usepackage{booktabs}
\usepackage[authoryear,longnamesfirst]{natbib}

\usepackage{amsmath,amssymb,amsfonts}
\usepackage{amsthm}
\usepackage{fancyvrb}
\usepackage{float}
\usepackage{bbm}
\usepackage{amsfonts}
\usepackage{graphicx}
\usepackage{url}
\usepackage{tikz}
\usepackage{listings}
\usepackage{matlab-prettifier}
\usepackage{mathtools}
\usepackage[ruled,vlined,linesnumbered]{algorithm2e}
\usepackage[noend]{algpseudocode}
\usepackage{adjustbox}
\usepackage{hyperref}
\hypersetup{
    colorlinks=true,
    linkcolor=cyan,
    citecolor=cyan,
    urlcolor=blue
}
\usepackage{cleveref}
\usepackage[figuresright]{rotating}
\usepackage{float}
\usepackage{longtable}
\usepackage{caption}

\DeclareMathOperator*{\argmax}{arg\,max}

\newtheorem{example}{Example}
\newtheorem{definition}{Definition}
\newtheorem{corollary}{Corollary}
\newtheorem{theorem}{Theorem}
\newtheorem{lemma}{Lemma}
\newtheorem{remark}{Remark}
\newcommand{\revnew}[1]{{\leavevmode\color{black}#1}}

\usepackage{geometry}
\usepackage{fancyhdr}
\begin{document}

\title{A Branch-Price-Cut-And-Switch Approach for Optimizing Team Formation and Routing for Airport Baggage Handling Tasks with Stochastic Travel Times}

\author{
  Andreas Hagn\thanks{Corresponding author. \texttt{andreas.hagn@tum.de}} \and
  Rainer Kolisch\thanks{\texttt{rainer.kolisch@tum.de}} \and
  Giacomo Dall'Olio\thanks{\texttt{giacomo.dallolio@tum.de}} \and
  Stefan Weltge\thanks{\texttt{weltge@tum.de}}
  \\[0.5em]
  \normalsize Technical University of Munich
}

\date{}

\maketitle

\begin{abstract}
In airport operations, optimally using dedicated personnel for baggage handling tasks plays a crucial role in the design of resource-efficient processes. Teams of workers with different qualifications must be formed, and loading or unloading tasks must be assigned to them. Each task has a time window within which it can be started and should be finished. Violating these temporal restrictions incurs severe financial penalties for the operator. In practice, various components of this process are subject to uncertainties. We consider the aforementioned problem under the assumption of time-dependent stochastic travel times across the apron. We present two binary program formulations to model the problem at hand and propose a novel solution approach that we call Branch-Price-Cut-and-Switch, in which we dynamically switch between two master problem formulations. Furthermore, we use an exact separation method to identify violated rank-1 Chvátal-Gomory cuts and utilize an efficient branching rule relying on task finish times. We test the algorithm on instances generated based on real-world data from a major European hub airport with a planning horizon of up to two hours, 30 flights per hour, and three available task execution modes to choose from.
Our results indicate that our algorithm is able to significantly outperform existing solution approaches. Moreover, an explicit consideration of stochastic travel times allows for solutions that utilize the available workforce more efficiently, while simultaneously guaranteeing a stable service level for the baggage handling operator.

\end{abstract}

% Main text
\input{sections/introduction}

\input{sections/problem_description}

\input{sections/literature_review}

\input{sections/models}

\input{sections/solution_approach}

\input{sections/computational_study}

\input{sections/conclusion}

\input{sections/acknowledgements}

\clearpage

% Credit authorship details can be added here if desired

\appendix
\input{sections/appendix}

\bibliographystyle{plainnat}

\bibliography{cas-refs}



\end{document}

%% file: sections/introduction.tex
\section{Introduction}
Baggage handling tasks, namely loading and unloading containers and bulk luggage, play an important role in aviation and airport operations. They are one of several \textit{ground handling tasks} \citep{evler2021airline}. Baggage handling is usually performed by dedicated personnel using dedicated equipment. Furthermore, there are no dependencies between baggage handling and other ground handling tasks, making it possible to consider baggage (un)loading in isolation from all other ground handling tasks. Typically, a central planner needs to assemble worker teams and decide which aircraft they (un)load and in which sequence they do so.\\
Workers have different skill levels, i.e., qualifications for the types of equipment, such as tractors or high cargo loaders, they are allowed to operate. Depending on the airplane model and type, various equipment compositions can be used to execute the (un-)loading task, leading to different workforce requirements. For instance, large-sized airplanes typically have at least two cargo holes. These can be unloaded either sequentially or in parallel. Each aircraft can be (un)loaded in several such modes, where the use of additional equipment implies the requirement of additional workers and typically leads to shorter execution times. Such equipment compositions are also called \textit{modes} or \textit{(team) formations} and are derived from standardized (un)loading procedures, which are defined in the respective aircraft manual. Each flight is assigned a time window, within which the baggage handling needs to take place. A central planner can reduce the execution time of a task by selecting a mode that either requires more workers in general or that requires higher-skilled workers. Because workforce is a scarce resource in baggage handling, the selection of the fastest possible mode for each incoming and outgoing flight is typically not feasible. At the same time, selecting slow modes too frequently can lead to the violation of time windows, causing departure delays and delayed baggage pick-up for arriving flights. Because such disruptive delays can significantly reduce customer satisfaction and incur financial penalties for the baggage handling operator, proper team formation and sequencing decisions by the central planner are crucial. Thus, a well-informed and efficient assignment of workers to flights and sequencing of flights can greatly benefit customer satisfaction and reduce fines paid by the operator. In the following, whenever loading tasks are discussed, both loading and unloading processes are encompassed.\\
\cite{coda2023q2} state that around 30\% of all aircraft delays are caused by airline-related issues, to which baggage handling processes are attributed. Furthermore, depending on the airport and the considered planning horizon, baggage handling is subject to significant uncertainties regarding travel times of ground vehicles. Delays while traveling from one parking position to another, e.g., because of planes crossing the apron, are quite frequent and typically accumulate to a significant extent. In the following, we consider loading times to be deterministic, while travel times are assumed to be stochastic with known probability distributions. Moreover, to limit the potential financial penalties for the baggage handling operator, it is reasonable to demand that each task's time window is satisfied with at least a predefined probability.\\
If the decision of how to form teams and how to assign them to incoming or outgoing flights is done improperly, one might end up with a shortage of workers or unnecessarily long execution times. This in turn causes delays within the baggage handling process, which frequently propagate for several hours and lead to increased waiting times at the baggage claim area or departure delays. Both of these inefficiencies significantly impact passenger satisfaction.  Optimizing the planning and processing of baggage (un)loading tasks can help to avoid excessively large delays or even eliminate them completely. For this reason, we consider the problem of forming and routing worker teams for baggage handling tasks under stochastic travel times, such that each task is executed by exactly one team and the available workforce is not exceeded. The objective function minimizes the sum of weighted expected finish times of all tasks and a penalty term for exceeding time windows. While the former aims to introduce buffers that can be used to compensate unforeseen delays during baggage handling, the latter considers financial fines paid by the baggage handling operator in case of delays, as well as an increasing chance of passengers missing connecting flights.\\
% Furthermore, we assume that travel times across the apron are stochastic to obtain a model that can be used to depict processes as realistically as possible.\\
Our work builds upon \cite{dallolio_temp}, which addresses the deterministic team formation and routing problem. The main contributions of this paper are:
\begin{itemize}
    \item[i.] We extend previous works on baggage handling optimization by including stochastic travel times.
    \item[ii.] We propose a novel Branch-Price-Cut-and-Switch solution approach that dynamically switches between two master problem formulations, depending on the solution's characteristics.
    \item[iii.] We conduct extensive experimental studies to analyze the impact of stochastic information on optimal solutions and assess our solution method's efficiency.
\end{itemize}
The remainder of this paper is structured as follows. Section \ref{section:problem_description} provides a detailed problem description with the help of mathematical notation. Section \ref{section:literature_review} reviews the most relevant literature for our study. In Section \ref{section:models}, we propose two binary programs that are used to model the problem at hand. In Section \ref{section:solution_approach}, we develop a Branch-Price-Cut-and-Switch solution approach and elaborate on its core components. Section \ref{section:experimental_study} presents computational experiments aiming at assessing the proposed algorithm's performance and the impact of stochasticity on optimal solutions. We summarize our findings in Section \ref{section:conclusion} and present several areas of future research.

%% file: sections/problem_description.tex
\section{Problem Description}\label{section:problem_description}
In the following section, we provide a detailed description of the problem considered in this publication with the help of mathematical notation, which will be used throughout the following sections. An overview over all used notation can be found in \Cref{appendix:notation}.

\paragraph{\textbf{General Notation}}\text{ }\\
We consider a set $\mathcal{I} = \{1,\dots,\vert\mathcal{I}\vert\}$ of tasks that have to be performed within a specified planning horizon. The planning horizon is discretized into a set $\mathcal{T}$ of equidistant time steps. Each task corresponds to either a loading or unloading task of an incoming (or outgoing) flight. We note that, in the context of VRPs, tasks are typically referred to as customers or customer visits. A time window $[\mathrm{ES}_i, \mathrm{LF}_i]$ is associated with each task $i\in\mathcal{I}$, describing the earliest and latest point at which service on an airplane can be started and should be finished, respectively.\\
Initially, a workforce of prespecified size is located at a central depot denoted by $d$. In general, workers can be subdivided into \textit{skill levels} $\mathcal{K}\coloneqq\{1,\dots,K\}$. Depending on their skill level, workers are only allowed to operate certain equipment. For instance, aircraft typically carry bulk luggage, containers, and/or pallets. While skill level $1$ only encompasses basic apron clearance, skill level $2$ allows for the operation of loader-transporters to unload containers. Heavy equipment used to unload multiple containers or pallets at once, such as high-loaders, requires a qualification of level $3$.
Skill levels are ordered hierarchically, i.e., a worker with skill level $k\in\mathcal{K}$ is able to operate any equipment requiring level $l$ not greater than $k$. Moreover, the amount of workers with skill level greater than or equal to $k$ that are available at the worker depot is fixed and defined as $N_k \coloneqq \underset{l = k}{\overset{K}{\sum}} N_l^{\mathrm{D}}$, where $N_k^{\mathrm{D}} \in\mathbb{N}_{>0}$ is the number of available workers with skill level $k\in\mathcal{K}$. We note that in practical instances, higher-skilled workers tend to be rather scarce and hard to obtain as they require lengthy and expensive training. A route is a sequence $r\coloneqq\left(d,v_1,\dots,v_l,d\right)$ with $l\geq 1$ and $v_i\in\mathcal{I}$ for all $i=1,\dots,l$. The set of tasks executed by route $r$ is denoted by $\mathcal{I}^r \coloneqq \{v_1,\dots,v_l\}$. Furthermore, each route $r$ is assigned a profile $q = \left(\xi_{q,k}\right)_{k\in\mathcal{K}}\in\mathcal{Q}$ which is used to execute the route, where $\xi_{q,k}\in\mathbb{N}_{\geq0}$ denotes the number of required workers of skill level greater than or equal to $k$ and $\mathcal{Q}$ is the set of available profiles. Because each type of aircraft requires different prerequisites and equipment for baggage handling tasks, only a subset $\mathcal{Q}_i\subseteq\mathcal{Q}$ of profiles can be used to execute a given task $i\in\mathcal{I}$. For example, one can choose to either unload multiple cargo holes sequentially or in parallel, yielding different workforce requirements and thus different profiles. For larger aircraft, one can also choose between the usage of loader-transporters or high-loaders, leading to additional suitable profiles.
Depending on the profile $q\in\mathcal{Q}_i$, executing task $i$ takes $p_{i,q}\in\mathbb{N}_{>0}$ units of time. The execution time of a task depends on several external factors, such as the number of containers, pallets, and the amount of bulk luggage that are to be handled. Each team route $(r,q)\in\mathcal{R}$, where $\mathcal{R}$ is the set of tuples of routes and suitable profiles, has time steps $\mathrm{tl}^{r,q}\in\mathcal{T}$ and $\mathrm{tr}^{r,q}\in\mathcal{T}$ at which the team leaves the depot and returns to it, respectively.
\vspace{0.7em}\paragraph{\textbf{Profiles and Skill Compositions}}\text{ }\\
By construction, a profile $q\in\mathcal{Q}$ is well-defined by its aggregated workforce requirements, i.e., lower bounds for the skill levels of required workers, but does not contain any information about their actual skill levels. To include this information, we define a \textit{skill composition} as a vector $s\in\mathbb{N}^{\vert\mathcal{K}\vert}$ and say that a skill composition $s=(s_{q,k})_{k\in\mathcal{K}}$ can be assigned to a profile $q\in\mathcal{Q}$ if
\begin{align}
    \xi_{q,k^{*}} &= \underset{k\in\mathcal{K}}{\sum} s_{q,k} \quad \text{and}\label{definition_skill_comp_sum}\\
    \xi_{q,k'} &\leq \underset{k\in\mathcal{K}: k \geq k'}{\sum} s_{q,k} \quad \forall k'\in\mathcal{K}\label{definition_skill_comp_leq}
\end{align}
hold, where $k^{*} = 1$ is the skill level with the lowest qualification. Furthermore, the set
\begin{equation*}
    \mathcal{S}_q \coloneqq \left\{(s_{q,k})_{k\in\mathcal{K}}: \ (s_{q,k})_{k\in\mathcal{K}} \ \text{satisfies} \ (\ref{definition_skill_comp_sum})-(\ref{definition_skill_comp_leq})\right\}
\end{equation*}
consists of all skill compositions that can be assigned to profile $q\in\mathcal{Q}$. Hence, we define the set
\begin{equation*}
    \mathcal{Q}^{\mathrm{D}} \coloneqq \left\{(q,s): q\in\mathcal{Q}, s\in\mathcal{S}_q\right\}.
\end{equation*}
as the set of \textit{disaggregated profiles}, i.e., profiles enhanced by information on the skill levels of workers used to assemble a team with profile $q$. Given a route $r$, we define the set of profiles that can be used to execute $r$ as $\mathcal{Q}^r\coloneqq \underset{i\in\mathcal{I}^r}{\bigcap}\mathcal{Q}_i$.
\begin{example} 
Consider the following example of a team route $(r,q)\in\mathcal{R}$ with $r\in\mathcal{R}$, $q\in\mathcal{Q}^r$ . We define $\mathcal{K}\coloneqq\{1,2,3\}$ and profile $q$ by
\begin{align*}
    \xi_{q} \coloneqq \left(\xi_{q,k})\right)_{i=1,2,3} = \left(
\begin{array}{c}
3\\
2\\
1\\
\end{array}
\right).
\end{align*}
Furthermore, we assume that we have an unlimited workforce available for each skill level. In practice, a team of profile $q$ could operate as follows. First, one worker operates a high-loader, which unloads and moves containers onto dollies. This requires skill level $3$, as the handling of high-loaders requires lengthy prior training. Second, one worker moves a conveyor belt into position and unloads bulk luggage into dollies. Because conveyor belts are significantly easier to handle than high-loaders, a skill level of at least $2$ is required for this job. Third, one worker unloads bulk luggage from the aircraft onto the conveyor belt. This is a rather simple task and can be executed by any worker of skill level $1$ or higher.
\begin{table}[ht]
\centering
\caption{Set of possible skill compositions for profile $q$}
\label{table:example_skill_comps}
\begin{adjustbox}{width=0.3\textwidth}
\small
\begin{tabular}{llllll}
\toprule
                                 & $s_1$ & $s_2$ & $s_3$ & $s_4$ & $s_5$ \\ \midrule
$s^i_{q,1}$ & 1     & 0     & 0     & 1     & 0     \\
$s^i_{q,2}$ & 1     & 2     & 1     & 0     & 0     \\
$s^i_{q,3}$ & 1     & 1     & 2     & 2     & 3     \\ \bottomrule
\end{tabular}
\end{adjustbox}
\end{table}
Table \ref{table:example_skill_comps} contains the set $\mathcal{S}_q = \left\{s_1,s_2,s_3,s_4,s_5\right\}$ of skill compositions that can be used to assemble a team with profile $q$. We note that each skill composition $s_i$ is uniquely defined by its worker usage $\left(s^i_{q,k}\right)_{k=1,2,3}$ per skill level. For instance, the third column in Table \ref{table:example_skill_comps} describes skill composition $s_2$, which uses two workers of skill level 2 and one worker of skill level 3 to assemble a team with profile $q$. Note that this implies that one worker with level $2$ is assigned the job of unloading bulk luggage from the aircraft onto the conveyor belt, i.e., he is downgraded to a job below his actual skill level. Furthermore, a team with profile $q$ always occupies the same workforce on an aggregated level, i.e., such a team always requires 3 workers of at least level 1, 2 workers of at least level 2, and one worker of level 3. However, depending on the selected skill composition, the workforce occupation per individual level differs. For instance, while $s_4$ occupies two workers of level 3, $s_5$ occupies only one worker of level 3. From a modelling point of view, considering workforce requirements on an aggregated level is often sufficient to create a feasible plan. However, on some occasions, this aggregated consideration leads to an infeasible plan, which can be handled by explicitly specifying and considering skill compositions. In total, team route $(r,q)$ can be used to derive five unique team routes $\left\{(r,q,s_i): \ i=,1\dots, 5\right\}$ in $\mathcal{R}^{\mathrm{D}}$. Thus, the number of skill compositions typically significantly exceeds the number of profiles.
\end{example}

\vspace{0.7em}\paragraph{\textbf{Travel Times}}\text{ }\\
After finishing a task $i$, a team with profile $q$ can either return to the depot $d$ and become available for regrouping or directly continue with another task $j$. Traveling between locations $i,j \in \mathcal{I}\cup\left\{d\right\}$, where $i$ and $j$ are either two tasks or one task and the depot, takes at least $t_{i,j}^{\mathrm{det}}\in\mathbb{N}_{\geq0}$ units of time and does not depend on the profile. We assume that travel times are stochastic and time-dependent. For this purpose, we decompose the planning horizon into subsequent \textit{time bins} of equal length. Formally, we define a disjoint decomposition $\mathcal{B} \coloneqq \left\{B_1,\dots,B_m\right\}$ with $m\in\mathbb{N}$ into sets $B_i \subset \mathcal{T}$  of equal cardinality and length. For instance, given a 2 hour planning horizon $\mathcal{T} = \{1,\dots,120\}$ and a time bin length of 15 minutes, a decomposition is then given by $\mathcal{B}_i = \{t_{15i+1},\dots,t_{15i+15}\}$ for $i=0,\dots,7$. Given a time bin $\mathcal{B}_i\in\mathcal{B}$, we assume that travel times are independent. Note that travel times across different time bins are, in general, dependent. Practically speaking, a time bin encompasses a period of stable traffic across the apron, in which uncertainties on travel times predominantly arise from unsystematic causes that affect individual pathways. Such independent delays can, for instance, be caused by aircraft crossing the apron or local traffic congestion.
In such situations, a common approach is to solve a multi-stage stochastic problem, where an initial solution is constructed and, once more information on stochastic parameters such as travel times becomes available over time, is readjusted if needed. However, as ground vehicle traffic is rather unpredictable and an aircraft's movements are typically only determined moments before the plane starts moving, it is not possible to avoid the usage of congested parts of the apron by re-adjusting the previously derived solution, i.e., by changing the sequences of tasks that should be executed by the same team or assigning a new team to a given task. To formalize this, we define non-negative, finite supports for travel time delays for each time bin $\mathcal{B}_k\in\mathcal{B}$ and each pair $(i,j)\in E$ of tasks (or tasks and the depot) by
\begin{equation*}
    B_{i,j}^k\subset \mathbb{N}_{\geq0}, \ \vert B_{i,j}^k\vert < \infty
\end{equation*}
where $E\coloneqq \left\{(i,j): i,j\in\mathcal{I}\cup\{d\}, \ i \neq j\right\}$. For each $\mathcal{B}_k\in\mathcal{B}$, we then define the set of possible travel time delays by
\begin{equation*}
    \Omega^k \coloneqq \underset{(i,j)\in E}{\prod} B_{i,j}^k.
\end{equation*}
Then, the vector of stochastic travel times $ t^k: \ \Omega^k \rightarrow \mathbb{N}_{>0}^{\vert E \vert}$ is given by 
\begin{align*}
    t^k(\omega) &= t^{\mathrm{det}} + \omega \quad \forall\omega \in \Omega^k
\end{align*}
for all $(i,j)\in E$ and $\mathcal{B}_k\in\mathcal{B}$. Consistently, we can represent $t^k(\omega)$ as $t^k(\omega)=(t^k_{i,j}(\omega))_{(i,j)\in E} = (t_{i,j}^{\mathrm{det}}+\omega_{i,j})_{(i,j)\in E}$. Given $k\in\{1,\dots,\vert\mathcal{B}\vert$, we assume that events in $\Omega^k$ are independent, thus $(t^k_{i,j})_{(i,j)\in E}$ are independent, non-negative random variables with finite support and a known probability distribution $\Xi_{i,j}^k: \mathbb{N}_{\geq 0} \rightarrow [0,1]$ with $\Xi^k_{i,j}(t) \coloneqq \mathbb{P}(t^{det}_{i,j} + \omega = t)$ and $\underset{t\in \mathbb{N}_{\geq0}}{\sum} \Xi^k_{i,j}(t) = 1$. In the following, $t_{i,j}^{\mathrm{det}}$ is also called \textit{deterministic} or \textit{best-case travel time}.  Note that for time bins $k,\bar{k}\in\{1,\dots\vert\mathcal{B}\vert\}$ with $k\neq\bar{k}$, $t^k_{i,j}$ and $t^{\bar{k}}_{i^*,j^*}$ can be dependent for any $(i,j),(i^*,j^*)\in E$. Additionally, given an arbitrary $\gamma\in[0,1]$, we denote the $\gamma$-quantile scenario of distributions $(t_{i,j}^k)_{(i,j)\in E}$ by
\begin{equation*}
    \omega_{\gamma}^k \coloneqq \underset{(i,j)\in E}{\prod} \min\left\{\omega_{i,j}^k : \omega_{i,j}^k \in \Omega_{i,j}^k \ \land \ \mathbb{P}(t_{i,j}^k \leq \omega_{i,j}^k) \geq \gamma \right\},
\end{equation*}
Moreover, we define $\omega_{\gamma} \coloneqq \left(\omega_{\gamma}^k\right)_{k: \mathcal{B}_k\in\mathcal{B}}$ and the \textit{worst-case scenario} of travel times by
\begin{equation*}
    \omega_{\mathrm{max}} = \left(\max\{\omega_{i,j}^k: \omega_{i,j}^k\in B^k_{i,j}\}\right)_{(i,j)\in E, k=1,\dots,\vert \mathcal{B}\vert}
\end{equation*}
In the following, the finish times in scenario $\omega_y$ are also called \textit{$\omega_{\gamma}$-scenario finish times}
\vspace{0.7em}\paragraph{\textbf{Routes and Route Feasibility}}\text{ }\\
If a worker team arrives at task $i$ before its time window opens, it has to wait until its earliest start $\mathrm{ES}_i$ to start the service. Because travel times are stochastic, the start and finish times are random variables, which depend on the worker profile $q\in\mathcal{Q}$ used to execute the route, the sequence $r$ of tasks, and a realization of travel times. Formally, given a team route $(r,q)\in\mathcal{R}$ and its median finish time $\bar{F}_{i-1}^{r,q}$ at task $i-1\in\mathcal{I}$, we define the start and finish times $S_i^{r,q}: \Omega^{\bar{F}_{i-1}^{r,q}} \rightarrow \mathbb{N}$ and $F_i^{r,q}: \Omega^{\bar{F}_{i-1}^{r,q}} \rightarrow \mathbb{N}$ at the subsequent task $i\in\mathcal{I}$ as
\begin{align*}
    S_i^{r,q}(\omega) &\coloneqq \max\{F_{i-1}^{r,q}(\omega) + t_{i-1,i}^{\bar{F}_{i-1}^{r,q}}(\omega), \mathrm{ES}_i\}, \quad \text{and}\\
    F_i^{r,q}(\omega) &\coloneqq S_i^{r,q}(\omega) + p_{i,q}, 
\end{align*}
where $F_{d}^{r,q}(\omega) = \mathrm{tl}^{r,q}$ is equal to the depot leave time of team route $(r,q)$ and $i-1$ is the predecessor of task $i$ along $r$. Here, $\Omega^{\bar{F}_{i-1}^{r,q}}$ is the distribution of travel times in time bin $\mathcal{B}_k\in\mathcal{B}$ with $\bar{F}_{i-1}^{r,q} \in \mathcal{B}_k$. Note that the above definition of $S_i^{r, q}$ uses the same travel time distribution based on $\bar{F}_{i-1}^{r,q}$ for all events $\omega\in\Omega^{\bar{F}_{i-1}^{r,q}}$ rather than event-dependent distributions based on $F_{i-1}^{r,q}(\omega)$. While this choice poses a slight simplification of reality, it allows for efficient computation of distributions via convolutions and stronger comparability of task sequences. Allowing the current time bin to depend on the realized finish time would require tracking joint distributions over finish times and time bins, leading to a combinatorial explosion in the state space and rendering an exact solution approach as proposed in the subsequent sections intractable. Finally, we note that, in contrast to arbitrary quantiles of finish times, the usage of median finish times in the above context is unbiased towards earlier or later bins and thus reduces the probability of selecting a current time bin that is equal to the true underlying bin only for very few, i.e., outlier events.
Furthermore, we denote by $\mathrm{tr}^{r,q}$ return time of team route $(r,q)$, which is realized when case scenario $\omega_\gamma$ occurs.\\
As the actual travel times are random and hardly predictable beforehand, finishing tasks after the time window closes, i.e., after $\mathrm{LF}_i$, is sometimes inevitable. Nevertheless, baggage handling operators often have to guarantee a minimum service level, i.e., the probability of a flight being delayed due to improper baggage handling. Moreover, in order to limit the probability of passengers missing their connecting flights, airlines often specify maximum acceptable delays in case of disruptions. Therefore, we allow tasks to be finished after their time window closes, while we limit said violation to at most a fixed amount of time, leading to a new \textit{extended latest finish time} $\mathrm{LF}_{i}^{\mathrm{e}} \geq \mathrm{LF}_{i}$ for each task $i\in\mathcal{I}$. Furthermore, to limit potential financial damages and customer dissatisfaction, we introduce a minimum service level $\alpha\in [0,1]$ and demand that the finish time $F_i^{r,q}$ of each task $i\in\mathcal{I}$ on each route $r$ must satisfy
\begin{align}
    &\mathbb{P}(F_i^{r,q} \leq \mathrm{LF}_i) \geq \alpha\label{chance_constraint},\\
    &\mathbb{P}(F_i^{r,q} > \mathrm{LF}_i^{\mathrm{e}}) = 0 \label{finish_at_lf_v}.
\end{align}
Constraint (\ref{chance_constraint}), also called chance constraint or service level constraint, limits the probability of delays caused by the baggage handling operator, while constraint (\ref{finish_at_lf_v}) guarantees that potential delays do not exceed a prespecified limit.
A route $r$ with profile $q$ is called \textit{feasible} if (\ref{chance_constraint})--(\ref{finish_at_lf_v}) are satisfied for all $i\in\mathcal{I}^r$. We note that, for planning horizons of up to a few hours, the arrival times of aircraft are typically subject to very little uncertainty as there are no major causes for delay once a plane has departed from its origin airport. Moreover, delays of departing flights, i.e., loading tasks, are either known far in advance or realize shortly before the time window opens. In the former case, delays can be incorporated into the definition of time windows. In the latter case, one is typically unable to readjust the current plan as the loading has already begun and, if said delays were modeled as stochastic time windows, one might obtain overly robust solutions due to the low probability of such events happening. Therefore, the assumption of deterministic time windows is reasonable for short- to medium-term planning horizons. Finally, earliest start times for incoming aircraft and latest finish time for outgoing aircraft are determined by the arrival time at the gate and the scheduled departure time, respectively. Latest finish times for incoming aircraft are defined based on baggage transfer time requirements alongside maximum baggage claim durations, while earliest start times for outgoing flights depend on the aircraft's earliest readiness time.

\vspace{0.7em}\paragraph{\textbf{Route Costs and Objective Function}}\text{ }\\
Each task $i\in\mathcal{I}$ has an assigned weight $w_i\geq 0$ that indicates its importance in the flight schedule. This can be, for instance, the number or percentage of passengers that have to reach a connecting flight at the destination airport. The expected cost of a team route $(r,q)\in\mathcal{R}$) is then given by
\begin{equation*}
    \mathbb{E}(c^r) = \underset{i\in\mathcal{I}^r}{\sum} w_i\cdot \mathbb{E}\left(\left(F_i^{r,q} - \mathrm{EF}_i\right) + P_i(F_i^{r,q})\right)
\end{equation*}
where $\mathrm{EF}_i \coloneqq \mathrm{ES}_i + \min\{p_{i,q}: \ q\in\mathcal{Q}_i\}$ is the earliest possible finish time of task $i\in\mathcal{I}$ and $P_i: (\mathrm{LF}_i,\mathrm{LF}_i^{\mathrm{e}}]\rightarrow \mathbb{R}_{>0}$ is a penalty function. The first part of the objective function aims to minimize weighted expected finish times in order to have as large safety time buffers as possible to absorb delays in other parts of the baggage handling process. The second part consists of a penalty for delaying flights. Because the probability of passengers missing connecting flights increases superlinearly in the delay of a given baggage handling task, the baggage handling operator aims to keep the variance of delays as low as possible. As the variance can be seen as a quadratic function of the delay, we use a quadratic penalty function
\begin{equation*}
    P_i(F_i^{r,q}(\omega)) \coloneqq \begin{cases}
    \begin{aligned}
        &(F_i^{r,q}(\omega) - \mathrm{LF}_i)^2 && \text{if} \ \ F_i^{r,q}(\omega) > \mathrm{LF}_i,\\
        &0 && \text{otherwise}
    \end{aligned}
    \end{cases}
\end{equation*}
for all $\omega\in\Omega$. In order to calculate the expected cost of a route $r = (d,v_1,\dots,v_l,d)$ with $l\geq 1$, it is necessary to have full information about the finish time distribution of each task assigned to $r$. Let $i,j\in \mathcal{I}$ be two tasks that are executed consecutively in route $r$, i.e., there exists an $h\in\{1,\dots,l-1\}$ with $v_h = i$ and $v_{h+1} = j$. If the distribution of $F_i^{r,q}$ is known, the start time distribution of the subsequent task $S_j^{r,q}$ can be calculated as
\begin{equation}
    \mathbb{P}(S^{r,q}_j = \tau) = \begin{cases}
        \begin{aligned}
            &\underset{z=0}{\overset{\mathrm{ES}_{j}}{\sum}} \mathbb{P}(F^{r,q}_{i} + t_{i,j}^{\bar{F}_i^{r,q}} = z) && \text{if} \ \ \tau = \mathrm{ES}_{j}\\
            &\mathbb{P}(F^{r,q}_i + t_{i,j}^{\bar{F}_i^{r,q}} = \tau) && \text{if} \ \ \tau > \mathrm{ES}_j\\
            &0 && \text{otherwise}
        \end{aligned}
    \end{cases}\label{calcul_start_time_distr}
\end{equation}
The finish time distribution $F_j^{r,q}$ can then be obtained because $F_j^{r,q}(\omega) = S_j^{r,q}(\omega) + p_{j,q}$ holds by definition for all $\omega\in\Omega^k$ and $k=1,\dots,\vert B\vert$.\\

%% file: sections/literature_review.tex
\section{Literature Review}\label{section:literature_review}
The problem considered in this paper can be seen as a variant of the technician scheduling and routing problem (TSRP). TSRPs typically consist of a routing part that can be seen as a vehicle routing problem (VRP) and a scheduling or team formation part. The solution approach proposed in the following is an exact approach and works with arbitrary distributional structures.
Section \ref{section:lit_rev_vrps} focuses on literature dealing with stochastic VRPs in a general context, with a particular emphasis on assumptions on the distributional structure of random variables and the assumption of independence.
Section \ref{section:lit_rev_tsrp} considers publications dealing with stochastic TSRP variants.

\vspace{0.5cm}
\subsection{Literature on Stochastic Vehicle Routing}\label{section:lit_rev_vrps}
Recently, stochastic formulations of the vehicle routing problem with time windows (VRPTW) have seen a noticeable interest increase. In the following, we provide an overview of literature dealing with stochastic variants of the VRPTW. Because distributional assumptions and assumptions regarding independence of random variables have a significant impact on the computational tractability of exact approaches, as well as the practical applicability of results, we especially focus on these two aspects. \cite{oyola2018stochastic} summarize the most relevant literature on common types of stochastic VRPs. A taxonomy and overview of the space of VRPs can be found in \cite{eksioglu2009vehicle}.\\
Generally, random variables in the stochastic VRPTW are assumed to be independent as stochastic dependency drastically complicates the calculation of distributions and their moments. While some publications focus on the VRPTW with stochastic demands (see \cite{lee2012robust}, \cite{zhang2016time}), stochastic travel and service times are far more common. Stochasticity can be addressed in different ways. A common approach is to include chance constraints, which limit the probability of violating time windows. While \cite{errico2018vehicle} connect chance constraints to the simultaneous satisfaction of all time windows, \cite{ehmke2015ensuring} as well as \cite{li2010vehicle} limit the probability of violating the time window of each customer individually.\\
In vehicle routing problems, the distributions of start times at a given customer depend on the sequence of customers previously visited by the same vehicle and the distributions of travel and service times along the route. Furthermore, the computational effort required to calculate said distributions heavily depends on the type of time windows.
\cite{tacs2014vehicle} assume that time windows are soft and travel times are independent, gamma distributed random variables. The authors propose an exact calculation of arrival time distributions by convolving finish time and travel time distributions and solve the problem using a Branch-Price-and-Cut approach. Because the convolution of gamma distributed random variables is gamma distributed, finish time distributions are again gamma distributed with known shape and scale parameter, which significantly simplifies their calculation.\\
If time windows are assumed to be hard, arrival time distributions need to be truncated at the start of each time window. Hence, distributional properties, such as moments or the existence of a closed-form expression, are not propagated along vehicle routes. Therefore, distributions can only be obtained by explicitly calculating the convolution of two distributions of arbitrary shapes, which is often not computationally tractable in reasonable time.
\cite{errico2018vehicle} assume service times to be stochastic, discretely distributed random variables and calculate start time distributions exactly. They use a Branch-Price-and-Cut algorithm to solve the problem and report computational results for four different variants of triangular distributions. \cite{miranda2016vehicle} assume normally distributed service and travel times. They approximate service and travel time distributions by discretizing them on a finite set of values, whose count and values depend on the underlying distribution's moments. This allows for an approximation of the finish time distribution of a given task by convolving the finish time distribution of the previous task with the aforementioned approximations of travel and service time distributions. They further improve their approach in \cite{miranda_revised} using improved lower bounds for the approximation intervals of service and travel times. Nevertheless, distributions are approximated, and thus the resulting algorithm is not exact.
\cite{ehmke2015ensuring} show that the first and second moment of start time distributions can be calculated with little effort if both travel times and start times at the previous customer are normally distributed. While the former is usually not the case, the authors present computational results that indicate that their approach provides suitable approximations for normal, shifted gamma, and shifted exponential travel time distributions. \cite{li2010vehicle} use stochastic simulation to derive estimates for start time distributions and solve the problem using tabu search. To the best of our knowledge, there is no publication that proposes an exact solution approach for medium- to large-sized VRPTW instances with stochastic travel times of arbitrary distributional shape.\\
We note that the distribution of travel times on airport aprons often does not follow standard distributional structures such as normal or gamma distributions. Furthermore, distributions evolve over time and thus change both their moments and their shape. Because this impedes the exploitation of distributional structures as done in the aforecited papers, the aim of this paper is to develop an exact solution approach that is able to handle travel time distributions of arbitrary shape, while also allowing for the consideration of temporal dependency of distributions.  To the best of our knowledge, the problem at hand has not been considered under the assumption of time-dependent travel times of arbitrary distributional shapes before. From the point of view of distributional assumptions and usage of chance constraints, our considerations can be seen as a combination and extension of \cite{errico2018vehicle} and \cite{li2010vehicle}, as we do not assume the knowledge of any distributional structures and calculate start time distributions exactly but interpret chance constraints as a non-route-interdependent property.

\vspace{0.5cm}
\subsection{Literature on the stochastic TSRP}\label{section:lit_rev_tsrp}
Unlike for the vehicle routing problem, there is no standard definition for the TSRP. In general, the TSRP consists of scheduling workers or assembling teams of workers and routing them across available tasks such that each task is executed by exactly one team (or worker). Depending on the context, additional requirements must be considered.\\
Due to the lack of standardized model formulations, several publications (cp. \cite{pereira2020multiperiod}, \cite{cakirgil2020integrated}, \cite{li2005manpower}) consider problems that are similar to the problem at hand. However, all of the aforementioned publications either make restrictive assumptions regarding team formation possibilities, such as the impossibility of downgrading or the existence of a single mode, or they focus on heuristic solution methods to solve medium- and large-sized instances. For an overview of further literature dealing with deterministic variants of the TSRP relevant to this study, we refer to \cite{dallolio_temp}.\\
While there is plenty of literature regarding the deterministic TSRP, very little research has been done on stochastic problem variants.
\cite{souyris2013robust} propose a robust formulation for the TSRP. Workers are assumed to be homogeneous and service times are stochastic. Each task must be started before a fixed deadline with a given probability. Moreover, each task must be executed by a single technician, thus team formation is not part of the model. The authors propose three different bounded uncertainty sets that limit the total service time delays per technician or client, respectively. The objective is to minimize the worst-case total delay and travel time. A branch-and-price approach is proposed and applied to real-world instances with 41 customers and 15 technicians.\\
\cite{yuan2015branch} address the TSRP in the context of health care workers and home health care services. Each task must be executed by a single worker. Furthermore, it is possible to downgrade workers to lower skill levels. Service times are assumed to be stochastic with a known probability distribution. The goal is to minimize the expected total travel costs, fixed costs of caregivers, service costs, and penalties for late arrival at customers. The authors use a branch-and-price approach to solve the problem exactly. Computational experiments assuming uniformly distributed service times and 25 to 50 customers indicate that the approach is able to provide very good results for small-sized instances. 
While the objective function and interpretation of stochasticity considered by \cite{yuan2015branch} is different from ours and several properties, such as the possibility of executing tasks using different modes and skill compositions, are missing, their solution approach exhibits similarities with the one proposed in the following and can be applied to the problem at hand. In Section \ref{section:influence_components}, we compare both methodologies on a set of test instances.\\
\cite{binart20162} consider a TSRP with mandatory and optional tasks, where time windows are hard and fixed, and predefined teams must be used to execute tasks. Although not explicitly done, their modeling framework would allow considering multiple skill levels, downgrading, and multiple modes for single tasks. They assume that travel and service times are discretely distributed according to triangular distributions. Furthermore, the time windows of each task must be satisfied with a given probability. A 2-stage approach is proposed to heuristically solve the problem. First, a feasible skeleton solution is obtained using a generic MIP solver by optimally covering all mandatory tasks. Second, the first-stage solution is refined by inserting optional customers into the existing routes such that time window restrictions are not violated. Computational results on instances with 5 to 9 mandatory and 30 to 50 optional tasks indicate that the approach can yield good results in most cases. The problem considered by \cite{binart20162} can be seen as a special case of the problem at hand where all technicians have the same skill level, each team formation represents one technician, and each task can be executed by each team formation. While the objective function and interpretation of stochasticity differ from ours, the models and solution approach proposed in the following can be applied to their problem.\\
Our work shows several similarities with \cite{binart20162} and \cite{yuan2015branch}, such as the stochasticity of travel times, the usage of exact solution methods, and the incorporation of several team properties such as multiple skill levels and modes. At the same time, our interpretation of service levels at individual tasks rather than entire solutions, the lack of assumptions on distributional properties, and the exact calculation of arrival times separate our considerations from previous approaches.\\
To conclude this section, it is worth noting that existing research primarily deals with heuristic solution procedures. Exact approaches are typically only efficient for small-sized instances, especially when stochasticity is considered.\\
The following considerations are widely based on \cite{dallolio_temp}, which consider the problem of forming and routing worker teams for baggage handling tasks in a deterministic setting. We extend their deterministic problem description by assuming stochastic travel times. Furthermore, we advance their proposed solution algorithm by introducing cutting planes and additional branching strategies, improving the runtime efficiency of the pricing step, and describing an efficient way to handle infeasible solutions that is based on dynamically switching between two master problem formulations.

%% file: sections/models.tex
\section{Two Binary Program Formulations}\label{section:models}
In the following sections, we present two set-covering formulations for the problem. Set-covering formulations are commonly used for vehicle routing problems in combination with column generation approaches. In such a formulation, each column represents a team performing a subset of tasks. The goal is to find a minimum cost subset of columns such that each tasks is visited (i.e., covered) exactly once, and resource or capacity constraints are satisfied. This modelling technique exploits the natural decomposable structure of routing problems. Furthermore, it allows for the consideration of the probabilistic, route-specific constraints (\ref{chance_constraint}) and (\ref{finish_at_lf_v}) in the subproblem without the need for linearization. In Section \ref{section:mp_aggregated_workforce}, we extend the model proposed in \cite{dallolio_temp} to incorporate stochastic travel times and a stochastic objective function. Besides the construction of routes, said model only decides on the profiles of teams, i.e., on lower bounds for the actual qualifications of required workers. Therefore, skill levels are implicitly aggregated, i.e., they are only considered on an aggregate level. Hence, one might obtain an integer solution that is operationally infeasible, i.e., it can not be implemented in practice. Whenever such a solution is found, we switch to an alternative formulation that considers workers based on individual skill levels. This alternative formulation, which always returns operationally feasible solutions but is considerably harder to solve, is presented in Section \ref{section:mp_disaggregated_workforce}, alongside additional insights into operational feasibility. Recall that all notation used in the following is summarized in \Cref{appendix:notation}.

\vspace{0.5cm}

\subsection{Master Problem with Aggregated Workforces}\label{section:mp_aggregated_workforce}
The following considerations assume an underlying finite time grid given by discrete time points $\tau\in\mathcal{T}$. We first introduce general notation and several auxiliary parameters.\\
Let $(r,q)\in\mathcal{R}$ be a team route. We define by $b_{k,\tau}^{r,q}(\omega_{\gamma})\in\left\{0,\xi_{q,k}\right\}$ the number of workers of skill level not greater than $k$ that are conducting route $r$ and are occupied at time $\tau$, given the $\gamma$-quantile scenario $\omega_\gamma$. In the following, $\gamma$ is also called the \textit{workforce travel time quantile}. This value is equal to the worker requirement $\xi_{q,k}$ of profile $q\in\mathcal{Q}$ for all $\tau\in [\mathrm{tl}^{r,q},\mathrm{tr}^{r,q}]$ and $0$ else. Binary variables $\lambda^r_q$ indicate if team route $(r,q)$ is part of the solution.
The problem can then be described using the following \textit{aggregated master problem}, short \textit{AMP}:
\begin{align}
    \min & \sum_{(r,q)\in \mathcal{R}} \mathbb{E}(c^r)\lambda^r_q \label{model:obj_aggregated_mp}\\
    \textrm{s.t.} \quad & \underset{(r,q)\in\mathcal{R}: \ i\in\mathcal{I}^r}{\sum} \lambda^r_q \geq 1 \quad \forall i\in\mathcal{I}\label{model:tasksdone_aggregated_mp}\\
    & \underset{(r,q)\in\mathcal{R}}{\sum}b_{k,\tau}^{r,q}(\omega_{\gamma})\lambda^r_q\leq N_k \quad \forall k\in\mathcal{K},\ \forall \tau\in \mathcal{T} \label{model:workerconstr_aggregated_mp}\\
    & \lambda^r_q\in\mathbb{N} \quad \forall (r,q)\in\mathcal{R}\label{model:lambda_binary_aggregated_mp}
\end{align}
Constraints (\ref{model:tasksdone_aggregated_mp}) ensure that each task is part of at least one route. It is easy to see that covering a task more than once can not be part of an optimal solution. Inequalities (\ref{model:workerconstr_aggregated_mp}) guarantee that the workforce required at any given time $\tau$ does not exceed the total available workforce on an aggregate level. The objective function (\ref{model:obj_aggregated_mp}) aims to minimize the expected finish times of tasks and incurred penalties in order to maximize buffer times. Furthermore, the AMP with only a subset of team routes $\bar{\mathcal{R}}\subset \mathcal{R}$ considered and integrality conditions (\ref{model:lambda_binary_aggregated_mp}) relaxed to $\lambda^r_q\geq 0$ is called \textit{aggregated reduced master problem (ARMP)}. However, for the sake of simplicity, we will still be referring to the set of columns of the ARMP as $\mathcal{R}$. We note that, unlike in standard set-covering formulations, we only require the variables to be integral rather than binary. However, for any feasible instance of the AMP, there exists an optimal solution which is binary. A proof for this can be found in \Cref{appendix:binary_sol_existence}.
\\
We note that chance constraints (\ref{chance_constraint}) are not route-interdependent, thus they can be fully embedded in the pricing problem.

\vspace{0.5cm}

\subsection{Master Problem with Skill-Level Specific Workforces}\label{section:mp_disaggregated_workforce}
In the following, we provide a more detailed master problem that considers the workforce on an individual skill level basis. Furthermore, we provide some insights into the dominance relation between the two proposed master problem formulations.\\
Let $(r,q)\in\mathcal{R}$ be a team route. As described in Section \ref{section:problem_description}, $q$ is well-defined by the amount $\xi_{q,k}$ of required workers of skill level greater or equal than $k$ for all $k\in\mathcal{K}$. This information suffices to ensure that the available workforce is never exceeded on an aggregate level, but does not guarantee feasibility in terms of allocation of workers to individual tasks.
To be precise, the ARMP implicitly allows workers to migrate to another team while executing a task along their route without incurring any travel time. As this is neither realistically possible nor feasible in practical applications, this can be seen as an artifact of the AMP formulation. For an example, the reader is referred to Section 4.2 of \cite{dallolio_temp}.
To deal with such undesirable solutions, the aforementioned authors propose an integer problem called \textit{feasibility check}, which tries to find a feasible allocation of available workers to teams and regrouping strategies by solving a network flow problem on an appropriate graph. If said problem is infeasible, a cut is added to the master problem and the ARMP is re-solved. In preliminary studies, we observed that there are several instances in which a large number of cuts have to be added, degrading the algorithm's performance by a large margin. In such instances, there is a large cardinality of binary solutions to the ARMP that all share the property of being operationally infeasible in the previously described sense and have the same (or almost the same) objective function value. This makes it necessary to consecutively forbid these solutions one by one. Furthermore, each additional cut introduces a dual value into the pricing problem, whose impact on the reduced cost of a column is only observable after a column has been generated. Therefore, it is possible that during the pricing procedure, columns that are part of a forbidden solution are re-generated. Thus, at the end of each pricing step, such columns need to be identified and discarded, increasing the computational effort required to solve the pricing problem. Additionally, the ARMP has to be re-solved every time without improving the solution quality.\\
We mitigate this issue by utilizing the concept of skill compositions to develop an alternative formulation to the AMP. For this, we define a disaggregated team route $(r,q, s)$ as a tuple consisting of a team route $(r,q)\in\mathcal{R}$ and a disaggregated profile $(q,s)\in\mathcal{Q}^{\mathrm{D}}$ and denote the set of disaggregated team routes by $\mathcal{R}^{\mathrm{D}}$.\\
% We use these concepts to develop a model analogous to the AMP. 
For each $(r,q,s)\in\mathcal{R}^{\mathrm{D}}$, we define by $\beta_{k,\tau}^{r,s}(\omega_{\gamma})\in\left\{0, s_{q,k}\right\}$ the number of workers of skill level $k$ required by route $r$ at time $\tau$, assuming scenario $\omega_\gamma$ occurs. These parameters take values $\beta_{k,\tau}^{r,s} = s_{q,k}$ for $\tau\in [\mathrm{tl}^{r,q}, \mathrm{tr}^{r,q}]$ and $\beta_{k,\tau}^{r,s} = 0$ otherwise. Using the previous definitions, we introduce the \textit{disaggregated master problem}, abbreviated as \textit{DMP}:
\begin{align}
    \min & \sum_{(r,q,s)\in \mathcal{R}^{\mathrm{D}}} \mathbb{E}(c^r)\lambda^r_{q,s} \label{model:obj_disaggregated_mp}\\
    \textrm{s.t.} \quad & \underset{(r,q,s)\in\mathcal{R}^{\mathrm{D}}: \ i\in\mathcal{I}^r}{\sum} \lambda^r_{q,s} \geq 1 \quad \forall i\in\mathcal{I}\label{model:tasksdone_disaggregated_mp}\\
    & \underset{(r,q,s)\in\mathcal{R}^{\mathrm{D}}}{\sum}\beta_{k,\tau}^{r,s}(\omega_{\gamma})\lambda^r_{q,s}\leq N_k^{\mathrm{D}} \quad \forall k\in\mathcal{K},\ \forall \tau\in \mathcal{T} \label{model:workerconstr_disaggregated_mp}\\
    & \lambda^{r}_{q,s}\in\mathbb{N} \quad \forall (r,q,s)\in\mathcal{R}^{\mathrm{D}}\label{model:lambda_binary_disaggregated_mp}
\end{align}
Unlike in Section \ref{section:mp_aggregated_workforce}, constraints (\ref{model:workerconstr_disaggregated_mp}) consider workforces on an individual rather than an aggregated level. Similar to the ARMP, we call the DMP when only a subset of routes is considered and (\ref{model:lambda_binary_disaggregated_mp}) is replaced by $\lambda^r_{q,s}\geq 0$ the \textit{aggregated reduced master problem (ARMP)}. For ease of reading, we will be referring to the columns of the DRMP as the set $\mathcal{R}^{\mathrm{D}}$. Moreover, we denote a solution $(\hat{\lambda}^r_q)_{(r,q)\in\mathcal{R}}$ to the ARMP, which can not be extended to a solution $(\hat{\lambda}^r_{q,s})_{(r,q,s)\in{\mathcal{R}^{\mathrm{D}}}}$ of the DRMP as a \textit{disaggregated-infeasible solution}. We note that this can be checked by solving the feasibility check, which has been proposed by \cite{dallolio_temp} and is included in \Cref{appendix:feas_check_model} of this publication. Analogous to Section \ref{section:mp_aggregated_workforce}, one can prove that there exists a binary solution for every feasible DMP instance.\\
Each disaggregated team route $(r,q,s)\in\mathcal{R}^{\mathrm{D}}$ corresponds to one column of the DRMP. It is easy to see that $\vert\mathcal{R}^{\mathrm{D}}\vert$ is considerably larger than $\vert \mathcal{R} \vert$. A formal proof of this is provided in \Cref{appendix:dominance_dmp}. Additionally, each column in $\mathcal{R}$ can be obtained by convex combinations of columns in $\mathcal{R}^{\mathrm{D}}$. Thus, vertices of the feasible region of the linear relaxation of the ARMP may correspond to higher-dimensional faces in the DRMP. This makes it significantly harder to cut off non-integer solutions from the feasible region; therefore, solving the DRMP using a branch-and-cut algorithm becomes substantially harder. Experimental studies have shown that solving the DRMP is, on average, around 30\% slower than solving the ARMP. Therefore, we only resort to solving the DRMP once a disaggregated-infeasible solution has been identified.

%% file: sections/solution_approach.tex
\section{Solution Approach}\label{section:solution_approach}
In this section, we present a Branch-Price-Cut-and-Switch algorithm to solve the problem at hand. We initially start the algorithm at the root node with a subset of columns, denoted by $\bar{\mathcal{R}} \subset \mathcal{R}$, consisting of single-task tours $(d,i,d)$ for each task $i\in\mathcal{I}$, each profile $q\in\mathcal{Q}_i$ and the earliest possible leave time from the depot $tl \coloneqq \left\{t \in \mathcal{T}: \ t + t_{d,i}^{k(t)}(\omega_{\max}) \geq ES_i\right\}$, where $k(t)$ is the time bin such that $t\in\mathcal{B}^{k(t)}$ holds. To ensure feasibility, we include a column that finishes each task at the latest possible time $\mathrm{LF}_i^{\mathrm{e}}$ and uses up the entire available workforce.  We then search the branching tree using a Branch-Price-and-Cut approach. If we obtain an integer solution, we perform the feasibility check proposed by \cite{dallolio_temp} to check if the solution is disaggregated-feasible. If the solution fails the feasibility check, we mark the current node and its sibling node as disaggregated-infeasible and restart the procedure. We note that children of marked nodes are also marked by default. Whenever a disaggregated-infeasible node is encountered with the search tree, we solve the DRMP instead of the ARMP. The selection of unexplored nodes is done via a best-first search.

\begin{figure}
\includegraphics[scale=0.55]{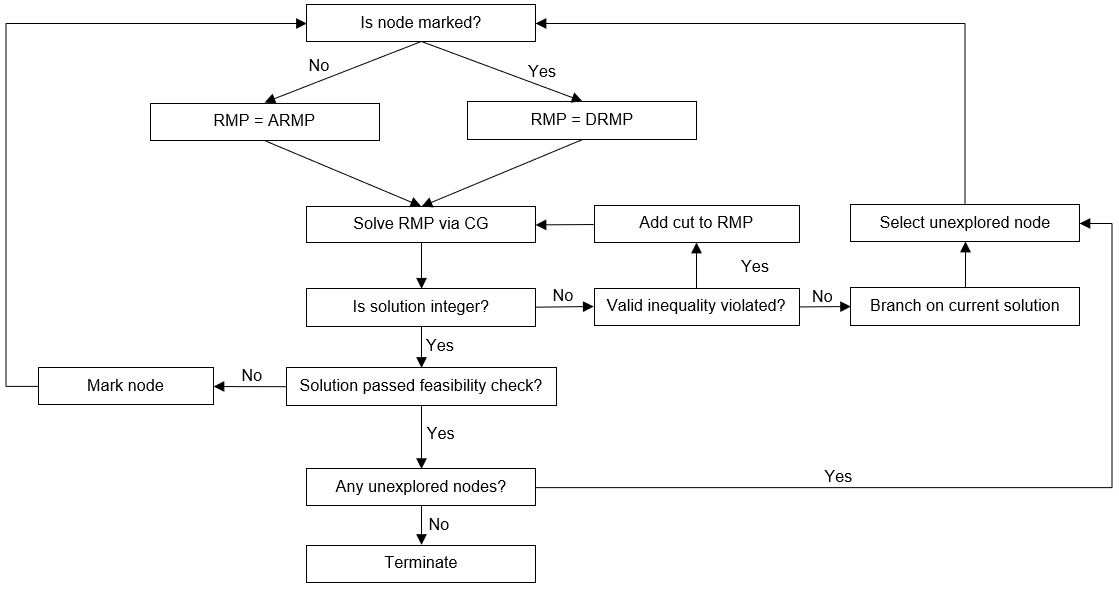} % BPCS_flowchart.jpg
\caption{Branch-Price-Cut-and-Switch}
\label{graphics:bpcs_framework}
\end{figure}
Figure \ref{graphics:bpcs_framework} provides an overview over the proposed Branch-Price-Cut-and-Switch framework. In Section \ref{section:pricing_step}, we describe the pricing problem for the ARMP. Section \ref{section:pricing_step_amp} focuses on the labeling algorithm used to solve the pricing problem. In Section \ref{section:pricing_step_dmp}, necessary adjustments to solve the pricing problem associated with the DRMP are explained. Section \ref{section:acceleration_strategies} elaborates on several acceleration strategies used to speed up the solution process. Unless otherwise stated, all considerations can be transferred to the DRMP and its set $\mathcal{R}^{\mathrm{D}}$ of disaggregated team routes.

\vspace{0.5cm}
\subsection{Pricing Problem for the ARMP}\label{section:pricing_step}
Whenever a solution to a reduced master problem (ARMP or DRMP) is obtained, we solve the corresponding pricing problem to check if there is a column with negative reduced costs. In the ARMP, the reduced cost of a column corresponding to a team route $(r,q)\in\mathcal{R}$ is equal to
\begin{equation*}
    \mathbb{E}(c^r) - \underset{i\in\mathcal{I}^r}{\sum} \mu_i - \underset{k\in\mathcal{K}}{\sum} \ \underset{\tau=\mathrm{tl}^{r,q}}{\overset{\mathrm{tr}^{r,q}}{\sum}}\delta_{k,\tau}b_{k,\tau}^{r,q}(\omega_{\gamma})\label{reduced_cost_formula_amp}
\end{equation*}
where $\mu = (\mu_i)_{i\in\mathcal{I}}$ and $\delta = (\delta_{k,\tau})_{k\in\mathcal{K},\tau\in\mathcal{T}}$ are the dual variables corresponding to constraints (\ref{model:tasksdone_disaggregated_mp}) and (\ref{model:workerconstr_disaggregated_mp}), respectively. We note that $\mu \geq 0$ and $\delta\leq 0$ holds.\\
The pricing problem of the ARMP corresponds to a set of elementary shortest path problems with resource constraints (ESPPRCs), one for each profile $q\in\mathcal{Q}$. These problems are typically solved using pricing networks, i.e., graphs where each path represents a column and its length is equal to the respective column's reduced cost. We use a non-time expanded graph with time-dynamic arc weights, i.e., weights whose value depends on the current finish time distribution, to model and solve the associated pricing problems. We note that time-expanded graphs can also be used in this context, however they scale rather poorly for more granular time grids. Thus, they are less suited for the stochastic problem formulation at hand.\\
We now turn our focus to the construction of the graph used to solve the ESPPRC.
For each profile $q\in\mathcal{Q}$, we define a directed graph $\mathcal{G}^q \coloneqq \left(\mathcal{V}^q, \mathcal{A}^q, b\right)$, where $b = \left(b_{k,t}^q(\omega_{\gamma})\right)_{k\in\mathcal{K},t\in\mathcal{T}}$ are the aggregated workforce requirements of a team with profile $q$, as follows: $\mathcal{V}^q$ contains two nodes $o, o'$ (called origin and destination) representing the depot and one node for each task $i\in\mathcal{I}_q$. Due to constraints (\ref{chance_constraint})--(\ref{finish_at_lf_v}), task $j$ can not be executed after task $i$ if, for any finish time of task $i$, (\ref{chance_constraint}) or is (\ref{finish_at_lf_v}) violated. Formally, task $j$ can not be executed after task $i$ if 
\begin{equation}
    t + t_{i,j}^{k(t)}(\omega_{\alpha}) > LF_j - p_{j,q} \quad \text{or} \quad t + t_{i,j}^{k(t)}(\omega_{\mathrm{max}})  > LF_j^e - p_{j,q} \label{arc_pruning_condition}
\end{equation}
holds for all $t\in[ES_i + p_{i,q}, LF_i^e + p_{i,q}]$, where $k(t) = \{k: t \in \mathcal{B}_k, \mathcal{B}_k\in\mathcal{B}\}$ is the time bin of time period $t\in\mathcal{T}$. Condition (\ref{arc_pruning_condition}) implies that, for any finish time distribution of task $i\in\mathcal{I}$, an extension along $(i,j)\in E$ would either violated the chance constraint (\ref{chance_constraint}) or the extended time window (\ref{finish_at_lf_v}). Therefore, we define the arc set $\mathcal{A}^q$ by
\begin{align*}
    \mathcal{A}^q \coloneqq \{&(o,i): \ i\in\mathcal{I}_q\} \ \cup \ \{(i,o'): \ i\in\mathcal{I}_q\} \ \cup \ \{(i,j)\in\mathcal{I}_q\times\mathcal{I}_q: (i,j) \ \text{violates (\ref{arc_pruning_condition})}\\
    & \text{for at least one } t\in [ES_i +p_{i,q}, LF_i^e + p_{i,q}]\}
\end{align*}
We can further reduce the size of $\mathcal{A}^q$ by removing arcs $(i,j)$ for which the following holds:
\begin{align}
    \mathrm{ES}_j - \mathrm{LF}_i^{\mathrm{e}} &\geq t_{i,d}^k{(\omega_{\gamma})} + t^k_{d,j}{(\omega_{\gamma})} \quad \forall k=1,\dots,\vert \mathcal{B}\vert \label{giacomo _arc_removal_1}
\end{align}
If inequality (\ref{giacomo _arc_removal_1}) holds, replacing arc $(i,j)$ in a feasible route $r$ with arcs $(i,d)$ and $(d,j)$ splits $r$ into two new routes, which are also feasible and have the same joint objective function value as $r$. Furthermore, they do not occupy more workforce than $r$ at any time $t\in\mathcal{T}$. Hence, we can remove arc $(i,j)$ from $\mathcal{A}^q$.
Let $P$ be a path in $\mathcal{G}^q$ with depot leave time $\mathrm{tl}^P \geq 0$. Because the reduced costs of an arc $(i,j)\in\mathcal{A}^q$ are time-dependent and $\mathcal{G}^q$ does not contain any temporal information, arc weights are time-dynamic and specifically depend on the distribution of finish times $F_i^P$ and the $\omega_{\gamma}$-scenario finish time $F_i^P(\omega_{\gamma})$ at the previous node $i$ along $P$. For this, we set $w_{o'} \coloneqq 0$, $\mu_{o'} \coloneqq 0$ and $F_{o}^P(\omega_{\gamma}) \coloneqq \mathrm{tl}^P - 1$. Then, the weight $W_{i,j}\left(F_i^P(\omega_{\gamma})\right)$ of an arc $(i,j)\in\mathcal{A}^q$ is given by
\begin{equation*}
    W_{i,j}\left(F_i^P(\omega_{\gamma})\right) \coloneqq w_{j} \cdot \mathbb{E}(F_{j} + P_{j}(F_{j})) -\mu_{j} - \underset{k\in\mathcal{K}}{\sum} \ \underset{\tau=F_{i}^{P}(\omega_{\gamma})+1}{\overset{F_{j}^P(\omega_{\gamma})}{\sum}} \delta_{k,\tau} b_{k,\tau}^{P,q}(\omega_{\gamma}).
\end{equation*}
Then, finding a minimum-cost $o-o'-$path that satisfies constraints (\ref{chance_constraint}) and (\ref{finish_at_lf_v}) is equivalent to finding a feasible team route $(r,q)\in\mathcal{R}$ (i.e., a new column) with minimum reduced cost.

\vspace{0.5cm}
\subsection{Labeling Algorithm for the ARMP Pricing Problem}\label{section:pricing_step_amp}
We solve the pricing problem using a labeling algorithm with a customized dominance rule, enhanced with several acceleration strategies that reduce the graph size and dimension of the label space.
In order to check if a path $P$ in $\mathcal{G}^q$ violates constraints (\ref{chance_constraint}) or (\ref{finish_at_lf_v}) at a task $i\in\mathcal{I}_q$, full information about the distribution of $F_i^P$ and $F_i^{P,\mathrm{\gamma}}$ needs to be available. Thus, for each partial path $P = (o,v_1,\dots,v_l,v)$ in $\mathcal{G}^q$, we define a label $L$ as a tuple $L\coloneqq \left(\mathrm{tl}^L, v, P, T_{v}^{\mathrm{cost}}, (T_{v,i}^{\mathrm{perf}})_{i\in\mathcal{I}_q}, F_{v}^L\right)$, where $\mathrm{tl}^L$ is the depot leave time, $T_{v}^{\mathrm{cost}}$ is the reduced cost of the path, $T_{v,i}^{\mathrm{perf}}\in\{0,1\}$ indicate if a task node $i$ can still be visited and $F_{v}^L$ is the distribution of finish times at the current last node $v$. We say that $L$ is \textit{feasible} for $\mathcal{G}^q$ if its associated path $P = (o,v_1,\dots,v_l,v)$ satisfies constraints (\ref{chance_constraint}) and (\ref{finish_at_lf_v}) for all task nodes visited by $P$ and $v_i \neq v_j$ holds for all $i,j\in\{1,\dots,l\}, i\neq j$.\\
%\vspace{0.7em}\paragraph{\textbf{Label Extension}}\text{ }\\
When extending a label $L$ along arcs $(v,v')$, we obtain a new label $L' = \left(\mathrm{tl}^{L'}, v', P', T_{v'}^{\mathrm{cost}}, (T_{v',i}^{\mathrm{perf}})_{i\in\mathcal{I}_q}, F_{v'}^{L'}\right)$ by using the following resource extension function:
\begin{align*}
    \mathrm{tl}^{L'} &= \mathrm{tl}^{L}\\
    P' &= (o,v_1,\dots,v_l,v,v')\\
    T_{v'}^{\mathrm{cost}} &= T_{v}^{\mathrm{cost}} + W_{v,v'}(F_v^{L}(\omega_{\gamma}))\\
    T_{v',i}^{\mathrm{perf}} &= \begin{cases}
                    \begin{aligned}
                        &T_{v,i}^{\mathrm{perf}} - 1 && \text{if} \ \  i = v' \\
                        &1 && \text{if} \ \  F_{v'}^{L'}(\omega_{max}) + t^{k}_{v',i}(\omega_{\mathrm{max}}) > \mathrm{LF}_{i}^{\mathrm{e}}-p_{i,q} \ \forall k=1\,\dots\mathcal{B}\\
                        &1 && \text{if} \ \  F_{v'}^{L'}(\omega_{max}) + t^{k}_{v',i}(\omega_{\alpha}) > \mathrm{LF}_i-p_{i,q} \ \forall k=1\,\dots\mathcal{B}\\
                        &T_{v,i}^{\mathrm{perf}} && \text{otherwise}
                    \end{aligned}
\end{cases}\\
    F_{v'}^{L'}(\tau) &= \begin{cases}
        \begin{aligned}
        &\underset{z=0}{\overset{\mathrm{ES}_{v'}+ p_{v',q}}{\sum}} \mathbb{P}(F_{v}^{L} + t^{\bar{F}^{L,\gamma}}_{v,v'} + p_{v',q}= z) && \text{if} \ \ \tau = \mathrm{ES}_{v'} + p_{v',q}\\
        &\mathbb{P}(F_{v}^{L} + t^{\bar{F}^{L,\gamma}}_{v,v'} + p_{v',q} = \tau)&& \text{if} \ \ \tau > \mathrm{ES}_{v'} + p_{v',q}\\
        &0 && \text{otherwise}
        \end{aligned}
    \end{cases}
\end{align*}
where we define $p_{o',q} = 0$. We note that we reset the resources $T_{v',i}^{\mathrm{perf}}$ of tasks that cannot be visited anymore without violating time windows or chance constraints to 1, as this strengthens the dominance relations between labels. Furthermore, $\mathrm{tl}^{L}, v$ and $P$ are properties necessary for a label $L$ to be well-defined, however they are not resources in the classical sense.\\
We say that the extension of label $L$ along $(v,v')$ is \textit{feasible} if the resulting label $L'$, which is obtained from $L$ using the above resource extension function, is feasible for $\mathcal{G}^q$.\\
\begin{example}\label{ex:label_extension}
Consider the graph $\mathcal{G}^q$ for a given profile $q\in\mathcal{Q}$ depicted in Figure \ref{graphics:example_label_extension}. Let $L^1$ be a label describing the subpath $(o, 1)$ with depot leave time $tl^1  = 2$. We assume $\mathcal{K} = \{1\}$, $\xi_{q,1} = 1$ and $\delta_{1,\tau} = -1$ for all $\tau\in[2,10]$. Furthermore, assume $LF^e_1 = 10$, $ES_1 = 0$, $ES_3 = 3$, $LF_3 = LF^3_e = 15$, $w_1 = w_3 = 1$, $\mu_1 = 2$, and $\mu_1 = 2$. Moreover, we define $p_{1,q} = p_{3,q} = 3$. Finally, define $\gamma = 0.95$ and time bins $\mathcal{B} = \{\mathcal{B}_1,\mathcal{B}_2\}$ with $\mathcal{B}_1 = \{0,\dots 5\}$ and $\mathcal{B}_2 = \{6,\dots,11\}$.
\begin{figure}[ht]
\centering
\includegraphics[scale=0.38]{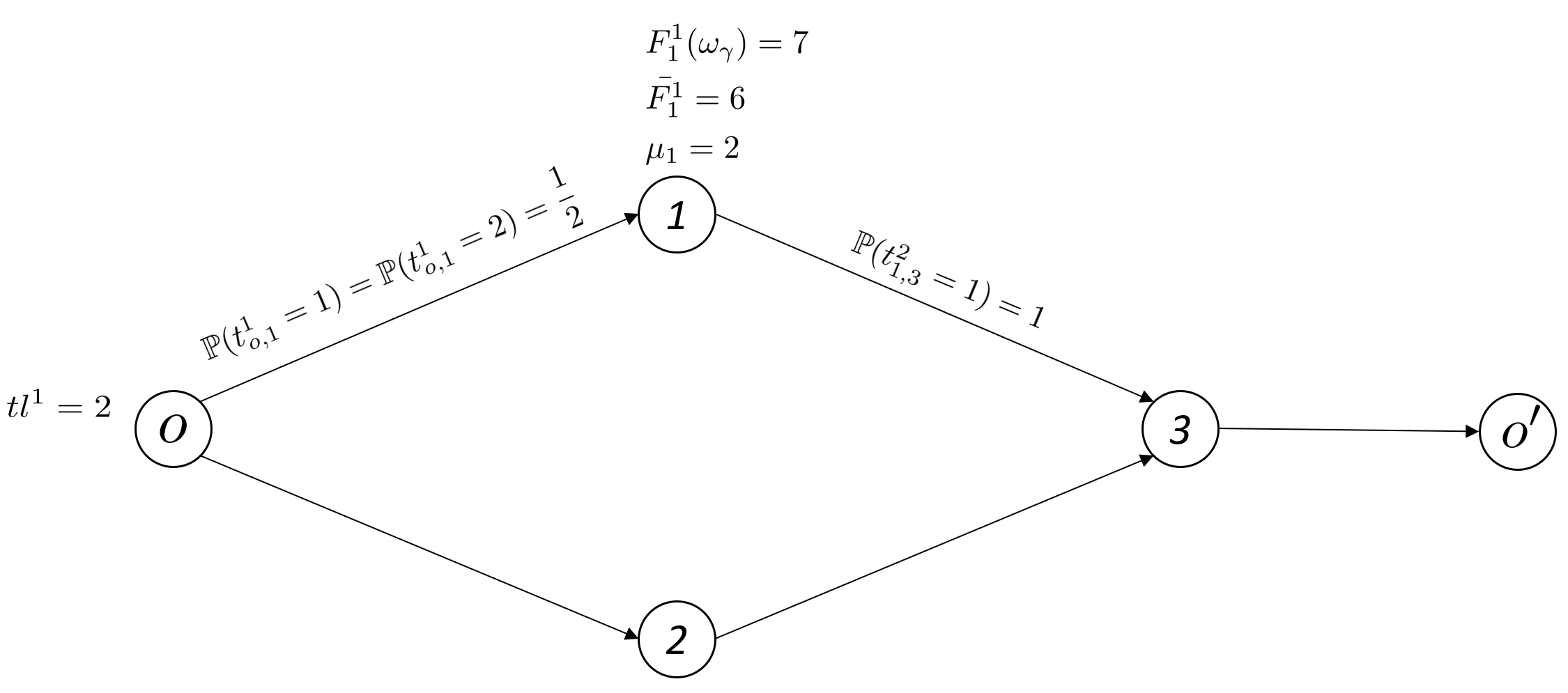} % ex_labelExtension.PNG
\caption{Label extension and time-dynamic arc weights}
\label{graphics:example_label_extension}
\end{figure}\\
Because $tl^1 = 2$ and $p_{1,q} = 3$ hold, the finish time distribution $F_1^1$ is then equal to $P(F_1^1 = 6) = P(F_1^1 = 7) = \dfrac{1}{2}$. Hence, the median finish time is $\bar{F_1^1} = 6$. Furthermore, in scenario $\omega_{\gamma}$, we have $t^1_{o,1}(\omega_{\gamma}) = 2$. Thus, one can conclude that the finish time of $L^1$ at task $1$ in scenario $\omega_{\gamma}$ is equal to $F_1^1(\omega_{\gamma}) = \max\{tl^1+t_{o,1}^1(\omega_{\gamma}),ES_1\}+p_{1,q} = 2+2+3 = 7$. The reduced cost of label $L^1$ is then equal to
\begin{equation*}
    T_1^{cost} = \mathbb{E}(F_1^1) - \mu_1 - \underset{t=2}{\overset{7}{\sum}}\delta_{1,\tau} = \dfrac{1}{2}\cdot (6 + 7) - 2 - \underset{t=2}{\overset{7}{\sum}}\delta_{1,t} = 4.5 + 6 = 10.5
\end{equation*}\\
Assume one wants to extend $L^1$ along the edge $(1,3)$. Because $\bar{F}_1^1 = 6\in\mathcal{B}_2$ holds, the finish time distribution $F_1^3$ thus depends on the travel time distribution of edge $(1,3)$ in time bin $\mathcal{B}_2$, i.e., on the distribution of $t_{1,3}^2$. Furthermore, the time-dynamic arc weight $W_{1,3}\left(F_1^1(\omega_{\gamma})\right)$ also depends on the distribution of $t_{1,3}^2$, as well as on the finish time distribution $F_1^1$. The finish time distribution at task $3$ is then given by $\mathbb{P}(F_1^3 = 10) = \mathbb{P}(F_1^3 = 11) = \dfrac{1}{2}$. Hence, the reduced cost of the resulting label is equal to
\begin{equation*}
    T_3^{cost} = T_1^{cost} + W_{1,3}\left(F_1^{1}(\omega_{\gamma})\right) = 10.5 + \dfrac{1}{2} (10 + 11) - \underset{t=8}{\overset{10}{\sum}}\delta_{1,t} = 10.5 + 10.5 + 3 = 24
\end{equation*}
Moreover, task resources $T_{3,3}^{perf}$ and $T_{3,1}^{perf}$ are updated to values $0$ and $1$, respectively. While the former is an immediate result of the extension along edge $(1,3)$, the latter can be reset to $1$ because 
\begin{equation*}
    F_{3}^{1}(\omega_{max}) + t^2_{3,1}(\omega_{max}) = 11 + t^2_{3,1}(\omega_{max}) > 12 = 15 - 3 = LF_1^e - p_{1,q}
\end{equation*}
holds for any non-negative travel time $t^2_{3,1}(\omega_{max})$ and thus, condition 2 of the task resource update in the resource extension function is satisfied. All remaining resources can be updated canonically according to the resource extension function.\\
\end{example}

A core component of every labeling algorithm is its dominance rule, which allows the discarding of labels that can not be part of an optimal $o$-$o'$ path.
\begin{definition}[Dominance Rule]\label{dominance_rule_amp}
Let $L^1\coloneqq \left(\mathrm{tl}^1, v, P^1, T_{v}^{1,\mathrm{cost}}, (T_{v,i}^{1,\mathrm{perf}})_{i\in\mathcal{I}_q}, F_{v}^{1}\right)$ and \\ $L^2\coloneqq \left(\mathrm{tl}^2, v, P^2, T_{v}^{2,\mathrm{cost}}, (T_{v,i}^{2,\mathrm{perf}})_{i\in\mathcal{I}_q}, F_{v}^{2}\right)$ be labels in $\mathcal{G}^q$. We say that $L^1$ dominates $L^2$ if the following properties hold:
\begin{align}
    T_{v}^{1,\mathrm{cost}} &\leq T_{v}^{2,\mathrm{cost}} \label{dominance:red_cost} \\
    T_{v,i}^{1,\mathrm{perf}} &\geq T_{v,i}^{2,\mathrm{perf}} \quad \forall i\in\mathcal{I}_q \label{dominance:task_resources}\\ 
    \mathbb{P}(F_{v}^1 \leq \tau) &\geq \mathbb{P}(F_{v}^2 \leq \tau) \quad \forall \tau\in [\mathrm{ES}_v+p_{v,q}, \mathrm{LF}_v^{\mathrm{e}}] \label{dominance:stochastic_dom}\\
    \bar{F}_v^{1} &= \bar{F}_v^2 \label{dominance:identical_median_finishes}\\
    F_v^{1}(\omega_{\gamma}) &= F_v^{2}(\omega_{\gamma}) \label{dominance:quantile_case_finishes}
\end{align}
where $\bar{F}_v^1$ and $\bar{F}_v^2$ are the median finish times of labels $L^1$ and $L_2$ at task $v$, respectively.
\end{definition}
Properties (\ref{dominance:task_resources}) and (\ref{dominance:stochastic_dom}) ensure that any feasible extension of $L^2$ is also feasible for $L^1$. Condition (\ref{dominance:identical_median_finishes}) ensures that stochastic dominance (\ref{dominance:stochastic_dom}) propagates along label extensions. Jointly with constraints (\ref{dominance:red_cost}), (\ref{dominance:stochastic_dom}) and (\ref{dominance:quantile_case_finishes}), this guarantees that, after extending both $L^1$ and $L^2$ along $(v,v')$, the reduced cost of $L^{1'}$ is still less or equal to the reduced cost of $L^{2'}$.\\
At the beginning of each iteration of the labeling algorithm, we create one label for each task $i\in\mathcal{I}_q$ and each depot leave time for which the corresponding label satisfies (\ref{chance_constraint})--(\ref{finish_at_lf_v}).
These labels are also called \textit{initial labels}. We then extend these labels using the previously described resources extension function, discard dominated labels, and repeat the same procedure for the oldest label until no feasible extensions can be found anymore. We note that the calculation of the start time distribution consumes the majority of runtime during label extensions. When the support of travel time distributions is small enough, these calculations can be done exactly. We refer to \cite{errico_recourse} and \cite{errico2018vehicle} for an extensive description of such an algorithm.

\vspace{0.5cm}

\subsection{Peculiarities for the DRMP}\label{section:pricing_step_dmp}
Though the DRMP is structurally very similar to the ARMP, several characteristics of the former must be considered when solving the pricing problem of the DRMP. 
Because the node and arc set does not depend on the skill composition $s$, we can define the graph of a disaggregated profile $(q,s)\in\mathcal{Q}^{\mathrm{D}}$ as $\mathcal{G}^{q,s}\coloneqq (\mathcal{V}^q,\mathcal{A}^q, \beta)$, where the only difference lies in the coefficients $\beta_{k,t}^s(\omega_{\gamma}))$ replacing $b_{k,t}^q(\omega_{\gamma})$ during the calculation of the time-dynamic arc weights. We note that $\beta \coloneqq \left(\beta_{k,t}^s(\omega_{\gamma})\right)_{k\in\mathcal{K},t\in\mathcal{T}}$ denotes the disaggregated workforce requirements of a team with profile $q$ and skill composition $s$.\\
A significant difference lies in the number of pricing networks. While there is exactly one pricing network for each profile $q\in\mathcal{Q}$ for the ARMP, there is one pricing network for each disaggregated profile $(q,s)\in\mathcal{Q}^{\mathrm{D}}$. Typically, the cardinality of $\mathcal{Q}^{\mathrm{D}}$ grows exponentially in $\vert\mathcal{Q}\vert$. Thus, creating and solving a unique pricing network for each disaggregated profile could render a column generation approach highly impractical due to the vast number of networks to be solved. In the following, we show that multiple pricing networks can be solved simultaneously when the used dominance rule is slightly altered.\\
Let $(q,s)\in\mathcal{Q}^{\mathrm{D}}$ be a disaggregated profile. We then introduce an adjusted dominance rule:
\begin{definition}[Dominance rule for the DRMP]\label{dominance_rule_dmp}
Let 
\begin{equation*}
    L^k\coloneqq \left(\mathrm{tl}^k, v, P^k, T_{v}^{k,\mathrm{cost}}, (T_{v,i}^{k,\mathrm{perf}})_{i\in\mathcal{I}_q},  (T_{v,g}^k)_{g\in\mathcal{G}}, F_{v}^{k}\right)
\end{equation*}
be labels for $k=1,2$. We say that $L^1$ dominates $L^2$ in $\mathcal{G}^{q,s}$ if the following properties hold:
\begin{align}
    T_{v}^{1,\mathrm{cost}} + \underset{k\in\mathcal{K}}{\sum} \underset{\tau=\mathrm{tl}^1}{\overset{F_v^{1}(\omega_{\gamma})}{\sum}}\delta_{k,\tau} \beta_{k,\tau}^{P^1,s,1}(\omega_{\gamma}) &\leq T_{v}^{2,\mathrm{cost}} + \underset{k\in\mathcal{K}}{\sum} \underset{\tau=\mathrm{tl}^2}{\overset{F_v^{2}(\omega_{\gamma})}{\sum}}\delta_{k,\tau} \beta_{k,\tau}^{P^2,s,2}(\omega_{\gamma}) \label{dominance:red_cost_dmp}\\
    T_{v,i}^{1,\mathrm{perf}} &\geq T_{v,i}^{2,\mathrm{perf}} \quad \forall i\in\mathcal{I}_q\label{dominance:task_resources_dmp}\\ 
    \mathbb{P}(F_{v}^1 \leq \tau) &\geq \mathbb{P}(F_{v}^2 \leq \tau) \quad \forall \tau\in [\mathrm{ES}_v + p_{v,q}, \mathrm{LF}_v^{\mathrm{e}}]\label{dominance:stochastic_dom_dmp}\\
    \bar{F}_v^{1} &= \bar{F}_v^2 \label{dominance:identical_median_finishes_dmp}\\
    F_v^{1}(\omega_{\gamma}) &= F_v^{2}(\omega_{\gamma})\label{dominance:quantile_case_finishes_dmp}\\
    \mathrm{tl}^1 &\geq \mathrm{tl}^2 \label{dominance:depot_leave_dmp}
\end{align}
where $\bar{F}_v^1$ and $\bar{F}_v^2$ are the median finish times of labels $L^1$ and $L_2$ at task $v$, respectively.

\end{definition}
Dominance rule \ref{dominance_rule_dmp} offsets the reduced costs of $L^1$ and $L^2$ by the workforce penalty of each path and imposes an additional restriction (\ref{dominance:depot_leave_dmp}) on their depot leave times. It is easy to see that if $L^1$ dominates $L^2$ in $\mathcal{G}^{q,s}$, it also dominates $L^2$ with respect to the dominance rule described in Definition \ref{dominance_rule_amp}. However, the converse is not true.
\begin{figure}[ht]
\centering
\includegraphics[scale=0.38]{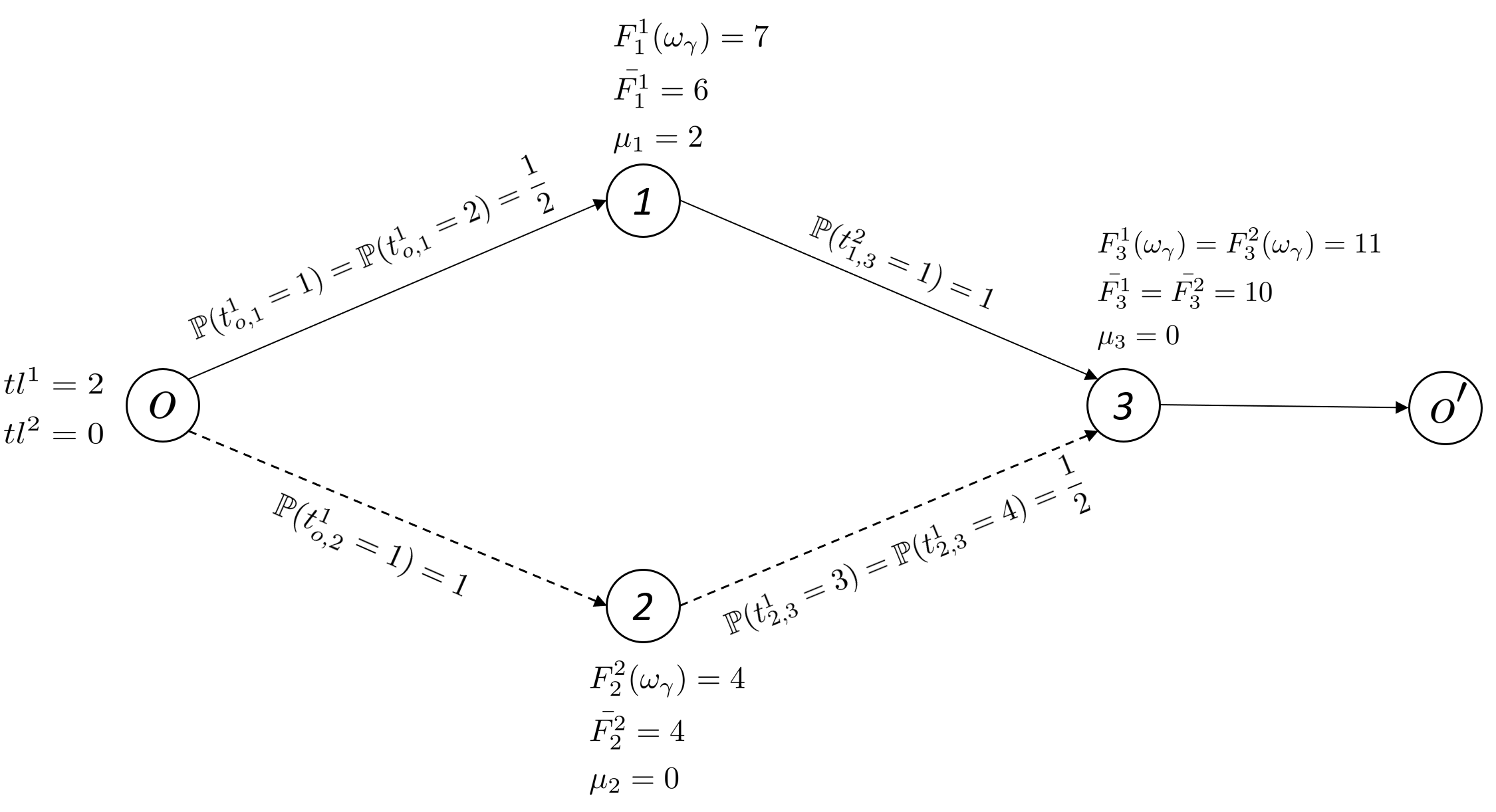} % ex_dominance.PNG
\caption{Labels satisfying dominance rule \ref{dominance_rule_amp} and violating rule \ref{dominance_rule_dmp}}
\label{graphics:example_dominance_rules}
\end{figure}\\

\begin{example} Figure \ref{graphics:example_dominance_rules} visualizes the pricing network of Figure \ref{graphics:example_label_extension} alongside an additional path $L^2$. Its path is denoted by dotted lines and equals the subtour $(o,2,3)$. In addition to the assumptions made in Example \ref{ex:label_extension}, we set $p_{2,q}=3$, $tl^2 = 0$, $\mu_2 = 0$, $\delta_{1,\tau} = - 2$ for $\tau=0,1$, $s_{q,1} = 1$, $w_2 = 1$, and $LF_2^e = 10$.
The reduced costs for labels $L^1$ and $L^2$ are then equal to $T_1^{1,\mathrm{cost}} = 24$ and $T_2^{2,\mathrm{cost}} = 27.5$.
Therefore, it is clear that $L^1$ dominates $L^2$ in the sense of Definition \ref{dominance_rule_amp}. However, if we offset the reduced cost of each label by the workforce penalties, we obtain
\begin{equation*}
    T_{1}^{2,\mathrm{cost}} + \underset{\tau=0}{\overset{10}{\sum}}\delta_{1,\tau} \beta_{k,\tau}^{P^2,s,2}(\omega_{\gamma}) = 27.5 - 13 = 14.5 < 15 = 24 - 9 = T_{1}^{1,\mathrm{cost}} + \underset{\tau=2}{\overset{10}{\sum}}\delta_{1,\tau} \beta_{k,\tau}^{P^1,s,1}(\omega_{\gamma})
\end{equation*}
Hence, $L^1$ does not dominate $L^2$ in $\mathcal{G}^{q,s}$, i.e., in the sense of Definition \ref{dominance_rule_dmp}. The dominance rule presented here tends to be more strict, leading to around 25\% less dominated labels.
\end{example}

As the only difference between the pricing graphs $\mathcal{G}^{q,\tilde{s}}$ and $\mathcal{G}^{q,s}$ for $q\in\mathcal{Q}$ and $s,\tilde{s}\in\mathcal{S}_q$ lies in the workforce penalty for the time-dynamic arc weights, we can transfer any feasible label $L = \left(\mathrm{tl}^L, v, P, T_{v}^{\mathrm{cost}}, (T_{v,i}^{\mathrm{perf}})_{i\in\mathcal{I}_q}, F_{v}^L\right)$ in $\mathcal{G}^{q,s}$ to a different pricing network $\mathcal{G}^{q,\tilde{s}}$ and obtain a new feasible label $\tilde{L}$ in $\mathcal{G}^{q,\tilde{s}}$ with
\begin{equation*}
    \tilde{L} = \left(\mathrm{tl}^L, v, P, \tilde{T}_{v}^{\mathrm{cost}}, (T_{v,i}^{\mathrm{perf}})_{i\in\mathcal{I}_q}, F_{v}^L\right)
\end{equation*}
that is, all properties besides the reduced cost $T_{v}^{\mathrm{cost}}$ of $L$ remain the same.
As the dominance rule on $\mathcal{G}^{q,s}$ merely considers reduced costs after offsetting them by the workforce penalty, it is clear that if a label $L^1$ dominates another label $L^2$ in $\mathcal{G}^{q,s}$ if and only if it dominates $L^2$ in $\mathcal{G}^{q,\tilde{s}}$ for all $\tilde{s}\in\mathcal{S}_q$. For a formal proof of these relations, see \Cref{appendix:proof_equiv_pricing_networks}.\\
Therefore, for a profile $q\in\mathcal{Q}$, the column with minimum reduced cost for the set $\left((q,s)\right)_{s\in\mathcal{S}_q}$ of disaggregated profiles can be calculated as follows: we select an arbitrary skill composition $\tilde{s}\in\mathcal{S}_q$ and solve the ESPPRC on $\mathcal{G}^{q,\tilde{s}}$. After no further labels can be created, the following optimization problem on the set $\mathcal{L}$ of non-dominated labels present at the sink $o'$ is solved to obtain a route with profile $q$, an optimal skill composition $s^{*}$ and minimal reduced cost:
\begin{equation*}
    \min\left\{T^{L,\mathrm{cost}}_{o'} + \underset{\tau=\mathrm{tl}^L}{\overset{\mathrm{tr}^L}{\sum}}\delta_{k,\tau} \beta_{k,\tau}^{L,\tilde{s}} - \underset{\tau=\mathrm{tl}^L}{\overset{\mathrm{tr}^L}{\sum}}\delta_{k,\tau} \beta_{k,\tau}^{L,s}: \ s\in\mathcal{S}_q, \ L\in\mathcal{L}\right\}
\end{equation*}
Because the number of labels at the sink and the number of possible skill compositions is usually small, this problem can be solved by explicit enumeration. In total, the runtime savings by solving one pricing network (instead of multiple) for each profile exceeds the additional runtime caused by a weaker dominance rule by several orders of magnitude.\\
We note that the above approach is also applicable to general SPPRCs if several properties are satisfied. Let $\mathcal{N}$ be a set of pricing networks and $\mathcal{N} = \underset{i=1}{\overset{n}{\bigcap}}\mathcal{N}_i$ with $n \in\mathbb{N}$ be a disjunct partition of $\mathcal{N}$. Then, one can apply the above methodology to a subset $\mathcal{N}_i$ of pricing networks for an arbitrary $i\in\mathcal{N}$ if the following properties are satisfied:
\begin{itemize}
    \item[i)] a path $P$ induced by a label $L$ in a pricing network $\mathcal{G}$ in $\mathcal{N}_i$ is feasible in $\mathcal{G}$ if and only if it is feasible in all pricing networks in $\mathcal{N}_i$,
    \item[ii)] there exists an adjusted dominance rule that only depends on properties shared among all pricing networks in $\mathcal{N}_i$, i.e., it is agnostic of the specific pricing network in $\mathcal{N}_i$ that is considered, and
    \item[iii)] if a label $L$ dominates a label $\tilde{L}$ with respect to the adjusted dominance rule, one can prove that it dominates $\tilde{L}$ in all pricing networks in $\mathcal{N}_i$ with respect to the original dominance rule.
\end{itemize}
In the application at hand, the set of pricing networks is given by $\mathcal{N} \coloneqq \left\{\mathcal{G}^{q,s}: \ q\in\mathcal{Q}, s\in\mathcal{S}^q\right\}$ and its partition is $\mathcal{N}_q \coloneqq \left\{\mathcal{G}^{q,s}: \ s\in\mathcal{S}^q \right\}$ for all $q\in\mathcal{Q}$. Furthermore, path feasibility is defined by the satisfaction of constraints (\ref{chance_constraint})--(\ref{finish_at_lf_v}), which do not depend on the skill composition, i.e., (i) is satisfied by the definition of $\mathcal{N}_q$. Moreover, it is clear to see that the adjusted dominance rule (\ref{dominance:red_cost_dmp})--(\ref{dominance:depot_leave_dmp}) does not depend on the skill composition, which is the only property that is not shared among all pricing networks within $\mathcal{N}_q$.
Finally, the usage of offset reduced costs (\ref{dominance:red_cost_dmp}), jointly with inequality (\ref{dominance:depot_leave_dmp}), allows for the proof of dominance with respect to the original dominance rule (\ref{dominance:red_cost})--(\ref{dominance:quantile_case_finishes}) in all pricing networks in $\mathcal{N}_q$ if dominance with respect to (\ref{dominance:red_cost_dmp})--(\ref{dominance:depot_leave_dmp}) is satisfied in an arbitrary pricing network in $\mathcal{N}_q$. 
For additional details on how the equivalence of pricing networks can be derived using properties i)--iii), we refer to the proof presented in \Cref{appendix:proof_equiv_pricing_networks}.

\vspace{0.5cm}
\subsection{Acceleration Strategies}\label{section:acceleration_strategies}
As the pricing step is by far the most runtime-intensive component in the proposed solution approach, we employ several strategies that allow us to generate columns with negative reduced costs more quickly. In the following, we describe these techniques in detail.
\paragraph{\textbf{Graph Size Reduction}}\text{ }\\
We use two heuristics that initially reduce the graph size and the set of initial labels and gradually increase them when necessary. The first heuristic consists of separating the set of possible depot leave times into \textit{containers} \{1,\dots,C\} of equal size and sorting these containers in an ascending order with respect to the minimum reduced cost of all labels inside the container. We then solve the ESPPRC considering only the initial labels in container $1$. If no column with negative reduced costs has been found, we continue with the next container until we either find a negative column or solve the final container $C$. In our algorithm, we set the value $C$ such that each container $\{1,\dots,C-1\}$ has a size of $5$.\\
The second heuristic is similar to the one proposed by \cite{desaulniers2006tabu}. For each task node $i$, we sort the set of outgoing arcs $(i,j)\in\mathcal{A}^q$ with $j\in\mathcal{I}_q$ based on their task duals $\mu_j$. Initially, we only allow extensions along $(i,j)$ if $\mu_j$ is among the $\Delta$ largest dual values of task nodes adjacent to $i$. If we do not find a negative column, we solve the pricing problem while allowing extensions along all arcs.\\
In our experiments, we set $\Delta = 4$ as this returned the best results. Furthermore, we first iterate through all initial label containers and then switch to allowing extensions along all arcs only if no negative column has been found in any of the containers.
\vspace{0.7em}\paragraph{\textbf{Decremental State Space Relaxation}}\text{ }\\
State space relaxation has first been proposed by \cite{christofides_state} and relies on reducing the dimension of the state space, i.e., the number of resources tracked. Hence, the ESPPRC is solved using labels with fewer resources, allowing for generally stronger dominance relations. We adopt the approach on decremental state space relaxation proposed by \cite{righini2008new}. In each pricing step, we initially replace task resources $\left(T_{v,i}^{\mathrm{perf}}\right)_{i\in\mathcal{I}}$ by the length $l_P$ of the path $P$ corresponding to $L$.Dominance rules \ref{dominance_rule_amp} and \ref{dominance_rule_dmp} are then adjusted by replacing inequality (\ref{dominance:task_resources}) or (\ref{dominance:task_resources_dmp}) with $l_1 \leq l_2$. Whenever a negative column that visits a task $i\in\mathcal{I}$ multiple times is found, we consider this vertex to be critical and re-activate the corresponding task resources $T_{v,i}^{\mathrm{perf}}$ by re-inserting it into the dominance rule. This step is repeated until the optimal column does not contain any cycles or has non-negative reduced costs. We note that in each pricing step, the set of critical vertices is initially empty.\\
Furthermore, \cite{righini2008new} propose directional labeling, during which labels are also extended backwards and eventually joined with forward labels, which often greatly reduces the number of explored labels. In order to apply this technique to the problem at hand, when backwards-extending a label $L\coloneqq \left(\mathrm{tl}^L, v, P, T_{v}^{cost}, (T_{v,i}^{perf})_{i\in\mathcal{I}_q}, F_{v}^L, \right)$ along $(v',v)$, it needs to be possible to derive the finish time distribution $F_{v'}^{L'}$ based on $F_{v}^{L}$ and the distribution of travel times $\left(\mathbbm{P}(t^{k}_{v',v} \leq \tau)\right)_{\tau\in B_{i,j}}$. However, when distributions are left-truncated as in the case of hard time windows, this is generally not the case (see \Cref{appendix:backward_label}). Alternatively, one can extend the node set of the graph such that distributional properties that are relevant for the stochasticity considered are embedded in the nodes. For instance, \cite{GAUVIN2014141} consider stochastic demand and minimize routing costs that consist of deterministic travel costs and expected costs of failure, which in turn depend on the cumulative demand distribution at each customer. The authors assume independent, Poisson-distributed demands. Hence, the cumulative demand distribution is again Poisson-distributed. This allows them to define a graph that contains nodes for each pair of customer and distributional parameter, i.e., the expected cumulative demand at the current customer. We note that this approach heavily relies on the possibility of propagating distributional properties, in this case, a Poisson distribution, along vehicle routes. Once distributions are truncated, as is the case for hard time windows, this is generally no longer possible.\\
Moreover, \cite{righini2008new} and \cite{BOLAND200658} present several other methods to solve the ESPPRC, which rely on relaxing task resources and only adding them if needed. \cite{righini2008new} develop a branch-and-bound framework, while \cite{BOLAND200658} propose several algorithms to forbid node repetitions. We note that in the test instances studied, time windows are often so tight that we rarely observed cycles, hence implementing such advanced methods to deal with non-elementary paths does not have a significant impact on the algorithm's performance for the problem at hand.

\vspace{0.5cm}
\subsection{Branching Strategies}\label{section:branching_strategies}
We use three different branching rules to cut off fractional solutions of the ARMP or DRMP. In the following, we summarize these strategies and their technicalities.
\paragraph{\textbf{Branching on Task Finish Times}} \text{ }\\
\cite{gelinas1995new} were among the first ones to branch on resource windows, more specifically, time windows. While this strategy is not novel, the process of selecting a task and a time instant to branch on has a significant impact on the branching rule's efficiency. Let $(\bar{\lambda}^r)_{r\in\mathcal{R}}$ be an optimal solution to the ARMP and let $\mathcal{R}_{\mathrm{F}}\subseteq\mathcal{R}$ be the set of columns for which $\bar{\lambda}^r$ is fractional. For each task $i\in\mathcal{I}$, we define
\begin{align*}
    &\mathcal{B}^{\mathrm{F}}(i) \coloneqq \left\{F_i^{r,q}(\omega_{\gamma}): \ (r,q)\in\mathcal{R}_{\mathrm{F}}: \ i\in r \right\},\\
    &\mathcal{B}^{\mathrm{F}}_U(i) \coloneqq \left\{F_i^{r,q}(\omega_{\gamma}): \ (r,q)\in\mathcal{R}_{\mathrm{F}}: \ i\in r, \ F_i^{r,q}(\omega_{\gamma}) \ \text{unique}\right\}
\end{align*}
as the set of all (unique) $\omega_{\gamma}$-scenario finish times of task $i$ within team routes with fractional values, respectively. We then select all tasks for which the cardinality of the latter set is maximal and denote it by $\mathcal{I}^{\mathrm{F}}$, i.e.,
\begin{equation*}
    \mathcal{I}^{\mathrm{F}} = \left\{i\in\mathcal{I}: \left\vert \mathcal{B}^{\mathrm{F}}_U(i) \right\vert = \max\left\{\ \left\vert \mathcal{B}^{\mathrm{F}}_U(j) \right\vert: \ j \in\mathcal{I}\right\} \right\}.
\end{equation*}
We then select the task $i^*$ for which the standard deviation of the  $\omega_{\gamma}$-scenario finish times of task $i^*$ in all routes in $\mathcal{B}^{\mathrm{F}}(i)$ is maximal, that is
\begin{equation*}
    i^{*} = \argmax\left\{\sigma\left(\left\{F_i^{r,q}(\omega_{\gamma}): \ (r,q)\in\mathcal{B}^{\mathrm{F}}(i)\right\}\right): \ i\in\mathcal{I}^{\mathrm{F}}\right\}
\end{equation*}
where $\sigma(\cdot)$ is the standard deviation of a finite set. We then branch on task $i^{*}$ and time instant $\tau^{*} = \left\lfloor M(\mathcal{B}^{\mathrm{F}}(i^{*}))\right\rfloor$,
where $M(\cdot)$ is the median of a finite set. We then create two child nodes and impose constraints
\begin{equation*}
    F_{i^{*}}^{r,q}(\omega_{\gamma}) \leq \tau^{*} \quad \forall (r,q)\in\mathcal{R}: \ i^{*}\in r \quad \text{or} \quad
    F_{i^{*}}^{r,q}(\omega_{\gamma}) > \tau^{*} \quad \forall (r,q)\in\mathcal{R}: \ i^{*}\in r,
\end{equation*}
respectively. Furthermore, we remove all tours that violate the constraints from the child nodes. In the pricing steps, we forbid extensions that would violate these task finish time constraints. If 
\begin{equation*}
    \max\left\{\sigma\left(\left\{F_i^{r,q}(\omega_{\gamma}): \ (r,q)\in\mathcal{B}^{\mathrm{F}}(i)\right\}\right): \ i\in\mathcal{I}^{\mathrm{F}}\right\} = 0
\end{equation*}
holds, we know by construction that in the current solution, $\left\vert\mathcal{B}^{\mathrm{F}}_U(i)\right\vert = 1$ holds for all $i\in\mathcal{I}$. Therefore, independent of the selection of $i^{*}$ and $\tau^{*}$, the above branching rule produces a trivial branch. If this is the case, a different branching strategy must be used. We note that the above branching strategy neither changes the master problems nor the pricing problem's structure.
\vspace{0.7em}\paragraph{\textbf{Branching on Number of Tours at a Given Time}}\text{ }\\
\cite{desrochers_branch_on_vehicle_count} introduced branching on tour counts, which has proven to be efficient for vehicle routing problems. Let $\tau^{*}$ be the time instant for which the number of tours with fractional $\lambda$ value that are active at time $\tau^{*}$ is maximal, and the sum $l_{\tau^{*}} = \underset{(r,q)\in\mathcal{R}}{\sum} g_{\tau^*}^r\lambda^r_q$ is closest to 0.5, where $g_\tau^r = 1$ if $\mathrm{tl}^{r,q} \leq g_\tau^r \leq \mathrm{tr}^{r,q}$ and $g_\tau^r = 0$ else. We then branch on the current solution by imposing
\begin{equation}
    \underset{(r,q)\in\mathcal{R}}{\sum}g_{\tau^{*}}^r\lambda^r_q \leq \lfloor l_{\tau^{*}}\rfloor \quad \text{or} \quad
    \underset{(r,q)\in\mathcal{R}}{\sum}g_{\tau^{*}}^r\lambda^r_q \geq \lceil l_{\tau^{*}}\rceil.\label{branching:vehicle_cnt}
\end{equation}
The pricing problem has to be adjusted in the sense that time-dynamic arc weights have to consider the dual cost of the additional inequalities. If a label $L$ is extended along an arc $(v,v')$ and $F_{v'}^L(\omega_{\gamma}) < \tau^* \leq F_{v'}^L(\omega_{\gamma})$ holds, the cost of arc $(v,v')$ must be adjusted by the dual cost of (\ref{branching:vehicle_cnt}). Similar to branching on task finish times, this branching strategy can potentially produce a trivial branch, making a fallback branching option necessary.
\vspace{0.7em}\paragraph{\textbf{Branching on Variables}}\text{ }\\
If both previously described branching rules fail to produce a nontrivial branching decision, we select the most fractional variable $\lambda^{r^{*}}_{q^*}$ and set it to $0$ or $1$ in the child nodes. When a team route $(r^{*},q^{*})$ is forced, i.e., $\lambda^{r^{*}}_{q^*} = 1$ is enforced, we remove all tasks which are visited by $r^{*}$ from the pricing networks, discard all team routes that share tasks with $r^{*}$ and adequately reduce the total available workforce $N_k$ and $N_k^{\mathrm{D}}$ for all $k\in\mathcal{K}$. When a team route $(r^{*},q^{*})$ is forbidden, we introduce an additional resource to the pricing problem that tracks how many arcs a label has used that correspond to segments of $r^{*}$. If this resource is fully consumed, we discard the label as it equals a forbidden tour. For the DRMP, we do not discard said labels but skip the skill composition $s$ that is used to execute the forbidden route. Furthermore, if a label is equal to a subpath of a forbidden tour, it can not dominate any other label as this might lead to cutting off optimal labels.

\vspace{0.5cm}
\subsection{Cutting Planes}\label{section:cutting_planes}
Whenever a fractional optimal solution is found at a node in the branching tree, we model the problem of finding a most violated rank-1 Chvátal-Gomory cut (CGC) as a mixed-integer problem and solve it using a generic solver with a very short time limit. If optimality could not be proven within the time limit, the best solution found so far is returned. If no solution is found, the cutting plane step is skipped and the algorithm proceeds with a branching step. This approach was first proposed by \cite{fischetti2007optimizing} and applied to a vehicle routing problem with time windows by \cite{petersen2008chvatal}. In the following, we base our considerations on the ARMP formulation. Let $(u_i)_{i\in\mathcal{I}} \in [0,1)$ and $(u_{k,t})_{k\in\mathcal{K},t\in\mathcal{T}}$ be the coefficients of the most violated cut. We then add the constraint
\begin{equation*}
    \underset{(r,q)\in\mathcal{R}}{\sum}\left\lfloor\underset{i\in\mathcal{I}^r}{\sum} u_i + \underset{k\in\mathcal{K}}{\sum} \ \underset{\tau=\mathrm{tl}^{r,q}}{\overset{\mathrm{tr}^{r,q}}{\sum}} b_{k,\tau}^{r,q}(\omega_{\gamma})u_{k,t}\right\rfloor \lambda^r_q \leq \left\lfloor\underset{i\in\mathcal{I}^r}{\sum} u_i + \underset{k\in\mathcal{K}}{\sum} N_k \cdot \left(\underset{\tau\in\mathcal{T}}{\sum} u_{k,t}\right)\right\rfloor
\end{equation*}
to the master problem and re-solve the current node.\\
Let $\mathcal{G}$ be the set of indices of CGCs and $\mathcal{C}^{\mathcal{G}} \coloneqq \{\left((u_i^g)_{i\in\mathcal{I}}, (u_{k,t}^g)_{k\in\mathcal{K},t\in\mathcal{T}}\right): \ g\in\mathcal{G}\}$ be the cut coefficients for all CGCs present at the current node. During the pricing step, we introduce an additional resource $T_{v,g}$ for each CGC $g\in\mathcal{G}$. When extending a label $L\coloneqq \left(\mathrm{tl}^L, v, P, T_{v}^{\mathrm{cost}}, (T_{v,i}^{\mathrm{perf}})_{i\in\mathcal{I}_q}, (T_{v,g})_{g\in\mathcal{G}}, \mathcal{F}_{F_{v}^L}\right)$ along an arc $(v,v')$, we update the resources $(T_{v,g})_{g\in\mathcal{G}}$ and $T_{v'}^{\mathrm{cost}}$ using the following resource extension function:
\begin{align*}
    T_{v'}^{\mathrm{cost}} &= T_{v}^{\mathrm{cost}} + \left\lfloor T_{v,g} + u_{v'}^g + \underset{k\in\mathcal{K}}{\sum} \ \underset{\tau = F_{v}^{L}(\omega_{\gamma})}{\overset{F_{v'}^{L'}(\omega_{\gamma})}{\sum}} b_{k,\tau}^{P,q}(\omega_{\gamma})u_{k,t}^g \right\rfloor \cdot \psi_g\\
    T_{v',g} &= T_{v,g} + u_{v'}^g + \underset{k\in\mathcal{K}}{\sum} \ \underset{\tau = F_{v}^{L}(\omega_{\gamma})}{\overset{F_{v'}^{L'}(\omega_{\gamma})}{\sum}} b_{k,\tau}^{P,q}(\omega_{\gamma})u_{k,t}^g - \left\lfloor T_{v,g} + u_{v'}^g + \underset{k\in\mathcal{K}}{\sum} \ \underset{\tau = F_{v}^{L}(\omega_{\gamma})}{\overset{F_{v'}^{L'}(\omega_{\gamma})}{\sum}} b_{k,\tau}^{P,q}(\omega_{\gamma})u_{k,t}^g \right\rfloor
\end{align*}
where $\psi_g \geq 0$ is the dual variable associated with CGC $g\in\mathcal{G}$. Dominance rule \ref{dominance_rule_amp} is then slightly adjusted and extended:
\begin{definition}[Dominance Rule]\label{dominance_rule_with_cgs}
Let 
\begin{equation*}
    L^k\coloneqq \left(\mathrm{tl}^k, v, P^k, T_{v}^{k,\mathrm{cost}}, (T_{v,i}^{k,\mathrm{perf}})_{i\in\mathcal{I}_q},  (T_{v,g}^k)_{g\in\mathcal{G}}, F_{v}^{k}\right)
\end{equation*}
be labels for $k=1,2$. Let $\mathcal{G}$ be the set of indices of CGCs and $\mathcal{G}^{>} \coloneqq \left\{g\in\mathcal{G}: \ T_{v,g}^1 > T_{v,g}^2\right\}$.
We say that $L^1$ dominates $L^2$ if constraints (\ref{dominance:task_resources})--(\ref{dominance:quantile_case_finishes}) and
\begin{align}
    T_{v}^{1,\mathrm{cost}} + \underset{g\in\mathcal{G}^{>}}{\sum} \psi_g &\leq T_{v}^{2,\mathrm{cost}} \label{dominance_cg:red_cost} \\
    T_{v,g}^1 &\leq T_{v,g}^2 \quad \forall g\in\mathcal{G}\setminus\mathcal{G}^{>}  \label{dominance_cg:coefficients_leq}
\end{align}
hold.
\end{definition}
Inequalities (\ref{dominance_cg:red_cost}) and (\ref{dominance_cg:coefficients_leq}) ensure that for any feasible extension $(v,v')$ of $L^2$, the reduced costs still remain dominated by $L^1$. For a proof of correctness of this approach, we refer to Section 4 of \cite{petersen2008chvatal}.\\
Identifying violated CGCs can be quite costly if repeated frequently. Furthermore, each CGC slightly weakens the dominance rule due to the additional constraints (\ref{dominance_cg:coefficients_leq}), and calculating the coefficients $T_{v,g}$ requires additional computational effort. Hence, limiting the maximum amount of CGC present at any node in the search tree can be beneficial. Finally, we note that the above CGCs remain feasible when the master problem formulation is switched to the DRMP.

\vspace{0.5cm}
\subsection{An Early Termination Heuristic}\label{section:heuristic_procedure}
Because several exterior factors, such as unexpected aircraft delays, can create the need for re-optimization, limiting the maximum runtime of the Branch-Price-Cut-and-Switch scheme is often necessary. Let $\bar{\mathcal{R}}$ and $\bar{\mathcal{R}^{\mathrm{D}}}$ be the sets of columns found during the branch-and-price procedure for the ARMP and DRMP, respectively. We then define the set $\mathcal{R}^{\mathrm{D},\mathrm{H}}$ as
\begin{equation*}
    \mathcal{R}^{\mathrm{D},\mathrm{H}} \coloneqq \left\{(r,q,\tilde{s}): (r,q) \in \bar{\mathcal{R}}, \tilde{s}\in\mathcal{S}_q\right\} \ \cup \ \left\{(r,q,\tilde{s}): \exists s\in\mathcal{S}_q: \ (r,q,s) \in \bar{\mathcal{R}^{\mathrm{D}}}, \tilde{s}\in\mathcal{S}_q \right\}
\end{equation*}
Whenever the time limit is reached, we solve the DMP (\ref{model:obj_disaggregated_mp})--(\ref{model:lambda_binary_disaggregated_mp}), including integrality constraints, on the column set $\mathcal{R}^{\mathrm{D},\mathrm{H}}$. If a feasible solution has been found within a prespecified time limit, we return the solution as the best integer solution found.

%% file: sections/computational_study.tex
\section{Experimental Study}\label{section:experimental_study}
In the following, we analyze the impact of cutting planes, switching between the DRMP and ARMP as
master problem formulations, and different branching strategies on solution quality and convergence speed. 
Furthermore, we compare stochastic and deterministic solutions and evaluate the impact of stochasticity on optimal strategies.\\
Section \ref{section:instance_set} elaborates on the generation of test instances. Section \ref{section:influence_components} summarizes the algorithm's performance for different configurations of the aforementioned components, as well as a performance comparison to the approach proposed by \cite{yuan2015branch}. In Section \ref{section:comparison_stoch_det}, stochastic and deterministic optimal policies are compared with respect to their practical feasibility. The proposed algorithmm as well as the benchmark approach by \cite{yuan2015branch}, were implemented using Python 3.12 and Gurobi 12.0.2. The codebase for said algorithms can be found under \url{https://github.com/andreashagntum/BC_APR}.
Furthermore, all computational studies were performed on a single machine equipped with an Intel® Core® i7-8700K 11th 12-core 3.7GHz processor, 32GB of RAM and running Windows 11. All instances and corresponding solutions can be found under \url{https://github.com/andreashagntum/StochasticTeamFormationRoutingAirport}.\\

\subsection{Instance Set}\label{section:instance_set}
We use the instance generator developed by \cite{dallolio_temp} to construct a set of test instances of different complexity. In short, a flight plan for a given time horizon is generated based on real flight schedules and aircraft usage. The workforce size depends on the instance and varies between 10\% and 90\% of the workforce required to trivially solve the instance (i.e., each task is started at its earliest possible time and is executed using the profile that requires the shortest task execution time). Profiles, skill compositions, and task execution times are defined based on aircraft manual data. For a detailed summary of the process of test instance generation, we refer to Section 6.1 of \cite{dallolio_temp}.
In the following, we briefly elaborate on parameters that have been varied throughout the generation.\\
In order to generate instances of different sizes, we vary the length of the planning horizon, as well as the frequency of flights. At Munich Airport, shift planning is typically done in three shifts (early, late, and night). Furthermore, the frequency of arrivals and departure can be categorized into low-frequency (i.e. \textit{off-peak}) and high-frequency (i.e. \textit{peak}) time periods. Each shift contains up to two such peaks, and each peak lasts for 1 to 2 hours and contains between 10 and 30 flights per hour. Because shift planning is done based on the workforce demand in peak times, the problem at hand becomes trivial to solve outside of peak periods due to the abundance of workforce. Hence, in the following, we focus on the aforementioned peak periods and create instances with a planning horizon between 60, 90, and 120 minutes and 10, 20, and 30 flights per hour. The underlying time grid consists of equidistant time steps with a length of 2 minutes. Each type of airplane served at Munich Airport can be loaded using one of up to 3 different team formations, called \textit{slow}, \textit{intermediate}, and \textit{fast mode}. Faster modes use up more workers but require less time to (un)load an aircraft. Not every plane can be loaded with all 3 modes, especially smaller planes typically only support slow or fast modes. We note that the same properties hold for unloading tasks.
We generate multiple instance sets by restricting the available team formations to only the intermediate mode, to the slow and the fast mode, or to all three modes, respectively. Furthermore, the available workforce ranges between 10\% and 90\% of the workers required when all tasks are started at the earliest possible time with the fastest possible mode. In the following, this factor is referred to as \textit{worker strength}. Lastly, the minimum service level is set to $\alpha = 0.9$ and the extended latest finish time is set to $\mathrm{LF}_i^{\mathrm{e}} = \mathrm{LF}_i + 5$, i.e., 5 time steps after the latest finish time. We generated five random flight schedules for each combination of the aforementioned parameters, leading to a total of $3\cdot 3\cdot 3\cdot 9 \cdot 5 = 1,215$ instances. We note that the largest test instances in this set, namely instances with a 2-hour planning horizon, 30 flights per hour, and three available modes, replicate the most complex instances one might encounter at Munich Airport, which are only observed during on-peak times. Furthermore, at Munich Airport, workforce planning is done based on 8-hour shifts, each of which includes up to two on-peak slots of at most 2 hours in length. Hence, during off-peak times, there is sufficient workforce available to decide on team formations and task assignments in a straightforward fashion, i.e., each task is executed by a team, which immediately returns back to the depot after finishing said task. We note that this is the approach currently taken at Munich Airport. However, during on-peak times, one needs to ensure that the available workforce is utilized efficiently in order to execute all scheduled tasks within their specified time windows. \\
All further properties, such as flight schedules and task execution times, are calculated as described by \cite{dallolio_temp}. For a precise description of the process, we refer to Section 6.1 and Appendices E and F of their paper.\\
We approximate the distribution of travel times between locations using a quadratic regression model that aims to predict travel times based on several exogenous factors. For this purpose, we collected real-world data on travel times of various ground operating vehicles, such as passenger buses and baggage handling vehicles, at Terminal 2 of Munich Airport for the entire year of 2024. These values, if needed, are adjusted to baggage handling vehicles by scaling them proportionally by their respective average speeds. We then categorized all datapoints based on several features, which in preliminary internal studies were identified to have a significant impact on travel times: weather conditions (cloudy, sunny, rainy, snowy), time during the day (on-peak or off-peak) and the week (weekday or weekend, holiday season yes or no), number of required apron crossings to get from origin to destination gate, distance, and best-case travel time. To estimate travel time distributions for a given gate pair under specified external conditions and a given time bin, we then apply the quadratic regression model to predict the percentual deviation from the shortest travel time. To account for variability, we then identify historical observations from other gate pairs in the same time bin with similar deterministic travel times, distances, and numbers of crossings. For these similar observations, we calculate the prediction errors of the percentual delay under their respective external conditions. Finally, we construct the travel time distribution for the target gate pair as follows. For each residual obtained from the aforementioned similar observations, we add the residual to the predicted percentual delay for the target gate pair and multiply the result by the best-case travel time. The resulting set of datapoints is then used to derive an empirical distribution for the given edge in the give time bin.\\
For the test instances used in this section, we assumed an intra-day on-peak time on a sunnyt weekday during holiday season. In practice, this setting resembles a typical high-demand day at the beginning of summer holiday season.

\subsection{Influence of the Algorithm’s Features}\label{section:influence_components}
From a methodological point of view, our solution approach differs from the ones typically used for variants of vehicle routing or team formation problems in three ways: dynamically switching between two master problem formulations, branching on task finish times and an exact separation of rank-1 CGCs. In the following, these are called the algorithm's \textit{features}. In this section, we compare a total of six solver configurations: we first enable and then disable all three features, resulting in what are called `full' and `basic' configurations (or algorithms), respectively. Furthermore, we implemented the branch-and-price approach proposed by \cite{yuan2015branch}, which differs from our approach in several ways. First, they do not use cutting planes and employ branching on arcs as their sole branching rule. Second, they apply a greedy and a neighborhood search heuristic to find negative columns before applying a labeling algorithm. Third, they only use a disaggregated master problem formulation, i.e., they do not switch between master problem formulations, and always return all negative columns found rather than just the most negative column. Fourth, while we aggregate multiple pricing networks into one by comparing offset reduced costs and adding an additional restriction (\ref{dominance:depot_leave_dmp}) on the depot leave times, they demand reduced cost dominance for all skill compositions. Thus, unlike us, they implicitly solve all pricing networks exactly. While \cite{yuan2015branch} consider a simplified variant of our problem and assume stochastic service times instead of travel times, their proposed approach can be adapted to solve the problem at hand.\\
Moreover, we disable exactly one feature to obtain three additional variants of our solution method called `no DRMP', `no CGCs' and `no branching on task finish times' (also abbreviated as `no branching'). For instance, the configuration `no DRMP' uses branching on task finish times and adds up to 12 CGCs at the root node, but does not dynamically switch to the DRMP master problem formulation. We note that we do not include solver configurations with exactly two features enabled in our analysis, as the results do not fundamentally differ from the ones presented in the following. Furthermore, except for paragraph \textit{Statistical Tests on Feature Impacts}, all results reported in this subsection are based on a test instance set with workforce travel time quantile $\gamma=0.9$. While we also conducted analogous analyses for the same test instances using $\gamma\in\left\{0,0.1,0.3,0.5,0.7,1\right\}$, the performance of the proposed algorithm does not fundamentally differ.
\paragraph{\textbf{Feature Design}}\text{ }\\
We configure the aforementioned features as follows. When branching, we first try to branch on task finish times by using the procedure described in Section \ref{section:branching_strategies}. If that is not possible, we look for branches on vehicle counts and, if this also fails, we use the fallback option of branching on variables. Furthermore, whenever a disaggregated-infeasible solution is found, we switch to solving the DRMP formulation at the current node and its sibling node. Additionally, child nodes of DRMP nodes inherit their master problem's type, i.e., they are also solved using the DRMP formulation. Finally, we only search for violated CGCs at the root node and add up to 12 cuts. The MIP required to identify such cuts, as described by \cite{fischetti2007optimizing} is solved using Gurobi with a CPU time limit of 0.3 seconds. For each instance, a hard time limit of 180 seconds is imposed. If no optimal solution has been found, an upper bound is obtained by the procedure described in Section \ref{section:heuristic_procedure}. We note that preliminary studies have shown that adding up to 12 cuts and imposing a CPU time limit of 0.3 seconds, on average, provides an optimal trade-off between lower bound improvements and an increase in computational complexity.
\paragraph{\textbf{Instance Set and Instance Classes}}\text{ }\\
Each of the aforementioned 5 configurations is used to solve the instance set generated as described in Section \ref{section:instance_set}, amounting to a total of 1,215 instances per configuration. In total, 614 instances are infeasible. We omit said instances and compare the ascribed feature configurations on the remaining 601 instances. \Cref{appendix:feasible_instance_count} contains additional details on the characteristics of all feasible instances.\\
In order to distinguish between easy and hard instances, we split the set of test instances into three categories. In preliminary studies, the worker strength has proven to have a significant impact on an instance's complexity. Therefore, we consider instances with a worker strength between 0.3 and 0.5 as `hard', while instances with a strength of 0.6 or 0.7 are seen as `medium' and 0.8 or 0.9 as `easy'. In fact, almost all instances in the latter category were solved within a few seconds, whereas instances in the hard category are frequently not solved to optimality within the set time limit. We note that all instances with a worker strength of 0.2 or less are infeasible, while only 3 instances with a worker strength of 0.3 are feasible.\\
Unless otherwise stated, all tables in this section, contain average values. The number of instances on which the following analyses are based can be seen in \Cref{appendix:instance_set_sizes}.
\paragraph{\textbf{Comparison to Benchmark Approaches}}\text{ }\\
Table \ref{tab:basic_vs_full} compares the results obtained using the full configuration with the basic configuration.

\begin{table}[ht]
    \centering
    \caption{Comparison of \cite{yuan2015branch} with basic and full solver configuration}
    \label{tab:basic_vs_full}
    
\begin{tabular}{llrrrrr}
    \toprule
\multicolumn{1}{c}{Complexity} & \multicolumn{1}{c}{Sample size} & \multicolumn{1}{c}{\% Opt} & \multicolumn{1}{c}{Gap \%} & \multicolumn{1}{c}{UB} & \multicolumn{1}{c}{LB} & \multicolumn{1}{c}{Runtime} \\ \midrule
Easy                           & Full                            & 99.23\%                    & 0.02\%                     & 2.77                   & 2.76                   & 6.12s                       \\ 
                               & Basic                           & 82.38\%                    & 6.51\%                     & 2.77                   & 2.06                   & 33.85s                      \\
                               & Yuan et al.                     & 90.04\%                    & 2.18\%                     & 2.77                   & 2.46                   & 33.18s                      \\  \midrule
Medium                         & Full                            & 92.49\%                    & 0.25\%                     & 14.95                  & 14.91                  & 31.16s                      \\
                               & Basic                           & 61.50\%                    & 6.60\%                     & 14.96                  & 13.81                  & 78.97s                      \\
                               & Yuan et al.                     & 62.44\%                    & 5.46\%                     & 14.97                  & 13.99                  & 90.90s                      \\  \midrule
Hard                           & Full                            & 66.14\%                    & 2.58\%                     & 68.08                  & 64.94                  & 86.89s                      \\
                               & Basic                           & 48.03\%                    & 4.26\%                     & 68.34                  & 64.44                  & 112.36s                     \\
                               & Yuan et al.                     & 34.13\%                    & 14.24\%                    & 67.63                  & 50.16                  & 138.52s                     \\  \midrule
All                            & Full                            & 89.85\%                    & 0.64\%                     & 20.89                  & 20.21                  & 32.06s                      \\
                               & Basic                           & 67.72\%                    & 6.07\%                     & 20.95                  & 19.40                  & 66.43s                      \\
                               & Yuan et al.                     & 68.50\%                    & 5.88\%                     & 20.72                  & 16.57                  & 75.79s   \\ \bottomrule                   
\end{tabular}
\end{table}

Column `\% Opt' contains the percentage of instances that have been solved to optimality.
It can be seen that the full configuration solves 22\% more instances to optimality than the basic configuration. Furthermore, it returns significantly better average optimality gaps for all instance classes, where this effect decreases with increasing worker strength. While the full configuration returns substantially better lower bounds for all instance classes, the upper bounds also increased slightly for harder instances. Moreover, runtimes for small and medium instances reduce drastically, while harder instances are solved around 26 seconds faster on average. Additionally, it can be observed that, while the algorithm by \cite{yuan2015branch} is able to outperform the basic configuration for easy instances, it struggles to do so for medium and hard instances. 
Finally, the proposed full configuration manages to return the best runtimes, bounds, and optimality percentages out of all compared approaches for all complexity classes.
\paragraph{\textbf{Statistical Tests on Feature Impacts}}\text{ }\\
To assess the statistical significance of performance differences between the full configuration and the configurations 'No DMP', 'No Branching', 'No CGCs', and 'Basic', respectively, we first computed the optimality gap percentages on all feasible instances for each of the aforementioned configurations. Then, we conducted a Friedman test at the significance level $\alpha_F = 0.05$ on the optimality gap percentages to assess whether there are significant performance differences for at least one pair of configurations. The resulting $p$-value is $<10^{-300}$ and thus indicates that, for at least one pair of configurations, there is a statistically significant difference in the solution gap. Hence, we conducted a sequence of pairwise permutation tests at the significance level $\alpha_P = 0.05$ to assess whether differences in performance between the full configuration and each of the four previously mentioned configurations are due to true over- or underperformance. For this, for each configuration and each instance, we computed a percentual gap difference $d_i \coloneqq gap_{other}^i - gap_{full}^i$, where $gap_{full}^i$ is the percentual optimality gap for instance $i$ obtained using the full configuration, and $gap_{other}^i$ is the corresponding value obtained using the compared-to configuration, i.e., 'No DMP' when comparing the full configuration to the configuration 'No DMP'. Then, a permutation test (cp. \cite{good2005permutation} with the null hypothesis 
\begin{equation*}
    H_0: \text{the distribution of gap differences has median }\theta=\theta_0
\end{equation*}
with $\theta_0 = 0$ is conducted. We note that, while permutation tests assume a symmetric underlying distribution, \cite{romano1990behavior} has shown that the test remains asymptotically exact for sufficiently large sample sizes, even when the underlying distribution is asymmetric. If $H_0$ is rejected, it is reasonable to assume that the full configuration exhibits a statistically significant improvement in performance over the configuration to which it is compared. For additional details on the test design, including multiplicity correction methods, and confidence interval and effect size computations, we refer to \Cref{appendix:perm_test}.\\
\begin{table}[!htbp]
\centering
\caption{Results of permutation test on median gap difference (all instances)}
\label{tab:stat_test_all}
\begin{tabular}{lrrrrr}
\toprule
\multicolumn{1}{c}{Configuration} & \multicolumn{1}{c}{Sample size} & \multicolumn{1}{c}{Confidence interval}  & \multicolumn{1}{c}{$p$-value} & \multicolumn{1}{c}{Effect size} \\ \midrule
Full vs. Basic                             & $4277$                             & $(-\infty, 0.524\%]$     & $1.0$                           & 0.9757                          \\
Full vs. No Branching                      & $4277$                             & $(-\infty, 0.472\%]$     & $1.0$                        & 0.9380                          \\f90\%
Full vs. No CGCs                           & $4277$                             & $[0.00\%, 0.327\%]$   & $1.0$                          & 0.7843                          \\
Full vs. No DMP                            & $348$                              & $(-\infty, 0.284\%]$   & $1.0$                           & 0.9759     \\     
\bottomrule
\end{tabular}
\end{table}
Table \ref{tab:stat_test_all} shows the results of the permutation test conducted on the set of all feasible test instances. It can be seen that $H_0$ is accepted in all cases with a $p$-value of 1.0. Furthermore, confidence intervals only slightly reach above zero. Therefore, it is reasonable to assume that the full configuration, when measured based on the median of gap differences, either does not outperform each of the other four configurations noticeably or only does so on a small scale. However, this does not mean that the full configuration is not a structural improvement over the remaining configurations for two reasons. First, structural outperformance must not be exclusively measured based on median gap differences. Instead, additional location parameters, such as the average optimality gap, can aid in providing a more sophisticated analysis of solver performances. Said value will be further evaluated in the following paragraphs of this section. Second, the median difference of gaps is heavily drawn towards zero because, depending on the configuration, between 60\% and 80\% of instances were solved to optimality, resulting in frequent observation of gap differences that equal zero. Most of these instances are rather easy to solve to optimality because they have a large worker strength, or their planning horizons and flight frequencies per hour are rather low.\\
Because the full configuration is designed to provide competitive optimality gap percentages and solution quality across the entire test instance set and, most importantly, improve the aforementioned metrics on difficult instances, we repeated the above permutation test on restricted instance sets. For this, for each pair of configurations, we only considered the set of instances that were not solved to optimality by at least one of the two configurations. The results are summarized in Table \ref{tab:stat_test_subopt}.\\
\begin{table}[!htbp]
\centering
\caption{Results of permutation test on median gap difference (instances not solved to optimality by at least one configuration)}
\label{tab:stat_test_subopt}
\begin{tabular}{lrrrrr}
\toprule
\multicolumn{1}{c}{Configuration} & \multicolumn{1}{c}{Sample size} & \multicolumn{1}{c}{Confidence interval}  & \multicolumn{1}{c}{$p$-value} &  \multicolumn{1}{c}{Effect size} \\ \midrule
Full vs. Basic                             & $1398$                             & $[8.389\%, 11.500\%]$     & $0.0$                          & 0.9757                          \\
Full vs. No Branching                      & $830$                             & $[3.123\%, 4.030\%]$     & $0.0$                        & 0.9380                          \\
Full vs. No CGCs                           & $673$                             & $[0.274\%, 0.531\%]$   & $0.0$                           &  0.7843                          \\
Full vs. No DMP                            & $123$                              & $[0.615\%, 1.496\%]$   & $0.0$                            & 0.9759     \\     
\bottomrule
\end{tabular}
\end{table}
In contrast to Table \ref{tab:stat_test_all}, it can be seen that $H_0$ is rejected for all configurations, with confidence intervals ranging between $0.281\%$ and $11.446\%$, depending on the configuration. For the basic configuration, median gap improvements lie between $8.389\%$ and $11.446\%$, indicating a noticeable improvement in performance. For the configuration 'No Branching', this improvement reduces to $3.122\%$ to $4.038\%$. For the configurations 'No CGCs' and 'No DMP', median gap percentage improvements decrease to below $2\%$. Nevertheless, in all cases, the hypothesis of a zero median difference is rejected with confidence intervals visibly above 0, indicating that the assumption of structural outperformance of the full configuration over the remaining four configurations is reasonable. We note that, while the results presented in Table \ref{tab:stat_test_subopt} do not allow for global statistical inference of significance, they still allow for a local statistical inference of significance for instances with a similar complexity as the ones that appear to be the most difficult ones.

\paragraph{\textbf{Solution Quality Robustness}}\text{ }\\
\begin{figure}[ht]
\centering
\includegraphics[scale=0.6]{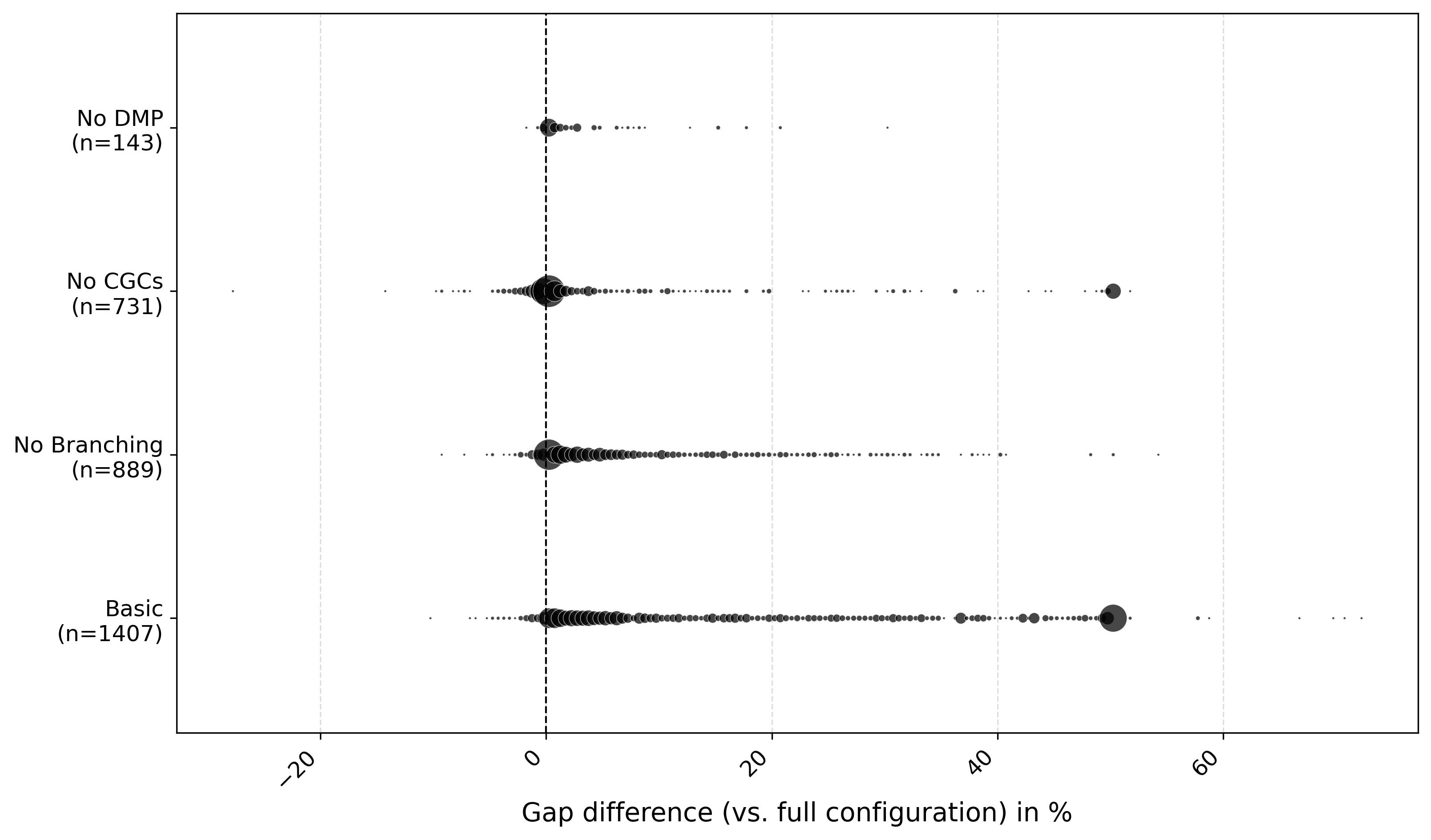} % Optimality gap differences dotstrip linear.jpg
\caption{Dot plot of optimality gap differences compared to the full configuration}
\label{graphics:optimality_gaps_dotplot}
\end{figure}
Figure \ref{graphics:optimality_gaps_dotplot} visualizes the distribution of per-instance differences in optimality gap percentages when comparing the configurations `No DMP', `No CGCs', `No Branching', and 'Basic' against the full configuration. Each dot represents an interval of length 1\%, with its area scaling linearly with the number of instances falling into that interval. Infeasible instances and instances solved to optimality by both compared configurations are excluded. For `No DMP', only instances where the DMP was invoked at least once are considered, as both configurations behave identically otherwise. The number of instances underlying each comparison is indicated on the y-axis.
Positive values imply that the full configuration achieves a smaller optimality gap than the respective alternative, i.e., the full configuration outperforms it on that instance. Note that, unlike for the remaining figures and tables presented in this section, we consider instances for all evaluated workforce travel time quantiles $\gamma\in\left\{0, 0.1, 0.3, 0.5, 0.7, 0.9, 1\right\}$ for this plot. It is clear that the vast majority of instances lie to the right of zero for all 4 comparisons, demonstrating that the full configuration is rarely outperformed. For the basic configuration, gap reductions of 20\%--60\% are common, undermining the substantial contribution of the combined algorithmic features. The configurations `No Branching' and `No CGCs' perform considerably better than the basic configuration, yet the full configuration still yields noticeable gap reductions of up to 40\% for the majority of instances, with reductions of up to 55\% observed in some cases. For `No DMP', gap differences are more modest yet noticeable, indicating that the DMP contributes less improvement than the other features. Overall, the full configuration is rarely outperformed by any of the partial configurations, confirming that it is the most robust choice in terms of optimality gap performance. Finally, we note that Figure \ref{graphics:optimality_gaps_dotplot} only analyzes the distribution of optimality gap differences, neglecting differences in other metrics that influence the performance of an algorithm. Hence, in the following, we evaluate additional metrics, including runtime, average upper and lower bounds, and the percentage of instances solved to optimality, to further assess the benefits of the proposed approach's features.

\vspace{1em}\paragraph{\textbf{Full vs. no Branching}}\text{ }\\
\begin{table}[H]
    \centering
    \caption{Comparison of full and no branching on task finish times configuration}
    \label{tab:basic_vs_no_branching}
        
    \begin{tabular}{llrrrrrrrrr}
    \toprule
\multicolumn{1}{c}{Complexity} & \multicolumn{1}{c}{Config.} & \multicolumn{1}{c}{\% Opt} & \multicolumn{1}{c}{Gap \%} & \multicolumn{1}{c}{UB} & \multicolumn{1}{c}{LB} & \multicolumn{1}{l}{Ropt} & \multicolumn{1}{c}{Nopt} & \multicolumn{1}{l}{Nexp} & \multicolumn{1}{l}{Rnode} & \multicolumn{1}{l}{Rpr} \\ \midrule
Easy                           & Full                        & 99.23\%                    & 0.02\%                     & 2.77                   & 2.76                   & 3.94s                    & 1.71                     & 2.51                     & 2.44s                     & 0.085s                  \\
                               & No Branching                & 96.17\%                    & 0.48\%                     & 2.77                   & 2.71                   & 3.81s                    & 1.53                     & 9.05                     & 1.18s                     & 0.092s                  \\ \midrule
Medium                         & Full                        & 92.49\%                    & 0.25\%                     & 14.95                  & 14.91                  & 9.29s                    & 8.35                     & 25.23                    & 1.24s                     & 0.138s                  \\
                               & No Branching                & 74.18\%                    & 2.63\%                     & 14.96                  & 14.38                  & 12.64s                   & 11.72                    & 43.99                    & 1.29s                     & 0.231s                  \\ \midrule
Hard                           & Full                        & 66.14\%                    & 2.58\%                     & 68.08                  & 64.94                  & 20.09s                   & 24.33                    & 48.68                    & 1.79s                     & 0.258s                  \\
                               & No Branching                & 51.97\%                    & 3.57\%                     & 68.13                  & 64.69                  & 25.82s                   & 28.67                    & 45.00                    & 2.30s                     & 0.411s                  \\ \midrule
All                            & Full                        & 89.85\%                    & 0.64\%                     & 20.89                  & 20.21                  & 7.91s                    & 6.99                     & 20.32                    & 1.58s                     & 0.140s                  \\
                               & No Branching                & 79.03\%                    & 1.90\%                     & 20.90                  & 19.94                  & 9.74s                    & 8.61                     & 29.03                    & 1.61s                     & 0.209s             \\    
           \bottomrule
    \end{tabular}
\end{table}
Table \ref{tab:basic_vs_no_branching} compares the results obtained by the full solver configuration with the configuration `no branching', i.e., branching on task finish times is disabled. Columns `Ropt' and `Nopt' describe the runtime and the number of explored nodes until optimality has been proven. We note that, unlike all other values in the above table, the baseline for these two metrics is not the entire instance set of a fixed instance class, but instead the set of instances that has been solved to optimality by both solver configurations. Column `Nexp' describes the number of explored nodes, while `Rnode' and `Rpr' contain the runtime per explored node and the runtime per pricing iteration, respectively. The average solution quality improves when the branching strategy is enabled, as 10\% additional instances are solved to optimality, and the average gap is reduced by 1.26\%. Furthermore, solving a single node is, on average, 25\% faster for harder instances. This can be explained by significant time savings in the pricing step, where the runtime decreases from 0.411 seconds per iteration to 0.258 seconds.\\
For medium instances, the full solver configuration requires less time to prove optimality and explores around 30\% fewer nodes to do so. For the set of hard test instances, using branching on task finish times improves the solution quality noticeably, while optimality gaps are reduced by 0.99\%. In summary, task finish times appear to be an efficient basis for branching decisions, as they significantly simplify the pricing step.
\paragraph{\textbf{Full vs. no DRMP}}\text{ }\\
\begin{table}[H]
    \centering
    \caption{Comparison of full and no DMP configuration}
    \label{tab:basic_vs_no_dmp}
    
    \begin{tabular}{llrrrrrrrrr}
    \toprule
\multicolumn{1}{c}{Complexity} & \multicolumn{1}{c}{Config.} & \multicolumn{1}{c}{\% Opt} & \multicolumn{1}{c}{Gap \%} & \multicolumn{1}{c}{UB} & \multicolumn{1}{c}{LB} & \multicolumn{1}{l}{Ropt} & \multicolumn{1}{c}{Nopt} & \multicolumn{1}{l}{ARMPn} & \multicolumn{1}{l}{DRMPn} & \multicolumn{1}{l}{NoAInf} \\ \midrule
Easy                           & Full                        & 100.00\%                   & 0.00\%                     & 10.82                  & 10.82                  & 53.95s                    & 14.75                    & 5.00                      & 9.80                      & 1.75                       \\
                               & No DMP                      & 100.00\%                   & 0.00\%                     & 10.82                  & 10.82                  & 36.59s                    & 13.00                    & 13.00                     & 0.00                      & 3.00                       \\ \midrule
Medium                         & Full                        & 94.74\%                    & 0.05\%                     & 17.68                  & 17.66                  & 33.62s                    & 27.92                    & 28.90                     & 12.50                     & 1.47                       \\
                               & No DMP                      & 68.42\%                    & 1.27\%                     & 17.68                  & 17.60                   & 48.79s                    & 37.62                    & 65.60                     & 0.00                      & 13.58                      \\ \midrule
Hard                           & Full                        & 76.92\%                    & 0.46\%                     & 63.63                  & 63.22                  & 86.53s                    & 103.25                   & 71.00                     & 9.00                      & 1.62                       \\
                               & No DMP                      & 61.54\%                    & 0.62\%                     & 63.63                  & 63.13                  & 78.04s                    & 108.12                   & 102.50                     & 0.00                      & 15.15                      \\ \midrule
All                            & Full                        & 88.89\%                    & 0.19\%                     & 33.51                  & 33.35                  & 53.81s                    & 49.92                    & 41.50                     & 10.90                     & 1.56                       \\
                               & No DMP                      & 69.44\%                    & 0.89\%                     & 33.51                  & 33.29                  & 56.20s                    & 56.24                    & 73.10                     & 0.00                        & 12.97      \\ \bottomrule                
    \end{tabular}
\end{table}

Table \ref{tab:basic_vs_no_dmp} combines the results obtained by the full solver configuration with the `no DRMP' configuration, i.e., whenever a disaggregated-infeasible solution is identified, the solution is forbidden explicitly using a constraint of type $\underset{r\in\mathcal{\bar{R}}}{\sum} \lambda^r \leq \vert \mathcal{\bar{R}} \vert - 1,$ where $\mathcal{\bar{R}}$ is the set of columns selected by an optimal, disaggregated-infeasible solution. Note that the baseline for Table \ref{tab:basic_vs_no_dmp} is the set of instances, for which either of the two configurations found at least one aggregated infeasible solution, which is the case for a total of 36 instances. Overall, 19.5\% more instances can be solved to optimality, and average gaps decrease by 0.7\% when using the DRMP formulation. Columns `ARMPn' and `DRMPn' contain the number of nodes solved using the ARMP and DRMP formulation, respectively. Additionally, column `NoAinf' contains the average number of disaggregated-infeasible solutions. The number of such solutions found is large for medium and hard instances, while fewer than 2 disaggregated-infeasible solutions are found when using the DRMP formulation. This, jointly with the small share of nodes solved using the DRMP, further solidifies our assumption that such solutions only occur on very few branches of the branching tree, underlining the reasonability of our approach of only switching to the DRMP within branches that returned such undesired solutions. While a gap improvement of around 0.7\% on average appears insignificant at first glance, since the problem at hand has to be solved frequently in practice, even a minor improvement in solution quality can result in significant financial savings and punctuality improvements in the long term.
\paragraph{\textbf{Full vs. no CGCs}}\text{ }\\
\begin{table}[ht]
    \centering
    \caption{Comparison of full and no Gomory cut configuration}
    \label{tab:basic_vs_no_gcs}
    \begin{tabular}{llrrrrrrrr}
    \toprule
\multicolumn{1}{c}{Complexity} & \multicolumn{1}{c}{Config.} & \multicolumn{1}{c}{\% Opt} & \multicolumn{1}{c}{Gap \%} & \multicolumn{1}{c}{UB} & \multicolumn{1}{c}{LB} & \multicolumn{1}{l}{Rnode} & \multicolumn{1}{c}{RneR} & \multicolumn{1}{l}{\% Root solved} & \multicolumn{1}{l}{RootLB} \\ \midrule
Easy                           & Full                        & 99.23\%                    & 0.02\%                     & 2.77                   & 2.76                   & 2.44s                     & 1.89s                    & 87.36\%                            & 2.68                       \\
                               & No CGCs                     & 96.55\%                    & 0.16\%                     & 2.77                   & 2.75                   & 0.58s                     & 0.60s                    & 45.98\%                            & 1.90                       \\ \midrule
Medium                         & Full                        & 92.49\%                    & 0.25\%                     & 14.95                  & 14.91                  & 1.24s                     & 0.98s                    & 47.89\%                            & 13.95                      \\
                               & No CGCs                     & 85.92\%                    & 0.75\%                     & 14.95                  & 14.85                  & 0.65s                     & 0.64s                    & 15.49\%                            & 12.87                      \\ \midrule
Hard                           & Full                        & 66.14\%                    & 2.58\%                     & 68.08                  & 64.94                  & 1.79s                     & 1.32s                    & 20.47\%                            & 61.79                      \\
                               & No CGCs                     & 66.14\%                    & 1.52\%                     & 67.92                  & 66.04                  & 1.17s                     & 1.07s                    & 4.72\%                             & 61.05                      \\ \midrule
All                            & Full                        & 89.85\%                    & 0.64\%                     & 20.89                  & 20.21                  & 1.58s                     & 1.19s                    & 59.23\%                            & 19.17                      \\
                               & No CGCs                     & 86.36\%                    & 0.66\%                     & 20.85                  & 20.41                  & 0.83s                     & 0.79s                    & 26.46\%                            & 18.29          \\ \bottomrule            
    \end{tabular}
\end{table}
Table \ref{tab:basic_vs_no_gcs} analyzes the results obtained by the full configuration and compares it to the `No CGCs' configuration, i.e., no CGCs are added. Column `rootLB' contains the lower bound obtained after solving the root node, while column `RneR' depicts the algorithm's runtime per node, excluding the root node. On average over all test instances, the usage of Gomory cuts improves the final lower bounds, allowing us to solve 3.49\% more instances to optimality and closing the optimality gap by an additional 0.02\%. Furthermore, the percentage of instances solved at the root node more than doubles from 26.46\% to 59.23\%, while it even increases five-fold for hard instances. Moreover, the lower bound at the root node improves by around 5\%. Additionally, column `RnerR' shows that CGCs significantly increase the runtime per non-root node, increasing from 0.79 seconds to 1.19 seconds on average. Altogether, Gomory cuts aid to provide excellent bounds early on during the solving procedure. However, the increase in computational complexity, especially during the pricing step, implies a trade-off between bound quality and computational complexity. Especially for larger instances, carefully separating and selecting CGCs plays a crucial role in the approach's efficiency.\\
To conclude this section, we showed that our developed solution strategy is able to significantly improve the `basic' configuration first described by \cite{dallolio_temp}, both with respect to optimality guarantees and bound quality. While the impact of every single configuration's component seems to be rather small, combining our three core features and using them simultaneously greatly benefits the resulting algorithm's performance.

\vspace{0.5cm}
\subsection{Stochastic and Deterministic Solutions}\label{section:comparison_stoch_det}
In the following, we compare the quality of stochastic solutions with that of deterministic ones. For that purpose, all 1,215 test instances are solved using our stochastic approach with varying workforce travel time quantiles $\gamma\in\{0,0.1,0.3,0.5, 0.7, 0.9, 1\}$ and four deterministic approaches assuming best-case, mean, median, or worst-case travel times. We note that we can adjust the AMP to use deterministic travel times $\tau_{i,j}^k$ for tasks $i,j\in\mathcal{I}$ and time bins $\mathcal{B}_k\in\mathcal{B}$ by assuming $P(t_{i,j}^k = \tau_{i,j}^k) = 1$ when constructing the respective master problem. Thus, when deterministic travel times are assumed, $\omega_{\gamma}^k = (\tau_{i,j}^k)_{(i,j)\in E}$ holds. Therefore, the underlying deterministic travel times are also used to verify the satisfaction of workforce constraints (\ref{model:workerconstr_aggregated_mp}) and (\ref{model:workerconstr_disaggregated_mp}). We obtained a total of up to 11 solutions for each instance, one for each type of deterministic travel time and one for each workforce travel time quantile $\gamma$, where stochastic travel times are assumed.
% Note that, due to (\ref{chance_constraint})--(\ref{finish_at_lf_v}), an instance that is feasible for stochastic travel times is not necessarily feasible for deterministic travel times. Similarly, feasibility with respect to best-, mean- or median-case travel times does not imply feasibility with respect to mean-, median- or worst-case travel times, respectively.\\
We then perform a simulation by sampling $N\in\mathbb{N}$ scenarios for each instance by randomly generating travel times for all pairs of parking positions and each time bin according to the respective empirical distributions. For each instance and each scenario, we analyze the finish times of all tasks for all 11 solutions. By aggregating these results for each instance and solution over all $N$ scenarios, we obtain empirical service levels per task, and an empirical objective function value (\ref{model:obj_disaggregated_mp}), which allows us to further assess the performance of deterministic and stochastic solutions. Furthermore, note that we impose that constraints (\ref{chance_constraint}) and (\ref{finish_at_lf_v}) need to be satisfied. For deterministic travel times, this is equivalent to using hard time windows $[\mathrm{ES}_i, \mathrm{LF}_i]$ for each task $i$ and disallowing delays.\\
Given a scenario of travel times, a plan (i.e., a solution obtained by solving (\ref{model:obj_aggregated_mp})--(\ref{model:lambda_binary_aggregated_mp})) is simulated as follows. Tours are sorted in ascending order with respect to their scheduled leave time and iteratively implemented. If no sufficient workforce is available to start a planned tour at the scheduled leave time, the tour is postponed until enough workers become available. In the following, we refer to the solution of an instance when using stochastic travel times as the \textit{stochastic solution} of an instance. For best-, median, and worst-case travel times, we analogously refer to \textit{best-, median,} and \textit{worst-case travel time solutions}, respectively. Furthermore, an instance is called \textit{model-feasible} for a given type of travel times and workforce travel time quantile $\gamma$ if the underlying problem (\ref{model:obj_aggregated_mp})--(\ref{model:lambda_binary_aggregated_mp}) is feasible. Moreover, a solution is called \textit{stochastic-feasible} if, when the solution is applied to all generated scenarios, service level constraints (\ref{chance_constraint}) and (\ref{finish_at_lf_v}) are satisfied for all tasks in the given instance.\\
The computation of an appropriate scenario count $N$ is done using Sample Average Approximation techniques. For additional details, we refer to \Cref{appendix:computation_scen_count}. In the following, all evaluations are performed based on $500$ scenarios per instance.

\begin{table}[ht]
    \centering
    \caption{Frequencies of stochastic (in)feasibility}
    \label{tab:feasible_instances}
    \begin{tabular}{lllrrrr}
    \toprule
                                                 \multicolumn{3}{c}{Travel Times}                   & \multicolumn{1}{c}{Model} & \multicolumn{3}{c}{Simulation}  \\
    \cmidrule(lr){1-3} \cmidrule(lr){4-4} \cmidrule(lr){5-7}
    \multicolumn{1}{l}{Obj.fctn.} & \multicolumn{1}{l}{Workforce cons.} & \multicolumn{1}{l}{Chance cons.} & \multicolumn{1}{r}{\#Feas.}       & \multicolumn{1}{r}{\#$\alpha$-feas.} & \multicolumn{1}{r}{\#$LF_e$-feas.} & \multicolumn{1}{r}{\#Stoch.-feas.} \\ \midrule
    Best & Best & Best                    & 720 & 16  & 79 & 13  \\
    Median & Median & Median                  & 657 & 263 & 530 & 246 \\
    Mean & Mean & Mean                 & 657 & 263 & 528 & 246 \\
    Worst & Worst & Worst                  & 582 & 582 & 582 & 582 \\ \midrule
    Stochastic  & $\gamma=0$ & Stochastic & 623 & 452 & 563 & 446 \\
    Stochastic & $\gamma=0.1$ & Stochastic & 622 & 475 & 579 & 468 \\
    Stochastic & $\gamma=0.3$ & Stochastic & 615 & 515 & 577 & 509 \\
    Stochastic & $\gamma=0.5$ & Stochastic & 611 & 548 & 590 & 540 \\
    Stochastic & $\gamma=0.7$ & Stochastic & 605 & 573 & 594 & 564 \\
    Stochastic & $\gamma=0.9$ & Stochastic & 601 & 599 & 601 & 594 \\
    Stochastic & $\gamma=1$ & Stochastic & 600 & 597 & 600 & 590 \\ \bottomrule
    \end{tabular}
\end{table}

Table \ref{tab:feasible_instances} summarizes the feasibility of instances and solutions under various travel time assumptions. Each row corresponds to one of the 11 travel time assumptions and corresponding solutions, as elaborated on at the beginning of this section. Recall that the proposed model requires two types of stochastic information, namely the travel time distributions for each combination of time bin and arcs, and the workforce travel time quantile $\gamma$. While the former is used to compute objective function values and ensure feasibility with respect to the chance constraints (\ref{chance_constraint}) and (\ref{finish_at_lf_v}), the latter controls the duration for which workers are occupied in the model and ensures the satisfaction of the workforce constraints (\ref{model:workerconstr_aggregated_mp}) and (\ref{model:workerconstr_disaggregated_mp}).\\
Columns `Obj.fctn', `Workforce cons', and `Chance cons.' of Table \ref{tab:feasible_instances} describe the travel time information used to compute objective values, to ensure satisfaction of the workforce constraints, and to satisfy the chance constraints, respectively. 

For each type of travel time and each workforce travel time quantile, column `\#Feas.' contains the number of model-feasible instances. 
Furthermore, column `\#$\alpha$-feas.' denotes the number of instances for which the solutions satisfy the desired service level $\alpha$ for each task. This requirement is equivalent to satisfying constraint (\ref{chance_constraint}). Column `\#$LF_e$-feas.' serves a similar purpose and contains the number of instances whose solutions guarantee no delays beyond extended latest finish times. This corresponds to satisfying equality (\ref{finish_at_lf_v}). Finally, column `\#Stoch.-feas.' contains the number of instances with stochastic-feasible solutions. This is equivalent to fulfilling constraints (\ref{chance_constraint}) and (\ref{finish_at_lf_v}) simultaneously. Note that the values in the latter three columns are derived by applying each solution to each scenario and validating the satisfaction of the respective constraint(s). In contrast, the entries in column \#Feas. do not depend on the simulation results. Only 16 out of 720 best-case travel time solutions satisfy the desired service level for all tasks, while only 13 solutions are feasible with respect to service level and maximum delay constraints. When median travel times are assumed, 246 out of 647, i.e., around 38\% of all solutions guarantee the desired service level and fulfill the maximum delay requirement (\ref{finish_at_lf_v}). Thus, if deterministic travel times are assumed, it is likely that either tasks are delayed with a probability of more than $1-\alpha$ or large delays of more than 10 minutes can occur, ultimately leading to frequent delays of multiple aircraft.
Moreover, it can be observed that a decreasing value for $\gamma$ only slightly increases the number of model-feasible instances. Hence, planning workforce availability in constraints (\ref{model:workerconstr_aggregated_mp}) and (\ref{model:workerconstr_disaggregated_mp}) using a quantile $\gamma\in[0,1]$ rather than worst-case travel times does not have a significant impact on model-feasibility. Instead, model-infeasibility is caused mainly by the service level constraints (\ref{chance_constraint}) and the presence of latest finish times (\ref{finish_at_lf_v}). At the same time, a decreasing value of $\gamma$ significantly reduces the number of instances whose solutions satisfy both the minimum service level (\ref{chance_constraint}) as well as maximum delays (\ref{finish_at_lf_v}), as can be seen in column `\#Stoch.-feas'. Therefore, reducing the quantile level $\gamma$ results in a trade-off between increasing model-feasibility and more frequent violation of service levels and maximum delays. Hence, if the satisfaction of the latter two properties is of utmost importance, using worst-case travel times in the workforce constraints appears to be an appropriate choice.\\

\begin{table}[ht]
    \centering
    \caption{Objective function values and service levels from simulation study}
    \label{tab:overview_stoch_comparison}
    \begin{tabular}{lllrrrrrrr}
    \toprule
                           \multicolumn{3}{c}{Travel Times}        & \multicolumn{2}{c}{Objective}                         & \multicolumn{3}{c}{Service Level}                                                                   & \multicolumn{1}{l}{}    & \multicolumn{1}{l}{}     \\  \cmidrule(lr){1-3} \cmidrule(lr){4-5}\cmidrule(lr){6-8}
\multicolumn{1}{l}{Obj.fctn.} & \multicolumn{1}{l}{Workforce cons.} & \multicolumn{1}{l}{Chance cons.} & \multicolumn{1}{c}{Obj.val.} & \multicolumn{1}{c}{Obj-Pen} & \multicolumn{1}{c}{$\mu_{SL}$} & \multicolumn{1}{c}{$\sigma_{SL}$} & \multicolumn{1}{c}{$\min_{SL}$} & \multicolumn{1}{c}{VSS} & \multicolumn{1}{c}{EVPI} \\ \midrule
Best & Best & Best                                                 & 195.01                  & 123.94                      & 88.04\%                       & 8.59                              & 0.00\%                          &                         &                          \\
Median & Median & Median                                                              & 43.64                   & 40.64                       & 98.18\%                       & 2.06                              & 1.60\%                          &                         &                          \\
Mean & Mean & Mean                                               & 43.69                   & 40.69                       & 98.18\%                       & 2.06                              & 1.60\%                          &                         &                          \\
Worst & Worst & Worst                                              & 21.58                   & 21.58                       & 99.99\%                       & 0.04                              & 88.20\%                         &                         &                          \\ \midrule
Stochastic  & $\gamma=0$ & Stochastic             & 22.22                   & 21.11                       & 99.51\%                       & 1.09                              & 3.00\%                          & 21.47                   & 4.94                     \\
Stochastic & $\gamma=0.1$    & Stochastic       & 20.85                   & 20.08                       & 99.61\%                       & 0.98                              & 0.40\%                        & 22.85                   & 4.02                     \\
Stochastic & $\gamma=0.3$ & Stochastic         & 19.28                   & 18.81                       & 99.78\%                       & 0.56                              & 24.00\%                         & 24.42                   & 2.99                     \\
Stochastic & $\gamma=0.5$  & Stochastic        & 18.43                   & 18.20                       & 99.87\%                       & 0.38                              & 38.20\%                         & 25.26                   & 2.45                     \\
Stochastic & $\gamma=0.7$  & Stochastic        & 18.36                   & 18.24                       & 99.92\%                       & 0.21                              & 74.40\%                         & 25.33                   & 2.38                     \\
Stochastic & $\gamma=0.9$  & Stochastic        & 18.58                   & 18.53                       & 99.95\%                       & 0.13                              & 88.20\%                         & 25.11                   & 2.38                     \\
Stochastic & $\gamma=1$   & Stochastic         & 19.10                   & 19.05                       & 99.95\%                       & 0.14                              & 88.20\%                         & 24.60                   & 2.55      \\ \bottomrule              
\end{tabular}
\end{table}

Table \ref{tab:overview_stoch_comparison} provides an overview of objective function values, service levels, and the value of the stochastic solutions (VSS) as well as the expected value of perfect information (EVPI) when applying all solutions to the simulated scenarios of the underlying instance. Similar to Table \ref{tab:feasible_instances}, each row corresponds to one of the 11 travel time assumptions and corresponding solutions. Furthermore, note that all values presented in Table \ref{tab:overview_stoch_comparison} are based on the results of the simulation. In order to have a stable foundation for comparison, all data visualized in Table \ref{tab:overview_stoch_comparison} is calculated based on the set of model-feasible instances when assuming worst-case travel times. This amounts to a total of 582 out of 1,215 instances. Moreover, for column `EVPI', we only consider instances that were solved to optimality for all values of $\gamma$. Note that model-feasibility with respect to worst-case travel times implies model-feasibility for all (deterministic and stochastic) types of travel times. Recall that all values depicted here are calculated based on a simulation of travel times. Columns `Obj' and `Obj-Pen' contain the average objective function value of deterministic solutions applied to stochastic travel times, with and without quadratic penalties for time window violations, respectively. Moreover, columns `$\mu_{SL}$', `$\sigma_{SL}$', and `$\min_{SL}$' represent the average service level for all tasks, its standard deviation, and the minimum service level of any task. 
On average, stochastic solutions return the smallest objective function values, both with and without penalties. 
Furthermore, the relations of the values in column `Obj.val.' show that stochastic solutions' objective function values are equal to around 10\% to 12\% of the objective function values of best-case travel time solutions, while this relation reduces to around 44\% to 51\% for mean and median travel times, depending on the quantile $\gamma$.
Practically speaking, the total safety time buffer accumulated by stochastic solutions is twice as large when using stochastic rather than mean or median-case travel time data. Column `VSS', which contains the average value of the stochastic solution over all instances, indicates a reduction of objective function values by up to 25.33 time periods when using stochastic instead of mean travel times during planning. At the same time, the EVPI ranges between 2.38 and 4.94, depending on the value of $\gamma$. Thus, one can conclude that most of the improvement comes from modeling stochasticity rather than from perfect information. Hence, explicitly accounting for stochastic travel times during planning can help realize the majority of the improvement gains from considering uncertainty and greatly improve overall service quality. Moreover, for every instance class, stochastic solutions guarantee a very high and stable service level, averaging over 99.5\% and having a standard deviation of 0.14 to 1.09.\\
Additionally, the solutions obtained for different workforce travel time quantiles $\gamma\in\{0, 0.1, 0.3, 0.5,0.7,0.9,1\}$ exhibit several noteworthy properties. Recall that, in constraint (\ref{model:workerconstr_aggregated_mp}), a smaller value of $\gamma$ implies earlier worker availability within the master problem. 
It can be seen that the objective function value, when measured as a function of the workforce travel time quantile $\gamma$, is monotonically decreasing in $[0, 0.7]$, and monotonically increasing in $[0.7, 1]$. At the same time, the average and minimum service levels are strictly monotonically increasing in $\gamma\in[0,1]$. Hence, if violating the minimum service level $\alpha$ is acceptable in individual cases, using a workforce travel time quantile of $\gamma=0.7$ appears to provide a superior trade-off between service level guarantees and efficient workforce utilization. If satisfying the service level is a key priority, selecting $\gamma=0.9$ can reduce the objective function by around 3\% without reducing service levels compared to planning with $\gamma=1$.
Furthermore, all deterministic travel times yield very poor service-level bounds for individual tasks, going as low as 1.6\% for mean and median travel times. Nevertheless, unlike best-case travel times, their solutions satisfy the prescribed minimum service level of $\alpha=0.95$ on average. Therefore, if a deterministic model is sought to solve the problem at hand, using mean or median travel times appears to be the most promising approach. Alternatively, worst-case travel times provide better objective function values and service levels, yet they frequently fail at finding a feasible solution in the first place. In either case, service levels for individual tasks can still be arbitrarily low and clearly lack a lower bound. Such guarantees can only be provided by stochastic solutions, which outperform deterministic travel time solutions across all previously mentioned metrics.\\
\begin{figure}[ht]
\centering
\includegraphics[scale=0.6]{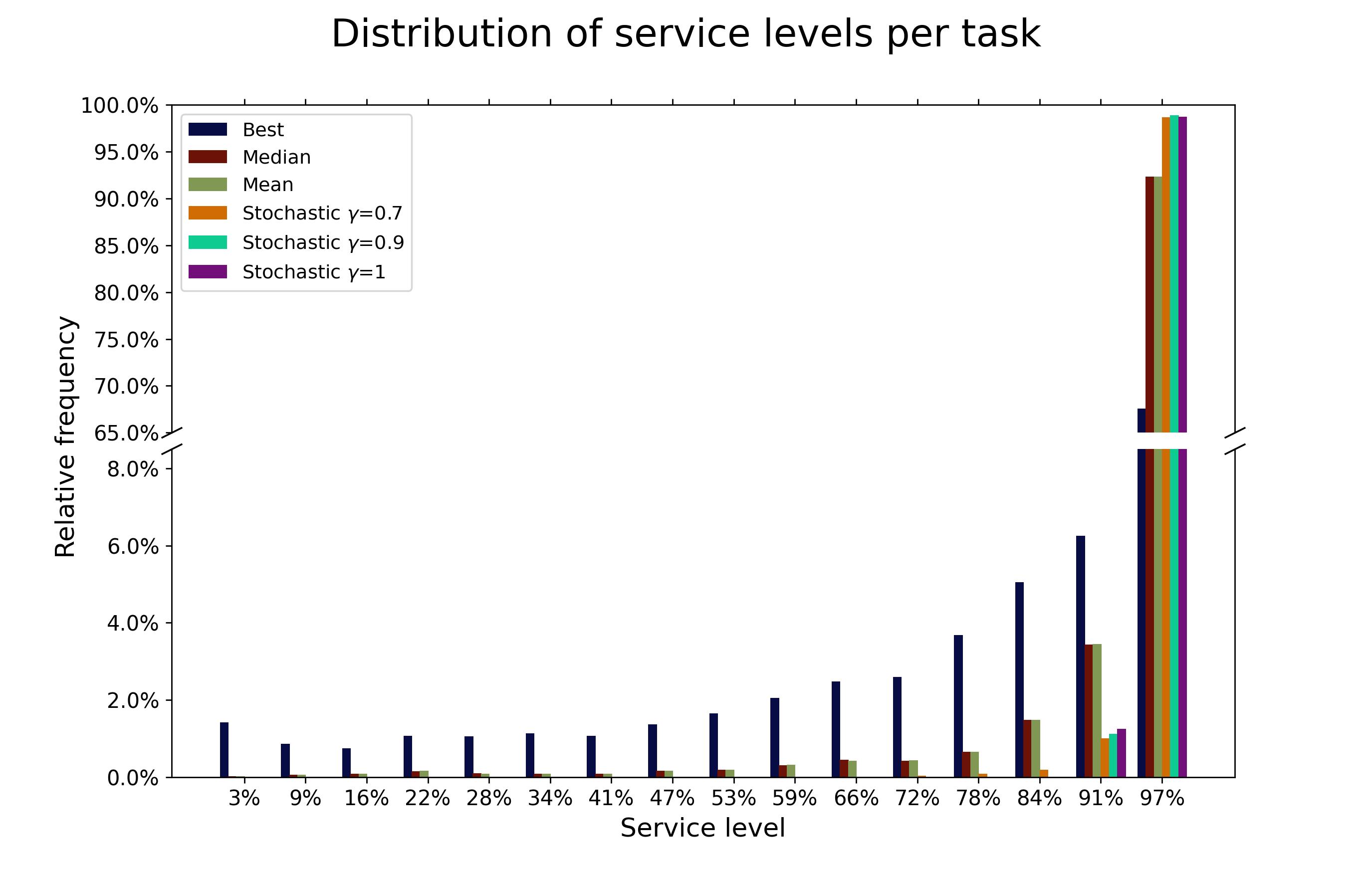} % service_level_distr.jpeg
\caption{Distribution of service levels}
\label{graphics:service_level_distribution}
\end{figure}
Figure \ref{graphics:service_level_distribution} shows the distribution of service levels per task. For this purpose, for each type of travel time, we collected all service levels for each individual task for every instance and plotted their empirical distribution. Because using a workforce travel time quantile of $\gamma < 0.7$ has shown to be not beneficial for either objective function values or service levels, in the following we only consider values $\gamma\in\left\{0.7,0.9,1\right\}$
While stochastic travel times guarantee a service level of at least 90\% at almost all times, median and best-case travel time solutions frequently provide significantly lower values, down to 1.6\% in some cases. Moreover, best-case travel time solutions violate the prescribed minimum service level for 34\% of tasks, while for median travel time solutions, this holds in around 8.5\% of cases. This emphasizes the unpredictability of minimum service levels when deterministic travel times are assumed, which ultimately has a significant impact on perceived service quality. Hence, explicitly considering stochastic travel times in the problem formulation yields solutions that seldom violate any time windows, resulting in a high and reliable quality of service.\\
\begin{figure}[ht]
\centering
\includegraphics[scale=0.6]{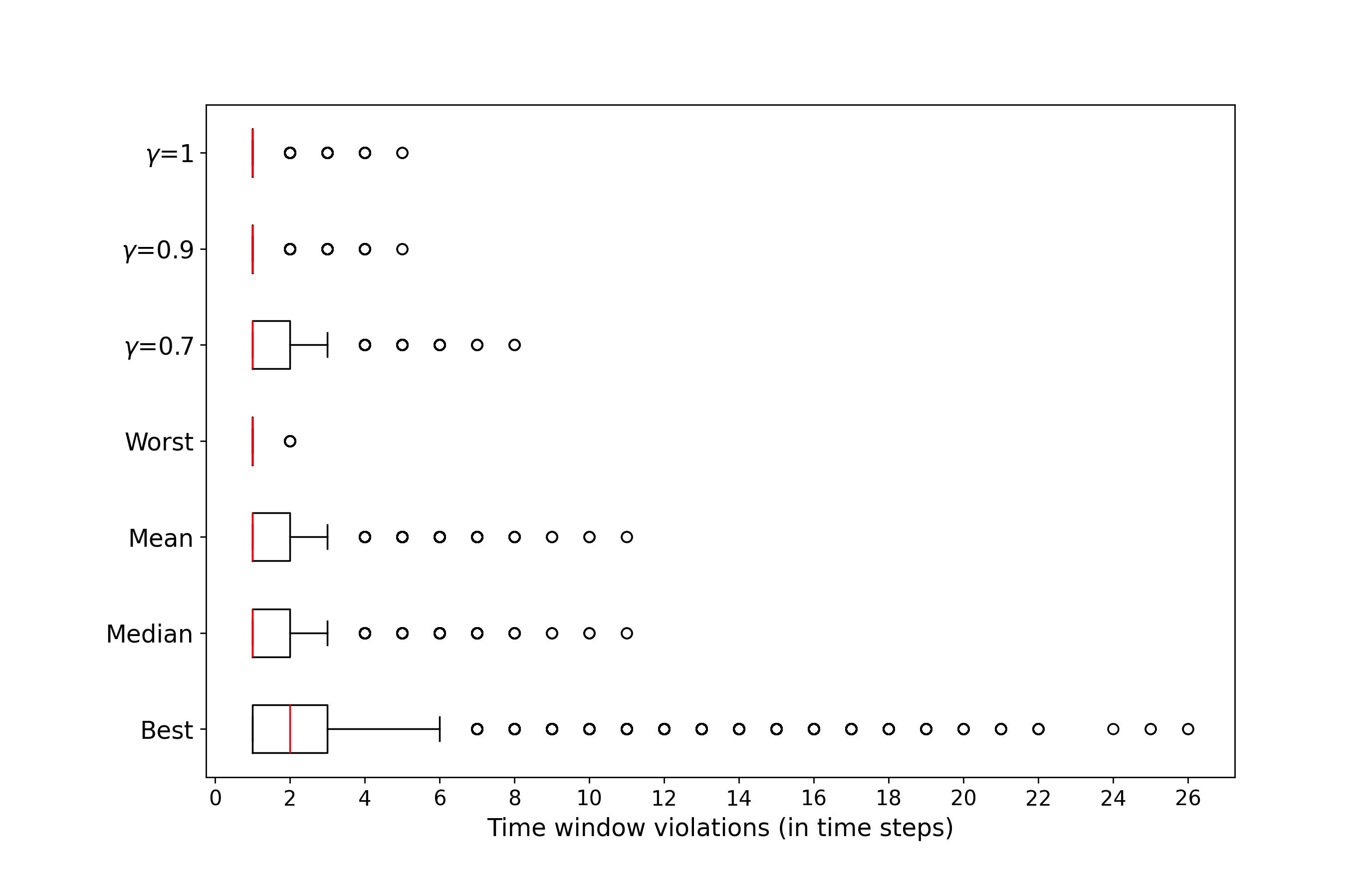} % boxplot_tw_violations.jpeg
\caption{Frequency of durations of time window violations}
\label{graphics:durations_tw_violations}
\end{figure}
Figure \ref{graphics:durations_tw_violations} visualizes the occurrence of time window violations, measured in time steps. Recall that for our purposes, each time step is 2 minutes long. As for Figure \ref{graphics:service_level_distribution}, we collected all potential time window violations and their lengths for each instance and each scenario.
Overall, best-case travel time solutions have higher median time window violations, larger 75\%-quantiles, and more outliers than stochastic solutions. At the same time, the median time window violations and 75\%-quantiles of median and stochastic solutions are almost identical. Nevertheless, the former solutions exhibit far larger delays, reaching up to 11 time steps, i.e., 22 minutes. In total, assuming deterministic travel times usually returns solutions that cause comparably long delays, are at risk of delaying tasks by significant amounts, and lead to highly volatile delays, which are several magnitudes larger than the maximum acceptable delay $\mathrm{LF}_i^{\mathrm{e}} - \mathrm{LF}_i$. These effects lead to passengers becoming dissatisfied and baggage handling operators having to pay financial penalties. Using stochastic travel times greatly benefits a solution's quality in these regards, typically causing few, small delays and ruling out undesirably long delays.\\
To summarize the above findings, we observe that solutions based on deterministic travel times, be they best-case, median, or worst-case, return strategies that do not live up to the set service quality. Explicitly incorporating time-dependent stochastic travel times into the model and ensuring workforce feasibility only in a percentage of scenarios allows for solutions that efficiently use the available workforce to increase safety time buffers, guarantee a stable service level, and prevent large delays.

%% file: sections/conclusion.tex
\section{Conclusion}\label{section:conclusion}
We have examined the problem of forming and routing worker teams, consisting of workers of different hierarchical skill levels. We extended the works of \cite{dallolio_temp} in several dimensions. From a modelling point of view, we introduce time-dependent travel times across the apron.
Moreover, we introduce a quantile-based workforce constraint formulation. Instead of assuming maximum travel times across the apron, we incorporate the $\gamma$-quantile scenario of each travel time distribution. This allows explicit control over the conservatism of workforce planning and can reduce workforce underutilization.\\
Furthermore, we extended the methodology by \cite{dallolio_temp} in three aspects. First, we introduced an improved branching strategy based on the \revnew{quantile-case finish times} of individual tasks, which was able to greatly improve the solution quality and the algorithm's convergence speed. Second, we applied an exact separation procedure for rank-1 Chvátal-Gomory cuts as proposed by \cite{petersen2008chvatal} and carefully designed the decision process of when and how to separate said cuts. Third, we introduce an alternative master problem formulation, abbreviated as DRMP. Based on said formulation and the ARMP formulation by \cite{dallolio_temp}, we propose a novel solution methodology which we term \textit{Branch-Price-Cut-and-Switch}. In this context, we switch between two master problem formulations, depending on the characteristics of the current (integer) solution and the position within the branching tree.\\
Our computational studies indicate that our proposed algorithm is able to outperform the standard Branch-Price-and-Check approach by \cite{dallolio_temp} as well as the approach by \cite{yuan2015branch}, solving more instances to optimality and significantly reducing optimality gaps. Additionally, we observed that including stochastic travel times into the model is a superior approach to assuming deterministic travel times, as the solutions obtained allow for earlier expected termination of tasks, less frequent and smaller delays, and a guaranteed, stable service level and maximum possible delay. Most importantly, stochastic planning and the introduction of quantile-based workforce constraints enable the operator to realize the majority of the benefit of modelling stochasticity, as well as to reduce idle times and thus derive efficient solutions.\\
One of our core assumptions is the separation of daily flight schedules into segments of 1 to 2 hours, within which optimal decisions can be made without considering what happens before or after this. While this is a realistic assumption for Munich Airport, it might not hold for other airports. Similarly, if longer time horizons are considered, time windows might also be subject to significant uncertainty. In order to derive reasonable team formation and routing decisions for such settings, one might consider an infinite time horizon, potentially applying multi-stage stochastic programming or online optimization techniques. 
Furthermore, as we consider task execution times and time windows to be deterministic, our model is still a simplification. As the exact solvability of a full-fledged stochastic model that incorporates all potential factors of stochasticity is questionable, further advances toward heuristic approaches can be made. For instance, the order in which the pricing networks are solved in each pricing step has a large impact on the algorithm's runtime. Moreover, pricing networks can be solved heuristically, potentially returning a very good, yet not optimal column, in a fraction of the runtime needed by an exact labeling approach. Such decision processes could be further improved by applying machine learning techniques tailored individually to the problem's structure. Finally, the applicability of the aggregation technique of multiple pricing networks introduced in Section \ref{section:pricing_step_dmp} to other problem types, such as different variants of the VRP, where the SPPRC arises as a pricing subproblem and several subproblems share many of their properties, can be further studied. In particular, multi-depot VRPs and VRPs with mixed fleets can be of special interest for this.

%% file: sections/acknowledgements.tex
\section*{Acknowledgements}
We thank Dr. Jan Evler and Aeroground GmbH for providing the data and key industry information. Andreas Hagn and Giacomo Dall'Olio have been funded by the Deutsche Forschungsgemeinschaft (DFG, German Research Foundation) - Project number 277991500.

%% file: sections/appendix.tex
\clearpage
\section{List of Symbols} \label{appendix:notation}

\small
\begin{longtable}{p{0.1\textwidth} p{0.88\textwidth}}

% ---- header for first page ----
\multicolumn{2}{l}{\textbf{Sets}} \\
\endfirsthead

% ---- header for continuation pages ----
\multicolumn{2}{l}{\textit{(List of Symbols, continued)}} \\[0.5em]
\endhead

% ---- footer for all but lastpairwi page ----
\multicolumn{2}{r}{\textit{Continued on next page\ldots}} \\
\endfoot

% ---- footer for last page ----
\endlastfoot

% Sets
$\mathcal{B}$      & set of time bins (disjoint decomposition of the planning horizon)\\
$B_{i,j}^k$        & (finite) support of travel time delays for tasks $i \in\mathcal{I}$ and $j\in\mathcal{I}$ (or tasks and the depot) and time bin $\mathcal{B}_k\in\mathcal{B}$\\
$\mathcal{I}$       & Set of tasks\\
$\mathcal{I}^r$     & Set of tasks executed by route $r\in\mathcal{R}$\\
$\mathcal{I}_q$     & Set of tasks that can be executed by profile $q\in\mathcal{Q}$\\
$\mathcal{G}$       & Set of indices of rank-1 CGCs present at given node\\
$\mathcal{K}$       & Set of worker skill levels\\
$\mathcal{Q}$       & Set of profiles\\
$\mathcal{Q}_i$     & Set of profiles compatible with task $i\in\mathcal{I}$\\
$\mathcal{Q}^{\mathrm{D}}$ & Set of disaggregated profiles $(q,s)$ for $q\in\mathcal{Q}, s\in\mathcal{S}$\\
$\mathcal{R}$       & Set of feasible team routes for the AMP\\
$\bar{\mathcal{R}}$ & Subset of all feasible team routes $\mathcal{R}$\\
$\mathcal{R}^{\mathrm{D}}$ & Set of feasible team routes for the DMP\\
$\mathcal{S}$       & Set of skill compositions\\
$\mathcal{S}_q$     & Set of skill compositions compatible with profile $q\in\mathcal{Q}$\\
$\mathcal{T}$       & Set of discrete time points\\
$\Omega$            & Set of all possible travel time realizations\\
\\
\multicolumn{2}{l}{\textbf{Deterministic Parameters}} \\
\\
$\alpha$            & Chance constraint probability (minimum service level requirement)\\
$\mathcal{A}^q$     & Arcs of networks $\mathcal{G}^q$ and $\mathcal{G}^{q,s}$\\
$\gamma$            & Travel time quantile level in workforce constraints\\
$d$                 & Central worker depot\\
$\delta_{k,t}$      & Value of the dual variable associated with workforce constraints\\
$\mathrm{ES}_i$     & Earliest start time of task $i\in\mathcal{I}$\\
$\mathcal{G}^q$     & Pricing network in the ARMP for profile $q\in\mathcal{Q}$\\
$\mathcal{G}^{q,s}$ & Pricing network in the DRMP for disaggregated profile $(q,s)$, $(q,s)\in\mathcal{Q}^D$\\
$\mu_i$             & Value of the dual variable associated with task covering constraints\\
$\mathrm{LF}_i$     & Latest finish time of task $i\in\mathcal{I}$\\
$\mathrm{LF}_i^{\mathrm{e}}$ & Extended latest finish time of task $i\in\mathcal{I}$\\
$N_k^{\mathrm{D}}$  & Number of available workers with skill level $k\in\mathcal{K}$\\
$N_k$               & Number of available workers with at least level $k\in\mathcal{K}$\\
$p_{i,q}$           & Processing time of task $i$ when undertaken with working profile $q\in\mathcal{Q}$\\
$\psi_g$            & Value of the dual variable associated with the rank-1 CGC with index $g\in\mathbb{N}$\\
$\mathrm{tl}^{r,q}$ & Leave time from the depot for the team with profile $q\in\mathcal{Q}$ executing route $r\in\mathcal{R}$\\
$\mathrm{tr}^{r,q}$ & Return time to the depot for the team with profile $q\in\mathcal{Q}$ executing route $r\in\mathcal{R}$\\
$u_i$               & rank-1 CGC coefficient for task constraint associated with task $i$\\
$u_{k,t}$           & rank-1 CGC coefficient for workforce constraint associated with skill level $k\in\mathcal{K}$ and time $t\in\mathcal{T}$\\
$\mathcal{V}^q$     & Nodes of networks $\mathcal{G}^q$ and $\mathcal{G}^{q,s}$\\
$w_i$               & Weight of task $i\in\mathcal{I}$\\
$\xi_{q,k}$         & Number of workers with at least level $k\in\mathcal{K}$ required by working profile $q\in\mathcal{Q}$\\
\\
\multicolumn{2}{l}{\textbf{Stochastic Parameters}} \\
\\
$b_{k,t}^{r,q}(\omega)$  & Number of workers with at least level $k$ required by team route $(r,q)$, $r\in\mathcal{R}$, $q\in\mathcal{Q}$ at time $t\in\mathcal{T}$ if $\omega\in\Omega$ occurs\\
$\beta_{k,t}^{r,s}(\omega)$ & Number of workers with skill level $k\in\mathcal{K}$ required by team route $(r,q,s)$, $r\in\mathcal{R}$, $(q,s)\in\mathcal{Q}^D$ at time $t\in\mathcal{T}$ if $\omega\in\Omega$ occurs\\
$\mathbb{E}(c^r)$        & Expected cost of route $r\in\mathcal{R}$\\
$F^{r,q}_i(\omega)$      & (Stochastic) finish time of task $i\in\mathcal{I}$ in tour $r$ and scenario $\omega\in\Omega$ with profile $q\in\mathcal{Q}$\\
$F^{r,q}_i(\omega_{\gamma})$ & Finish time of task $i\in\mathcal{I}$ in tour $r\in\mathcal{R}$ with profile $q\in\mathcal{Q}$ in scenario $\omega_\gamma$\\
$F_{i}^{L}(\omega_{\gamma})$ & Finish time of task $i\in\mathcal{I}$ in the path associated with label $L$ in scenario $\omega_\gamma$\\
$\omega$                 & (single) realization of travel times\\
$\omega_{\mathrm{max}}$  & worst-case realization of travel times\\
$P_i(\cdot)$             & Penalty function for time window violations at task $i\in\mathcal{I}$\\
$S^{r,q}_i$              & (Stochastic) start time of task $i\in\mathcal{I}$ in tour $r$ with profile $q\in\mathcal{Q}$\\
$t^k$                    & random variable of stochastic travel times in time bin $\mathcal{B}_k\in\mathcal{B}$\\
$t^k_{i,j}$              & random variable of stochastic travel times between nodes $i \in\mathcal{I}$ and $j\in\mathcal{I}$ in time bin $\mathcal{B}_k\in\mathcal{B}$\\
$W_{i,j}(F_i^{P}(\omega_{\gamma}))$ & time-dynamic weight of arc $(i,j)$ depending on the finish time $F_i^{P}(\omega_{\gamma})\in\mathbb{N}$ of the previous task in scenario $\omega_{\gamma}$\\
$\omega^k_{\gamma}$      & $\gamma$-quantile scenario of distributions $(t^k_{i,j})_{(i,j)\in E}$ for $\alpha\in[0,1],\mathcal{B}_k\in\mathcal{B}$\\
$\omega_{\gamma}$        & vector of $\gamma$-quantile scenarios $\omega_{\gamma}^k$ for all $\mathcal{B}_k\in\mathcal{B}$\\
\\
\multicolumn{2}{l}{\textbf{Decision Variables}} \\
\\
$\lambda^r_q$            & Binary variable, equals to 1 if feasible team route $(r,q)$ with $r\in\mathcal{R}$, $q\in\mathcal{Q}$ belongs to a solution of the AMP\\
$\lambda^r_{q,s}$        & Binary variable, equals to 1 if feasible team route $(r,q,s)$ with $r\in\mathcal{R}$, $(q,s)\in\mathcal{Q}^D$ belongs to a solution of the DMP\\

\end{longtable}

\clearpage

\section{Proof of existence of optimal binary AMP solutions}\label{appendix:binary_sol_existence}
\begin{theorem}\label{theorem:binary_sol_existence}
    For any feasible instance of the AMP, there exists at least one optimal solution $\left(\bar{\lambda}^r_q\right)_{(r,q)\in\mathcal{R}}$ such that $\bar{\lambda^r_q} \in \{0,1\}$ holds for all $(r,q)\in\mathcal{R}$:
\end{theorem}
\proof{Proof. }
Consider a feasible instance of the AMP. By definition, we know that 
\begin{equation}
    \mathbb{E}(c^r) \geq 0 \label{proof:costs_nonnegative}
\end{equation} holds for any route $(r,q)\in\mathcal{R}$. Hence, there exists an optimal solution. Let $\left(\bar{\lambda}^r_q\right)_{(r,q)\in\mathcal{R}}$ be an optimal solution. Define a new solution $\left(\tilde{\lambda}^r_q\right)_{(r,q)\in\mathcal{R}}$ by
\begin{align*}
    \tilde{\lambda}^r_q = \begin{cases}
        \begin{aligned}
            &\bar{\lambda}^r_q && \text{if} \ \ \bar{\lambda}^r_q \leq 1\\
            &1 && \text{otherwise}
            \end{aligned}
    \end{cases} 
\end{align*}
It is clear that $\left(\tilde{\lambda}^r_q\right)_{(r,q)\in\mathcal{R}}$ is also feasible for the AMP and
\begin{equation*}
    \sum_{(r,q)\in \mathcal{R}} \mathbb{E}(c^r)\tilde{\lambda^r_q} \overset{(\ref{proof:costs_nonnegative})}{\leq} \sum_{(r,q)\in \mathcal{R}} \mathbb{E}(c^r)\bar{\lambda^r_q}
\end{equation*}
holds. Therefore, $\left(\tilde{\lambda}^r_q\right)_{(r,q)\in\mathcal{R}}$ is an optimal solution to the AMP for which $\tilde{\lambda}^r_q \in \{0,1\}$ holds for all $(r,q)\in \mathcal{R}$.
$\square$\\
We note that the above proof can canonically be applied to an instance of the DMP:
\begin{corollary}
    Theorem \ref{theorem:binary_sol_existence} also holds for any feasible instance of the DMP.
\end{corollary}

\section{Feasibility Check Model}\label{appendix:feas_check_model}
Without loss of generality, let $(\bar{\lambda}^r_{q})_{(r,q)\in\mathcal{R}}$ be an integer solution to the AMP. We define by
\begin{equation*}
    \bar{\mathcal{R}} \coloneqq \left\{(r,q)\in\mathcal{R}: \ \bar{\lambda}^r_{q} = 1 \right\}
\end{equation*}
the set of selected  team routes. Furthermore, we denote by
\begin{align*}
    \bar{\mathcal{R}}^r_{-} &\coloneqq \left\{(\tilde{r}, \tilde{q}) \in\bar{\mathcal{R}}: \ \mathrm{tr}^{\tilde{r},\tilde{q}} \leq \mathrm{tl}^{r,q}\right\}\\
    \bar{\mathcal{R}}^r_{+} &\coloneqq \left\{(\tilde{r}, \tilde{q}) \in\bar{\mathcal{R}}: \ \mathrm{tl}^{\tilde{r},\tilde{q}} \geq \mathrm{tr}^{r,q}\right\}
\end{align*}
the set of predecessor and successor tours of team route $(r,q)\in\bar{\mathcal{R}}$, respectively. In the following, we will be referring to elements in $\bar{\mathcal{R}}^r_{-}$ and $\bar{\mathcal{R}}^r_{+}$ solely by their respective route $\tilde{r}$, because the actual profile and skill composition team routes are mostly irrelevant for the considerations to come. Furthermore, let $x_k^{r,\tilde{r}}$ be integer variables indicating the number of workers of skill level $k$ that are regrouped from tour $r$ into tour $\tilde{r}$ after $r$ has finished and $\chi_k^{r, \tilde{r}}$ be slack variables. Then, the feasibility check is defined as
\begin{align}
    \min & \underset{(r,q)\in\bar{\mathcal{R}}}{\sum}\chi_k^{o, r}\label{feas_check:obj}\\
    \textrm{s.t.} \quad & \underset{\rho\in\bar{\mathcal{R}}^r_{-}}{\sum} \ \underset{\kappa=k}{\overset{K}{\sum}} x_k^{\rho,r} + \underset{\kappa=k}{\overset{K}{\sum}} x_k^{o,r} + \chi_k^{o,r}\geq \xi_{q,k} \quad \forall (r,q)\in \bar{\mathcal{R}} \ \forall k\in \mathcal{K}\label{feas_check:worker_requirements}\\
    & x_k^{o,r} + \underset{\rho\in\bar{\mathcal{R}}^r_{-}}{\sum} x_k^{\rho, r} = x_k^{r,o'} +  \underset{\phi\in\bar{\mathcal{R}}^r_{+}}{\sum} x_k^{r, \phi} \quad \forall (r,q)\in\bar{\mathcal{R}} \ \forall k\in\mathcal{K} \label{feas_check:flow_conservation}\\
    & x_{k}^{o,o'} + \underset{(r,q)\in \bar{\mathcal{R}}}{\sum} x_k^{o,r} = N_k^{\mathrm{D}} \quad \forall k \in\mathcal{K}\label{feas_check:depot_leave}\\
    & x_{k}^{o,o'} + \underset{(r,q)\in \bar{\mathcal{R}}}{\sum} x_k^{r,o'} = N_k^{\mathrm{D}} \quad \forall k \in\mathcal{K}\label{feas_check:depot_return}\\
    & \chi_k^{o,r} \in \mathbb{Z}_{0} \quad \forall k\in\mathcal{K} \ \forall r \in \mathcal{\bar{R}} \label{feas_check:chi_integer}\\
    & x_k^{r,\phi} \in \mathbb{Z}_{0} \quad \forall (r,q)\in\bar{\mathcal{R}} \ \forall \phi\in\bar{\mathcal{R}^r_{+}} \ \forall k\in\mathcal{K} \label{feas_check:integer_tour}\\
    & x_k^{o,r},x_k^{r,o'} \in \mathbb{Z}_{0} \quad \forall (r,q)\in\bar{\mathcal{R}} \quad \forall k\in\mathcal{K}\label{feas_check:integer_depot}
\end{align}
Constraint (\ref{feas_check:worker_requirements}) ensures that each selected tour has enough qualified workers assigned to it. Constraint (\ref{feas_check:flow_conservation}) acts as a worker flow conservation constraint. Constraints (\ref{feas_check:depot_leave}) and (\ref{feas_check:depot_return}) ensure that the existing workforce size is not exceeded and each worker starts and ends their shift at the depot. The objective function (\ref{feas_check:obj}) aims at minimizing the sum over all slack variables, i.e., the number of additional workers required to make the current solution disaggregated-feasible. If this value is greater than 0, the available workforce is not sufficient, therefore the found solution is operationally infeasible. We note that the above model can easily be adjusted to solutions of the DMP by considering disaggregated team routes instead of team routes.\\
$\square$

\clearpage
\section{DMP Solutions and Feasibility Check}\label{appendix:feas_check_dmp_proof}
\begin{theorem}
    Let $(\bar{\lambda}^r_{q,s})_{(r,q,s)\in\mathcal{R}^{\mathrm{D}}}$ be an integer solution to the DMP (\ref{model:obj_disaggregated_mp})--(\ref{model:lambda_binary_disaggregated_mp}). Then, there exists a solution to the feasibility check (\ref{feas_check:obj})--(\ref{feas_check:integer_depot}) with objective function value 0, i.e., the solution to the DMP passes the feasibility check.
\end{theorem}

\proof{Proof. } We construct a solution for the feasibility check as follows:\\
\begin{algorithm}[H]
\label{construction_sol_feas_check_skill_comps}
  \SetAlgoLined
\caption{Construction of a solution for the feasibility check}
\For{$(r,q,s)\in\mathcal{\bar{R}}$ \text{with} $\mathcal{\bar{R}}_{-} = \emptyset$}
    {\For{$k\in\mathcal{K}$}
        {Set $x_k^{o,r} \coloneqq s_{q,k}$}
    }
Define $\mathcal{\bar{R}}_{>0} \coloneqq \left\{ (r,q,s)\in\mathcal{\bar{R}}: \ \mathcal{\bar{R}}_{-} \neq \emptyset \right\}$ and sort $\mathcal{\bar{R}}_{>0}$ ascending with respect to leave times $\mathrm{tl}^{r,q}$\\
\For{$(r,q,s)\in\mathcal{\bar{R}}_{>0}$}
    { Sort $\mathcal{\bar{R}}^{r}_{-}$ ascending with respect to $\mathrm{tl}^{\tilde{r},\tilde{q}}$\\
    \For{$k\in\mathcal{K}$}
        {\For{$(\tilde{r},\tilde{q},\tilde{s})\in\mathcal{\bar{R}}_{-}^r$}
            {\If{$s_{q,k} = \underset{\rho\in\mathcal{\bar{R}}_{-}^r}{\sum} x_k^{\rho,r}$ holds}{go to step 7}
            \Else{set
                \begin{equation}
                    x_{k}^{\tilde{r},r} \coloneqq \min\left\{ s_{q,k} - \underset{\rho\in\mathcal{\bar{R}}^r_{-}}{\sum} x_{k}^{\rho,\tilde{r}}, x_k^{o,\tilde{r}} + \underset{\rho\in\mathcal{\bar{R}}_{-}^{\tilde{r}}}{\sum} x_{k}^{\rho,\tilde{r}} - \underset{\phi\in\mathcal{\bar{R}}_{+}^{\tilde{r}}}{\sum} x_{k}^{\tilde{r},\phi} \right\}
                \end{equation}}
            }
        
        \If{$s_{q,k} > \underset{\rho\in\mathcal{\bar{R}}_{-}^r}{\sum} x_k^{\rho,r}$ holds}{ set
            \begin{equation}
                x_{k}^{o,r} \coloneqq \min\left\{s_{q,k} - \underset{\rho\in\mathcal{\bar{R}}^r_{-}}{\sum} x_{k}^{\rho,r}, N_k^{\mathrm{D}} - \underset{(\bar{r},\bar{q},\bar{s})\in\mathcal{\bar{R}}}{\sum} x_{k}^{o,\bar{r}}\right\}
            \end{equation}}
        }
    }

\For{$(r,q,s)\in\mathcal{\bar{R}}$}
    {\For{$k\in\mathcal{K}$}
        {set $x_k^{r,o'} \coloneqq x_k^{o,r} + \underset{\rho\in\mathcal{\bar{R}}_{-}^{r}}{\sum} x_{k}^{\rho, r} - \underset{\phi\in\mathcal{\bar{R}}_{+}^{r}}{\sum} x_{k}^{r,\phi}$}
        %}
    }
\For{$k\in\mathcal{K}$}
    {set $x_k^{o,o'} \coloneqq  N_k^{\mathrm{D}} - \underset{(r,q,s)\in\mathcal{\bar{R}}}{\sum} x_k^{o,r}$}
\end{algorithm}

We now show that the solution obtained from Algorithm \ref{construction_sol_feas_check_skill_comps} satisfies constraints (\ref{feas_check:worker_requirements})--(\ref{feas_check:integer_depot}).\\
We start by considering inequalities (\ref{feas_check:worker_requirements}). Let $(r,q,s)\in\mathcal{\bar{R}}$ be a disaggregated team route and $k\in\mathcal{K}$ be a skill level. Assume that $s_{q,k} = x_k^{o,r} + \underset{\rho\in\mathcal{\bar{R}}_{-}^r}{\sum} x_k^{\rho,r}$ holds. We can then conclude
\begin{equation*}
    \underset{\kappa=k}{\overset{K}{\sum}} \left(x_k^{o,r} + \underset{\rho\in\mathcal{\bar{R}}_{-}^r}{\sum} x_k^{\rho,r}\right) = \underset{\kappa=k}{\overset{K}{\sum}} s_{q,k} \overset{(\ref{definition_skill_comp_leq})}{\geq} \xi_{q,k}
\end{equation*}
holds and thus, constraint (\ref{feas_check:worker_requirements}) is satisfied. Hence, it is sufficient to show that after executing Algorithm 1, 
\begin{equation}
    s_{q,k} = x_k^{o,r} + \underset{\rho\in\mathcal{\bar{R}}_{-}^r}{\sum} x_k^{\rho,r} \label{equality_skill_comp}
\end{equation}
holds for all $(r,q,s)\in\mathcal{\bar{R}}$ and $k\in\mathcal{K}$.\\
Let $\mathcal{\bar{R}}_{tl}$ be the ordering of elements in $\mathcal{\bar{R}}$ in which they are treated in Algorithm \ref{construction_sol_feas_check_skill_comps}. Practically speaking, the first elements of $\mathcal{\bar{R}}_{tl}$ are equal to all disaggregated team routes without predecessor tours, i.e., all team routes whose inbound worker flow variables $x_k^{o,r}$ and $x_k^{\rho,r}$ are set in lines 1-4 of Algorithm \ref{construction_sol_feas_check_skill_comps}, followed by all routes with at least one predecessor tour, sorted ascending with respect to their depot leave time. We note that, for the satisfaction of (\ref{equality_skill_comp}), only the values of inbound worker flow variables $x_k^{o,r}$ and $x_k^{\rho,r}$ are relevant. Hence, we can prove the satisfaction of said constraint for all skill levels and disaggregated team routes via induction over the set $\mathcal{\bar{R}}_{tl}$.\\
Let $(r,q,s)\in\mathcal{\bar{R}}_{tl}$ be the first element in $\mathcal{\bar{R}}_{tl}$. From lines 1-5 of Algorithm \ref{construction_sol_feas_check_skill_comps}, we then know that $x_k^{o,r} = s_{q,k}$ holds for all $k\in\mathcal{K}$. Therefore, (\ref{equality_skill_comp}) is satisfied. We now assume that (\ref{equality_skill_comp}) is satisfied for all $k\in\mathcal{K}$ and all $(r,q,s)\in\mathcal{\bar{R}}_{tl}$ up to a certain index $i-1$. Let $(r,q,s)\in\mathcal{\bar{R}}_{tl}$ be the $i$-th element in $\mathcal{\bar{R}}_{tl}$. Furthermore, let $k\in\mathcal{K}$ be arbitrary and assume that, after executing loop 8-21 of Algorithm \ref{construction_sol_feas_check_skill_comps}, (\ref{equality_skill_comp}) is not satisfied for $(r,q,s)$ and $k$. By construction of the worker flow variables $x_k^{o,r}$ and $x_k^{\rho,r}$, we then know that
\begin{equation}
    s_{q,k} > x_k^{o,r} + \underset{\rho\in\mathcal{\bar{R}}_{-}^r}{\sum} x_k^{\rho,r} \label{inequality_skill_comp}
\end{equation}
holds. Now, consider the state of Algorithm \ref{construction_sol_feas_check_skill_comps} before executing loop 8-21 for $(r,q,s)$ and $k$. Then, from lines 15 and 19 of Algorithm \ref{construction_sol_feas_check_skill_comps} we know that
\begin{equation}
    \underset{\tilde{r}\in\mathcal{\bar{R}}_{-}^r}{\sum} \left(x_k^{o,\tilde{r}} + \left(\underset{\rho\in\mathcal{\bar{R}}_{-}^{\tilde{r}}}{\sum} x_k^{\rho, \tilde{r}} -  \underset{\phi\in\mathcal{\bar{R}}_{+}^{\tilde{r}}}{\sum} x_k^{\tilde{r}, \phi} \right)\right) + \left(N_k^{\mathrm{D}} - \underset{(\tilde{r},\tilde{q},\tilde{s})\in\mathcal{\bar{R}}}{\sum} x_k^{o,\tilde{r}}\right) < s_{q,k}\label{proof_not_enough_workers}
\end{equation} 
is satisfied. Additionally, we can reformulate the total available workforce $N_k^{\mathrm{D}}$ with skill level $k$ as
\begin{equation*}
    N_k^{\mathrm{D}} = N_k^{\mathrm{D}} + \underset{(\tilde{r},\tilde{q},\tilde{s})\in\mathcal{\bar{R}}}{\sum} x_k^{o,\tilde{r}} - \underset{(\tilde{r},\tilde{q},\tilde{s})\in\mathcal{\bar{R}}}{\sum} x_k^{o,\tilde{r}} = \left(N_k^{\mathrm{D}} - \underset{(\tilde{r},\tilde{q},\tilde{s})\in\mathcal{\bar{R}}}{\sum} x_k^{o,\tilde{r}}\right) + \underset{(\tilde{r},\tilde{q},\tilde{s})\in\mathcal{\bar{R}}}{\sum} x_k^{o,\tilde{r}}
\end{equation*}
Moreover, we know that for each $(\tilde{r},\tilde{q},\tilde{s})\in\mathcal{\bar{R}}$ and each predecessor tour $\rho\in\mathcal{\bar{R}}_{-}^{\tilde{r}}$, there exists an $(\bar{r},\bar{q},\bar{s})\in\mathcal{\bar{R}}$ and a successor tour $\phi\in\mathcal{\bar{R}}_{+}^{\bar{r}}$ such that $\tilde{r} = \phi$ and $\bar{r} = \rho$ hold, i.e., worker flow variables between tours cancel each other out. Therefore, 
\begin{equation}
    \underset{(\tilde{r},\tilde{q},\tilde{s})\in\mathcal{\bar{R}}}{\sum} \left(\underset{\rho\in\mathcal{\bar{R}}_{-}^{\tilde{r}}}{\sum} x_k^{\rho, \tilde{r}} -  \underset{\phi\in\mathcal{\bar{R}}_{+}^{\tilde{r}}}{\sum} x_k^{\tilde{r}, \phi} \right) = 0\label{proof_tour_flows_zero}
\end{equation}
holds. Hence, we conclude
\begin{equation*}
    N_k^{\mathrm{D}} = \left(N_k^{\mathrm{D}} - \underset{(\tilde{r},\tilde{q},\tilde{s})\in\mathcal{\bar{R}}}{\sum} x_k^{o,\tilde{r}}\right) + \underset{(\tilde{r},\tilde{q},\tilde{s})\in\mathcal{\bar{R}}}{\sum} \left(x_k^{o,\tilde{r}} + \left(\underset{\rho\in\mathcal{\bar{R}}_{-}^{\tilde{r}}}{\sum} x_k^{\rho, \tilde{r}} -  \underset{\phi\in\mathcal{\bar{R}}_{+}^{\tilde{r}}}{\sum} x_k^{\tilde{r}, \phi} \right)\right)
\end{equation*}
Moreover, the set $\mathcal{\bar{R}}$ can be partitioned into
\begin{equation*}
    \mathcal{\bar{R}} = \mathcal{\bar{R}}_{-}^r \cup \left\{(\tilde{r},\tilde{q},\tilde{s})\in\mathcal{\bar{R}}: \ \mathrm{tl}^{\tilde{r}} \leq \mathrm{tl}^{r,q} \leq \mathrm{tr}^{\tilde{r},\tilde{q}} \right\} \cup \left\{(\tilde{r},\tilde{q},\tilde{s})\in\mathcal{\bar{R}}: \ \mathrm{tl}^{r,q} < \mathrm{tl}^{\tilde{r}} \right\}
\end{equation*}
Because the last set of the above partition has only trivial worker inflow variables $x_k^{o,\tilde{r}}$ and $x_k^{\rho,\tilde{r}}$ associated with it, we can conclude
\begin{align*}
    N_k^{\mathrm{D}} &= \left(N_k^{\mathrm{D}} - \underset{(\tilde{r},\tilde{q},\tilde{s})\in\mathcal{\bar{R}}}{\sum} x_k^{o,\tilde{r}}\right) + \underset{\tilde{r}\in\mathcal{\bar{R}}_{-}^{r}}{\sum} \left(x_k^{o,\tilde{r}} + \left(\underset{\rho\in\mathcal{\bar{R}}_{-}^{\tilde{r}}}{\sum} x_k^{\rho, \tilde{r}} -  \underset{\phi\in\mathcal{\bar{R}}_{+}^{\tilde{r}}}{\sum} x_k^{\tilde{r}, \phi} \right)\right) +\\
    &+ \underset{\tilde{r}: \ \mathrm{tl}^{\tilde{r}} \leq \mathrm{tl}^{r,q} \leq \mathrm{tr}^{\tilde{r},\tilde{q}}}{\sum} \left(x_k^{o,\tilde{r}} + \left(\underset{\rho\in\mathcal{\bar{R}}_{-}^{\tilde{r}}}{\sum} x_k^{\rho, \tilde{r}} -  \underset{\phi\in\mathcal{\bar{R}}_{+}^{\tilde{r}}}{\sum} x_k^{\tilde{r}, \phi} \right)\right)
\end{align*}
Using (\ref{proof_not_enough_workers}) and (\ref{model:workerconstr_disaggregated_mp}), we then infer
\begin{align*}
    N_k^{\mathrm{D}} &\overset{(\ref{proof_not_enough_workers})}{<} s_{q,k} + \underset{\tilde{r}: \ \mathrm{tl}^{\tilde{r}} \leq \mathrm{tl}^{r,q} \leq \mathrm{tr}^{\tilde{r},\tilde{q}}}{\sum} \left(x_k^{o,\tilde{r}} + \left(\underset{\rho\in\mathcal{\bar{R}}_{-}^{\tilde{r}}}{\sum} x_k^{\rho, \tilde{r}} -  \underset{\phi\in\mathcal{\bar{R}}_{+}^{\tilde{r}}}{\sum} x_k^{\tilde{r}, \phi} \right)\right) \leq s_{q,k} + \underset{\tilde{r}: \ \mathrm{tl}^{\tilde{r}} \leq \mathrm{tl}^{r,q} \leq \mathrm{tr}^{\tilde{r},\tilde{q}}}{\sum} \left(x_k^{o,\tilde{r}} + \underset{\rho\in\mathcal{\bar{R}}_{-}^{\tilde{r}}}{\sum} x_k^{\rho, \tilde{r}}\right) =\\
    &= s_{q,k} + \underset{\tilde{r}: \ \mathrm{tl}^{\tilde{r}} \leq \mathrm{tl}^{r,q} \leq \mathrm{tr}^{\tilde{r},\tilde{q}}}{\sum} s_{\tilde{q},k} \leq \underset{(\tilde{r},\tilde{q},\tilde{s})\in\mathcal{\bar{R}}}{\sum} \beta_{k,\mathrm{tl}^{r,q}}^{\tilde{s}}(\omega_{\gamma}) \overset{(\ref{model:workerconstr_disaggregated_mp})}{\leq} N_k^{\mathrm{D}}
\end{align*}
which is a contradiction. Hence, (\ref{equality_skill_comp}) must be satisfied with equality. Therefore, constraint (\ref{feas_check:worker_requirements}) is satisfied for $(r,q,s)$ and $k$.\\
We now consider constraints (\ref{feas_check:flow_conservation})--(\ref{feas_check:depot_return}). From Loops 23-26 and 28-30, it is clear to see that (\ref{feas_check:flow_conservation}) and (\ref{feas_check:depot_leave}) are satisfied. Furthermore, we know from (\ref{proof_tour_flows_zero}) and lines 24-26 that
\begin{equation*}
    \underset{(r,q,s)\in\mathcal{\bar{R}}}{\sum} x_k^{r,o'} \ \overset{L.24-26}{=} \underset{(r,q,s)\in\mathcal{\bar{R}}}{\sum} \left( x_k^{o,r} + \left( \underset{\rho\in\mathcal{\bar{R}}_{-}^r}{\sum} x_k^{\rho, r} - \underset{\phi\in\mathcal{\bar{R}}_{+}^r}{\sum} x_k^{r, \phi}\right)\right) \overset{(\ref{proof_tour_flows_zero})}{=} \underset{(r,q,s)\in\mathcal{\bar{R}}}{\sum} x_k^{o,r}
\end{equation*}
holds. Because (\ref{feas_check:depot_leave}) holds, we can therefore infer that (\ref{feas_check:depot_return}) is also satisfied.\\
It is left to show that all worker flow variables are integral and non-negative. Using minima in lines 15 and 19 ensures that
\begin{equation*}
    x_k^{o,\tilde{r}} + \underset{\rho\in\mathcal{\bar{R}}_{-}^{\tilde{r}}}{\sum} x_{k}^{\rho,\tilde{r}} - \underset{\phi\in\mathcal{\bar{R}}_{+}^{\tilde{r}}}{\sum} x_{k}^{\tilde{r},\phi} \geq 0\label{proof_nonnegativity}
\end{equation*}
and
\begin{equation*}
    N_k^{\mathrm{D}} - \underset{\bar{r},\bar{q},\bar{s}\in\mathcal{\bar{R}}}{\sum} x_k^{o,\bar{r}} \geq 0
\end{equation*}
are always satisfied for all $(\tilde{r},\tilde{q},\tilde{s})\in\mathcal{\bar{R}}$. Hence, all variable updates preserve non-negativity and integrality. Therefore, Algorithm \ref{construction_sol_feas_check_skill_comps} returns a feasible solution to the feasibility check with an objective function value of 0.

$\square$
\clearpage

\section{Proof of Equivalent Pricing Networks}\label{appendix:proof_equiv_pricing_networks}
We first prove that transferring a label to a different pricing network that shares the same profile does not alter its feasibility.
\begin{lemma}\label{lemma_transferred_label_feasible}
    Let $(q,s)\in\mathcal{Q}^{\mathrm{D}}$ be a disaggregated profile. Furthermore, let 
    \begin{equation*}
        L\coloneqq \left(\mathrm{tl}^L, v, P^L, T_{v}^{L,\mathrm{cost}}, (T_{v,i}^{L,\mathrm{perf}})_{i\in\mathcal{I}_q}, F_{v}^{L},\right)
    \end{equation*}
    be a label in $\mathcal{G}^{q,s}$. Then, for any disaggregated profile $(q,\tilde{s})\in\mathcal{Q}^{\mathrm{D}}$,
    \begin{equation*}
        \tilde{L}\coloneqq \left(\mathrm{tl}^L, v, P^L, T_{v}^{\tilde{L},\mathrm{cost}}, (T_{v,i}^{L,\mathrm{perf}})_{i\in\mathcal{I}_q}, F_{v}^{L}\right)
    \end{equation*}
    is a label in $\mathcal{G}^{q,\tilde{s}}$ and $P^L$ is a feasible path in $\mathcal{G}^{q,\tilde{s}}$.
\end{lemma}

\proof{Proof. }
By definition, task execution times $p_{i,q}$ and travel times $t_{i,j}$ are independent of the underlying skill composition. Therefore, the finish time distributions and $\omega_{\gamma}$-scenario finish times of tasks in $P^L$ are identical in both pricing networks $\mathcal{G}^{q,s}$ and $\mathcal{G}^{q,\tilde{s}}$. By assumption, $P^L$ is feasible in $\mathcal{G}^{q,s}$. Thus, $P^L$ is also feasible in $\mathcal{G}^{q,\tilde{s}}$. Furthermore, the reduced cost of $\tilde{L}$ are equal to
\begin{equation}
    T_v^{\tilde{L}, \mathrm{cost}} = T_v^{L,\mathrm{cost}} + \underset{k\in\mathcal{K}}{\sum} \underset{\tau=\mathrm{tl}^L}{\overset{F_v^{L}(\omega_{\gamma})}{\sum}}\delta_{k,\tau} \beta_{k,\tau}^{L,s}(\omega_{\gamma}) -  \underset{k\in\mathcal{K}}{\sum} \underset{\tau=\mathrm{tl}^{\tilde{L}}}{\overset{F_v^{\tilde{L}}(\omega_{\gamma})}{\sum}}\delta_{k,\tau} \beta_{k,\tau}^{\tilde{L},\tilde{s}}(\omega_{\gamma}).\label{reduced_cost_transformation}
\end{equation}
Therefore, $\tilde{L}$ is a label in $\mathcal{G}^{q,\tilde{s}}$.
$\square$\\

We say that $\tilde{L}$ is the \textit{transferred label} of $L$ from $\mathcal{G}^{q,s}$ to $\mathcal{G}^{q,\tilde{s}}$. We now show that label dominance does not depend on the particular skill composition considered.
\begin{lemma}\label{lemma:dominance_equivalence}
    Let $(q,s)\in\mathcal{Q}^{\mathrm{D}}$ be a disaggregated profile. Furthermore, let
    \begin{equation*}
        L^j\coloneqq \left(\mathrm{tl}^{L^j}, v, P^{L^j}, T_{v}^{L^j,\mathrm{cost}}, (T_{v,i}^{L^j,\mathrm{perf}})_{i\in\mathcal{I}_q}, F_{v}^{L^j}\right)
    \end{equation*}
    be labels in $\mathcal{G}^{q,s}$ and $\tilde{L}^j$  their transferred labels from $\mathcal{G}^{q,s}$ to $\mathcal{G}^{q,\tilde{s}}$ for $j=1,2$. Then, the following holds:
    \begin{equation*}
        \text{$L^1$ dominates $L^2$ in $\mathcal{G}^{q,s}$ $\Leftrightarrow$ $\tilde{L}^1$ dominates $\tilde{L}^2$ in $\mathcal{G}^{q,\tilde{s}}$}
    \end{equation*}
\end{lemma}
\proof{Proof. }
Assume that $L^1$ dominates $L^2$ in $\mathcal{G}^{q,s}$. By construction of $\tilde{L}^1$ and $\tilde{L}^2$, (\ref{dominance:task_resources_dmp})--(\ref{dominance:quantile_case_finishes_dmp}) and (\ref{dominance:depot_leave_dmp}) hold. Furthermore, for the reduced cost offset by the workforce penalty, using Lemma \ref{lemma_transferred_label_feasible} we obtain
\begin{align*}
    T_v^{\tilde{L}^1, \mathrm{cost}} &+ \underset{k\in\mathcal{K}}{\sum} \underset{\tau=\mathrm{tl}^{1}}{\overset{F_v^{1}(\omega_{\gamma})}{\sum}}\delta_{k,\tau} \beta_{k,\tau}^{\tilde{L},\tilde{s},1} \overset{(\ref{reduced_cost_transformation})}{=} T_v^{L^1,\mathrm{cost}} + \underset{k\in\mathcal{K}}{\sum} \underset{\tau=\mathrm{tl}^1}{\overset{F_v^{1}(\omega_{\gamma})}{\sum}}\delta_{k,\tau} \beta_{k,\tau}^{L^1,s,1}(\omega_{\gamma}) \overset{Def. \ref{dominance_rule_dmp}}{\leq}\\
    &\leq T_v^{L^2,\mathrm{cost}} + \underset{k\in\mathcal{K}}{\sum} \underset{\tau=\mathrm{tl}^2}{\overset{F_v^{2}}{\sum}}\delta_{k,\tau} \beta_{k,\tau}^{L^2,s,2}(\omega_{\gamma}) = T_v^{\tilde{L}^2, \mathrm{cost}} + \underset{k\in\mathcal{K}}{\sum} \underset{\tau=\mathrm{tl}^{2}}{\overset{F_v^{2}(\omega_{\gamma})}{\sum}}\delta_{k,\tau} \beta_{k,\tau}^{L^2,\tilde{s},2}.
\end{align*}
Thus, $\tilde{L}^1$ dominates $\tilde{L}^2$ in $\mathcal{G}^{q,\tilde{s}}$. The converse statement can be proven analogously.
$\square$\\

It is left to show that each minimum reduced cost label in a pricing network can be obtain by solving a different pricing network that shares the same profile and then transferring all non-dominated sink labels to the original pricing network. For this, we first show that all sublabels of a transferred (optimal) sink label remain non-dominated.
\begin{lemma}\label{lemma_opt_label_sublabels_nondominated}
    Let $(q,s)\in\mathcal{Q}^{\mathrm{D}}$ be a disaggregated profile. Moreover, let 
    \begin{equation*}
        L^*\coloneqq \left(\mathrm{tl}^{L^*}, v, P^{L^*}, T_{v}^{L^*,\mathrm{cost}}, (T_{v,i}^{L^*,\mathrm{perf}})_{i\in\mathcal{I}_q}, F_{v}^{L^*}\right)
    \end{equation*} 
    be the sink label in $\mathcal{G}^{q,s}$ with minimum reduced cost, i.e.
    \begin{equation*}
        T^{L^*,\mathrm{cost}} \leq T^{L,\mathrm{cost}}
    \end{equation*}
    holds for all sink labels $L$ in $\mathcal{G}^{q,s}$. Additionally, let $\tilde{s}\in\mathcal{S}_q$ be arbitrary, $\tilde{L}^*$ be the transferred label of $L^*$ from $\mathcal{G}^{q,s}$ to $\mathcal{G}^{q,\tilde{s}}$ and $\tilde{P}^* = (o,v_1,\dots,v_l,o')$ be its associated path. Then, each sublabel $\tilde{L}_{v_j}^*$ obtained from $\tilde{L}^*$ by selecting a subpath $(o,v_1,\dots,v_j)$ with $j\in\{1,\dots,l\}$ is non-dominated in $\mathcal{G}^{q,\tilde{s}}$.
\end{lemma}
\proof{Proof. }
Let $\tilde{L}^*_{v_j}$ be a sublabel of $\tilde{L}^*$ and assume that there exists a label $\tilde{L}_{v_j}$ that dominates $\tilde{L}^*_{v_j}$. We know that  $(v_j,v_{j+1},..,o')$ is a feasible extensions of $\tilde{L}^*_{v_j}$. By Definition of the dominance rule, $(v_j,v_{j+1},..,o')$ is also a feasible extension for $\tilde{L}_{v_j}$. When extending both labels along $(v_j,v_{j+1},..,o')$, we obtain labels $\tilde{L}^*$ and $\tilde{L}$ both ending at $o'$ with reduced costs
\begin{align*}
    T^{\tilde{L}^*, \mathrm{cost}} &= T^{\tilde{L}^*_{v_j}, \mathrm{cost}} - \underset{i = j+1}{\overset{l}{\sum}}(w_{v_i} \cdot \mathbb{E}(F_{v_i}^{L^*} + F_{v_i}^P(F_{v_i}^{L^*})) - \mu_{v_i}) - \underset{k\in\mathcal{K}}{\sum}\underset{\tau = F_{v_j}^{L^*}(\omega_{\gamma})+1}{\overset{\mathrm{tr}^{L^*}}{\sum}} \delta_{k,\tau} \beta_{k,\tau}^{L^*,s}(\omega_{\gamma}) = \\
    &= T^{\tilde{L}^*_{v_j}, \mathrm{cost}} - \underset{i = j+1}{\overset{l}{\sum}}(w_{v_i} \cdot \mathbb{E}(F_{v_i}^{L} + F_{v_i}^P(F_{v_i}^{L})) - \mu_{v_i}) - \underset{k\in\mathcal{K}}{\sum}\underset{\tau = F_{v_j}^{L}(\omega_{\gamma})+1}{\overset{\mathrm{tr}^{L}}{\sum}} \delta_{k,\tau} \beta_{k,\tau}^{L,s}(\omega_{\gamma}) \geq \\
    &\geq T^{\tilde{L}_{v_j}, \mathrm{cost}} - \underset{i = j+1}{\overset{l}{\sum}}(w_{v_i} \cdot \mathbb{E}(F_{v_i}^{L} + F_{v_i}^P(F_{v_i}^{L})) - \mu_{v_i}) - \underset{k\in\mathcal{K}}{\sum}\underset{\tau = F_{v_j}^{L}(\omega_{\gamma})+1}{\overset{\mathrm{tr}^{L}}{\sum}} \delta_{k,\tau} \beta_{k,\tau}^{L,s}(\omega_{\gamma}) = T^{\tilde{L}, \mathrm{cost}} 
\end{align*}
which is a contradiction to the optimality of $\tilde{L}^*$.
$\square$\\

We can now prove the equivalence of pricing networks under the proposed dominance rule.
\begin{theorem}
    Let $(q,s)\in\mathcal{Q}^{\mathrm{D}}$ be a disaggregated profile. Then each sink label $L^*$ in $\mathcal{G}^{q,s}$ with minimum reduced cost can be obtained by transferring all non-dominated sink labels $\tilde{L}$ in $\mathcal{G}^{q,\tilde{s}}$ to $\mathcal{G}^{q,s}$.
\end{theorem}
\proof{Proof. }
    Let $\mathcal{L}^{\tilde{s}}$ be the set of non-dominated sink labels in $\mathcal{G}^{q,\tilde{s}}$. Furthermore, let $L^*$ be sink label in $\mathcal{G}^{q,s}$ with minimum reduced cost. From Lemma \ref{lemma_opt_label_sublabels_nondominated} we know that all its sublabels $\tilde{L}^*_{v_j}$ are non-dominated in $\mathcal{G}^{q,\tilde{s}}$. We can conclude that its transferred label $\tilde{L}^*$ is also non-dominated in $\mathcal{G}^{q,\tilde{s}}$, i.e., $\tilde{L}^* \in \mathcal{L}^{\tilde{s}}$ holds. Therefore, transferring $\tilde{L}^*$ back to $\mathcal{G}^{q,s}$ yields $L^*$.
$\square$
\clearpage

\section{Proof of Dominance of the DMP}\label{appendix:dominance_dmp}
\begin{theorem}
    For any feasible solution of the DMP, there exists an equivalent feasible solution for the AMP.
\end{theorem}
\proof{Proof. }
    Let $(\bar{\lambda}^r_{q,s})_{(r,q,s)\in\mathcal{R}^{\mathrm{D}}}$ be a feasible solution of the DMP. We define $(\hat{\lambda}^r_q)_{(r,q)\in\mathcal{R}}$ as follows:
    \begin{equation*}
        \hat{\lambda}^r_q \coloneqq  \underset{s\in\mathcal{S}^q}{\sum} \bar{\lambda}^r_{q,s}
    \end{equation*}
    By construction, constraint $(\ref{model:tasksdone_aggregated_mp})$ is satisfied. Furthermore, from the objective function (\ref{model:obj_disaggregated_mp}) and constraint (\ref{model:tasksdone_disaggregated_mp}), it is clear that $\hat{\lambda}^r_q\in\{0,1\}$ holds. Additionally,
    \begin{equation*}
        \sum_{(r,q)\in \mathcal{R}} \mathbb{E}(c^r)\hat{\lambda}^r_q = \sum_{(r,q,s)\in \mathcal{R}^{\mathrm{D}}} \mathbb{E}(c^r)\bar{\lambda}^r_{q,s}
    \end{equation*}
    holds. Thus, $(\bar{\lambda}^r_{q,s})_{(r,q,s)\in\mathcal{R}^{\mathrm{D}}}$ and $(\hat{\lambda}^r_q)_{(r,q)\in\mathcal{R}}$ can be considered as equivalent. It is left to show that $(\hat{\lambda}^r_q)_{(r,q)\in\mathcal{R}}$ satisfies $(\ref{model:workerconstr_aggregated_mp})$.\\
    Let $(r,q)\in\mathcal{R}$, $k\in\mathcal{K}$ and $t\in\mathcal{T}$ be arbitrary. Without loss of generality assume that $t\in[\mathrm{tl}^{r,q},\mathrm{tr}^{r,q}]$ holds. Moreover, let $s\in\mathcal{S}^q$ be a skill composition. By definition of $b_{k,t}^{r,q}(\omega_{\gamma})$, we know that
    \begin{equation}
        b_{k,t}^r(\omega_{\gamma}) = \xi_{q,k} \overset{(\ref{definition_skill_comp_leq})}{\leq} \underset{\kappa\geq k}{\sum} s_{q,\kappa} = \underset{\kappa\geq k}{\sum} \beta_{k,t}^{s}(\omega_{\gamma})\label{b_beta_inequality}
    \end{equation}
    is satisfied. We then conclude
    \begin{align*}
        &\underset{(r,q)\in\mathcal{R}}{\sum} b_{k,t}^r(\omega_{\gamma}) \hat{\lambda}^r_q = \underset{(r,q)\in\mathcal{R}}{\sum} \ \underset{s\in\mathcal{S}^q}{\sum} b_{k,t}^r(\omega_{\gamma}) \bar{\lambda}^r_{q,s} =  \underset{(r,q,s)\in\mathcal{R}^{\mathrm{D}}}{\sum} b_{k,t}^r(\omega_{\gamma}) \bar{\lambda}^r_{q,s} \overset{(\ref{b_beta_inequality})}{\leq} \\
        &\overset{(\ref{b_beta_inequality})}{\leq} \underset{(r,q,s)\in\mathcal{R}^{\mathrm{D}}}{\sum} \ \underset{\kappa\geq k}{\sum} \beta_{k,t}^s(\omega_{\gamma}) \bar{\lambda}^r_{q,s} = \underset{\kappa\geq k}{\sum} \ \underset{(r,q,s)\in\mathcal{R}^{\mathrm{D}}}{\sum} \beta_{k,t}^s(\omega_{\gamma}) \bar{\lambda}^r_{q,s} \overset{(\ref{model:workerconstr_disaggregated_mp})}{\leq} \underset{\kappa\geq k}{\sum} N_k^{\mathrm{D}} = N_k.
    \end{align*}
    Hence, $(\hat{\lambda}^r_q)_{(r,q)\in\mathcal{R}}$ also satisfies constraints (\ref{model:workerconstr_aggregated_mp}) and therefore constitutes a feasible solution to the AMP, while having the same objective function value as the respective solution to the DMP.
$\square$\\

\begin{corollary}
    In the above sense of equivalent solutions, the feasible region of the DMP is a subspace of the feasible region of the AMP.
\end{corollary}
\begin{remark}
    When relaxing binary conditions (\ref{model:lambda_binary_aggregated_mp}) to $\lambda^r_q\in [0,1]$, the above statement holds true for any feasible solution of LP relaxation of the DMP.
\end{remark}
\clearpage

\section{Example: Non-unique finish time distributions after label backwards-extension}\label{appendix:backward_label}
Let $q\in\mathcal{Q}$ be a profile and $L\coloneqq \left(\mathrm{tl}^L, v, P, T_{v}^{\mathrm{cost}}, (T_{v,i}^{\mathrm{perf}})_{i\in\mathcal{I}_q}, F_{v}^L, F_{v}^{L}(\omega_{\gamma})\right)$ be a label. Consider an extension of $L'$ of $L$ along $(v',v)$, i.e., a backward extension. We are interested in finding a label $L'$ and corresponding finish time distribution $F_{v'}^{L'}$ such that when $L'$ is extended along $(v,' v)$, we obtain $L$. This is equivalent to finding a function $F_{v'}^{L'}: [\mathrm{ES}_v, \mathrm{LS}_v^{\mathrm{e}}] \rightarrow [0,1]$ that satisfies the following linear (in)equalities: 
\begin{align}
    F_{v}^{L}(t+p_{v,q}) &= \underset{\tau\in B_{v',v}: \mathrm{ES}_{v'} \leq t - \tau - p_{v',q} \leq \mathrm{LS}_{v'}}{\sum} \mathbb{P}(t_{v',v} = \tau) \cdot F_{v'}^{L'}(t-\tau) \quad \forall t \in\left\{\mathrm{ES}_v+1,\dots,\mathrm{LS}_v^{\mathrm{e}}\right\} \label{eq:appendix_backwards_propagation_1}\\
    F_{v}^{L}(t+p_{v,q}) &= \underset{\tau\in B_{v',v}}{\sum} P(t_{i,j} = \tau) \cdot \underset{t'= \mathrm{ES}_{v'}}{\overset{\mathrm{LS}_{v'}^{\mathrm{e}}}{\sum}} F_{v'}^{L'}(t-\tau + p_{v',q}) \quad, t = \mathrm{ES}_v \label{eq:appendix_backwards_propagation_2}\\
    \underset{t = \mathrm{ES}_{v'}}{\overset{\mathrm{EF}_{v'}}{\sum}} F_{v'}^{L'}(t+p_{v',q}) &= 1\\
     F_{v'}^{L'}(t+p_{v',q}) &\geq 0 \quad \forall t\in [\mathrm{ES}_{v'}, \mathrm{EF}_{v'}] \label{eq:appendix_backwards_propagation_4}
\end{align}
We note that there can be infinitely many solutions to the above set of (in)equalities.
\begin{example}
Let $q\in\mathcal{Q}$ be an arbitrary profile, let $v',v\in\mathcal{I}_q$ be tasks and assume $p_{v',q} = p_{v,q} = 0$. Furthermore, let the time windows of $v$ and $v'$ be defined by $[\mathrm{ES}_v, \mathrm{LS}_v] = [10,12]$ and $[\mathrm{ES}_{v'}, \mathrm{LS}_{v'}] = [4,5]$, respectively. We note that start and finish time windows are identical due to the fact that execution times are assumed to be zero. Moreover, let the travel times $t_{v',v}$ follow the distribution
\begin{align*}
    \mathbb{P}(t_{v',v} = t) = \begin{cases}
        \begin{aligned}
            &0.5 && \text{if} \ \  t \in \{2,3\}\\
            &0 && \text{otherwise}
            \end{aligned}
    \end{cases}
\end{align*}
and assume that $v$ is always finished at its earliest possible finish time, i.e.
\begin{equation}
    F_{v}(t) = \begin{cases}
        \begin{aligned}
            &1 &&  \text{if} \ \ t=\mathrm{ES}_v\\
            &0 && \text{otherwise}
        \end{aligned}
        \end{cases} \label{eq:appendix_finish_time_distr}
\end{equation}
holds. Then, (\ref{eq:appendix_backwards_propagation_1})--(\ref{eq:appendix_backwards_propagation_4}) are equal to
\begin{align*}
    &1  \overset{!}{=}  \left(F_{v'}(3) + F_{v'}(4)\right) \cdot \left(\mathbb{P}(t_{v',v} = 2) + \mathbb{P}(t_{v',v}=3)\right) = F_{v'}(3) + F_{v'}(4)\\
    F_{v'}(3), F_{v'}(4) &\in [0,1]
\end{align*}
Practically speaking, any arbitrary finish time distribution of task $v'$ results, when extended along arc $(v',v)$, in the finish time distribution (\ref{eq:appendix_finish_time_distr}). Thus, when backwards extending a label along $(v',v)$, one generally needs to create infinitely many labels at $v'$, one for each possible finish time distribution that, when forward extended along $(v',v)$, would result in the desired finish time distribution (\ref{eq:appendix_finish_time_distr}) in $v$.
\end{example}

\clearpage

\section{Number of Feasible Instances by Instance Characteristics}\label{appendix:feasible_instance_count}
Column ``\#Inst" contains the number of instances for which the specific criteria are set to the respective value. In this context, ``hor" describes the length of the time horizon in minutes, ``fph" the number of flights per hour, ``rs" the workforce size as a multiple of the workforce needed to trivially solve the instance and ``modes" the set of available modes, i.e., (s)low, (i)ntermediate, and (f)ast. Columns ``\#Feas." and ''\%Feas." contain the number of feasible instances and the percentage of instances that are feasible. Columns ``\#Infeas." and ``\%Infeas." do the same for the number and percentage of infeasible instances.
\begin{table}[H]
    \centering
    \caption{Number of (in)feasible instances by instance characteristics for $\gamma=0.9$}
    \label{tab:feas_instances_per_characteristic}
    \begin{tabular}{llrrrrr}
    \toprule
    Criteria & Value & \multicolumn{1}{c}{\#Inst.} & \multicolumn{1}{c}{\#Feas.} & \multicolumn{1}{c}{\#Infeas.} & \multicolumn{1}{c}{\%Feas.} & \multicolumn{1}{c}{\%Infeas.} \\ \midrule
    hor      & 60    & 405                         & 199                          & 206                           & 49.14\%                      & 50.86\%                       \\
    hor      & 90    & 405                         & 205                          & 200                           & 50.62\%                      & 49.38\%                       \\
    hor      & 120   & 405                         & 197                          & 208                           & 48.64\%                      & 51.36\%                       \\ \midrule
    fph      & 10    & 405                         & 195                          & 210                           & 48.15\%                      & 51.85\%                       \\
    fph      & 20    & 405                         & 202                          & 203                           & 49.88\%                      & 50.12\%                       \\
    fph      & 30    & 405                         & 204                          & 201                           & 50.37\%                      & 49.63\%                       \\ \midrule
    rs       & 0.1   & 135                         & 0                            & 135                           & 0.00\%                       & 100.00\%                      \\
    rs       & 0.2   & 135                         & 0                            & 135                           & 0.00\%                       & 100.00\%                      \\
    rs       & 0.3   & 135                         & 3                            & 132                           & 2.22\%                       & 97.78\%                       \\
    rs       & 0.4   & 135                         & 46                           & 89                            & 34.07\%                      & 65.93\%                       \\
    rs       & 0.5   & 135                         & 78                           & 57                            & 57.78\%                      & 42.22\%                       \\
    rs       & 0.6   & 135                         & 100                           & 35                            & 74.07\%                      & 25.93\%                       \\
    rs       & 0.7   & 135                         & 113                          & 22                            & 83.70\%                      & 16.30\%                       \\
    rs       & 0.8   & 135                         & 128                          & 7                             & 94.81\%                      & 5.19\%                        \\
    rs       & 0.9   & 135                         & 133                          & 2                             & 98.52\%                      & 1.48\%                        \\ \midrule
    modes    & sif   & 405                         & 256                          & 149                           & 63.21\%                      & 36.79\%                       \\
    modes    & sf    & 405                         & 226                          & 179                           & 55.80\%                      & 44.20\%                       \\
    modes    & i     & 405                         & 119                          & 286                           & 29.38\%                      & 70.62\%                       \\
    \bottomrule
    \end{tabular}
\end{table}
\clearpage

\section{Size of Underlying Instance Sets for Computational Studies}\label{appendix:instance_set_sizes}
\begin{table}[ht]
    \centering
    \caption{Comparison of basic and full solver configuration for $\gamma=0.9$}
    \label{tab:instance_set_sizes}
    \begin{tabular}{llrrrrrr}
    \toprule
    Comparison           & \multicolumn{1}{c}{Instance Class} & \multicolumn{1}{c}{Ropt} & \multicolumn{1}{c}{Nopt} & \multicolumn{1}{c}{All other} \\ \midrule
    (Full, Basic)        & Easy       &                          &                                                                                                           & 261                           \\
    \multicolumn{1}{l}{} & Medium     &                          &                                                                                                           & 213                           \\
    \multicolumn{1}{l}{} & Hard       &                          &                                                                                                           & 127                           \\
                         & All        &                          &                                                                                                           & 601                           \\ \midrule
    (Full, No Branching) & Easy       & 251                      & 251                                                                                                       & 261                           \\
    \multicolumn{1}{l}{} & Medium     & 158                      & 158                                                                                                       & 213                           \\
    \multicolumn{1}{l}{} & Hard       & 64                       & 64                                                                                                        & 127                           \\
                         & All        & 473                      & 473                                                                                                       & 601                           \\ \midrule
    (Full, no DRMP)      & Easy       & 4                      & 4                                                                                                       & 4                           \\
    \multicolumn{1}{l}{} & Medium     & 13                      & 13                                                                                                       & 19                           \\
    \multicolumn{1}{l}{} & Hard       & 8                       & 8                                                                                                        & 13                           \\
                         & All        & 25                      & 25                                                                                                       & 36                           \\ \midrule
    (Full, No CGCs)      & Easy       &                          &                                                                                                           & 261                           \\
    \multicolumn{1}{l}{} & Medium     &                          &                                                                                                           & 213                           \\
    \multicolumn{1}{l}{} & Hard       &                          &                                                                                                           & 127                           \\
                         & All        &                          &                                                                                                           & 601                           \\ \bottomrule
    \end{tabular}
\end{table}

In Table \ref{tab:instance_set_sizes}, blank cells mean that the corresponding metric has not been evaluated for the respective configuration pair. Column ``All other" refers to the number of instances that were used to evaluate all metrics that have not been addressed in any other column of the above table.
\clearpage

\section{Details on Permutation Tests}\label{appendix:perm_test}
To account for the increasing probability of making at least one type I error when conducting multiple tests in sequence, multiplicity corrections such as the Holm-Bonferroni correction or Hochberg's step-up procedure are typically applied (cp. \cite{demvsar2006statistical}). However, in the following, all reported $p$-values are either $0.0$ or $1.0$, eliminating the necessity to apply such correction methods altogether. Furthermore, we also compute an effect size computed analogous to the rank-biserial correlation coefficient for Wilcoxon-signed ranked sum tests (see e.g. \cite{king2018statistical}) as $\dfrac{D^+ - D^-}{D^+ + D^-}$, where $D^+ \coloneqq \underset{i: d_i > 0}{\sum} d_i$ is the sum of positive gap differences (i.e., gap differences of instances where the full configuration yielded a strictly lower gap than the other configuration), and $D^-  \coloneqq \underset{i: d_i < 0}{\sum} -d_i$ is the inverted sum of negative gap differences. Typically, the larger this value is, the more gap differences lean towards an overperformance of the full configuration.\\
We compute confidence intervals by inverting the permutation test as described by \cite{good2005permutation}. For this, we start with the interval $CI=[-100, 100]$, which contains all theoretically possible gap differences, and iteratively bisect the interval until the smallest and largest values $\underline{\theta}$ and $\overline{\theta}$, for which the null hypothesis with $\theta_0 = \theta$ is accepted, are found. The confidence interval is then given by $[\underline{\theta}, \overline{\theta}]$. We note that, for some comparisons, $H_0$ is accepted for all $\theta_0 \leq 0$. This is mainly caused by a large point mass of the distribution of gap differences located at exactly 0. In these cases, the observations are not sufficient to reject the null hypothesis for any $\theta \leq 0$, and hence it is accepted even for practically impossible values below $-100$. To remain coherent, we still report the true confidence interval, which, in this case, is unbounded from below.\\

\section{Computation of Scenario Count}\label{appendix:computation_scen_count}
To determine a sufficient number of scenarios, we employ a Sample Average Approximation (SAA) procedure. 
Starting with $N=50$ scenarios, we generate $M=25$ independent scenario batches of size $N$ for each instance $i$. 
For each batch $m=1,\dots,M$, we compute the empirical objective value
\begin{equation*}
    \bar{v}_{N,i}^{m} = \frac{1}{N} \sum_{n=1}^{N} F_i(\lambda,\zeta_n^{m}),
\end{equation*}
where $F_i(\lambda,\zeta)$ denotes the objective value (\ref{model:obj_aggregated_mp}) of solution $\lambda$ for instance $i$ under scenario $\zeta_n^m$.
The average empirical objective value across batches is
\begin{equation*}
    \bar{v}_{N,i} = \frac{1}{M} \sum_{m=1}^{M} \bar{v}_{N,i}^{m}.
\end{equation*}
To quantify statistical stability, we compute the empirical standard deviation
\begin{equation*}
    s_{v,i} = 
    \sqrt{
        \frac{1}{M-1}
        \sum_{m=1}^{M}
        \left(
            \bar{v}_{N,i}^{m} - \bar{v}_{N,i}
        \right)^2
    }.
\end{equation*}
Then, for sufficiently large $N$, $\left(\bar{v}_{N,i}^m\right)_{m=1,\dots,M}$ is normally distributed and thus, the 95\% confidence interval half-width is
\begin{equation*}
    \mathrm{CI}_{i}
    =
    \phi(0.975)
    \frac{s_{v,i}}{\sqrt{M}},
\end{equation*}
where $\phi(\cdot)$ denotes the quantile function of the standard normal distribution. To ensure comparability across heterogeneous instances, we consider a scenario count $N$ as sufficient if the standard deviation and confidence interval are sufficiently small, i.e.,
\begin{equation}
    \max_{i} s_{v,i} \leq max\{0.05\cdot {\bar{v}_{N,i}}, 0.5\}
    \quad \text{and} \quad
    \max_{i} \mathrm{CI}_{i} \leq \max\{0.05 \cdot {\bar{v}_{N,i}}, 0.5\}. \label{saa_criterion}
\end{equation}
holds. Note that we utilize a hybrid criterion because the test instance set's objective functions range between 0 and 250. Thus, using only a percentual criterion (first term of the right-hand sides of (\ref{saa_criterion})) or only an absolute criterion (second term of (\ref{saa_criterion})) can lead to false-positive rejections of truly sufficiently large scenario counts. We then increase $N$ sequentially in steps of size $50$ until both criteria are satisfied. 
This condition was first met for $N=500$, which we therefore adopt in all subsequent computational experiments.